\documentclass[jcp, showkeys, twocolumn, floatfix, superscriptaddress, reprint, 10pt]{revtex4-1}
\usepackage[utf8]{inputenc}
\usepackage{graphicx}
\usepackage{hyperref}
\usepackage[version=4]{mhchem}
\ifdefined\DIRACREP
\else
\newcommand{\DIRACREP}{}

\usepackage{physics}
\usepackage{xparse}
\ifdefined\COSMOMATHS
\else
\newcommand{\COSMOMATHS}{}

\newcommand{\mbf}[1]{\ensuremath{\mathbf{#1}}}

\fi %

\NewDocumentCommand{\rep}{s d<| d|>}{%
\IfBooleanTF{#1}{
   \IfValueTF{#2}{
       \IfValueTF{#3}{\braket{#2}{#3}}{\bra{#2}}
       }{
       \IfValueTF{#3}{\ket{#3}}{}
       }
   }{
   \IfValueTF{#2}{
       \IfValueTF{#3}{\braket*{#2}{#3}}{\bra*{#2}}
       }{
       \IfValueTF{#3}{\ket*{#3}}{}
       }
   }
}

\NewDocumentCommand{\rbra}{sm}{\IfBooleanTF{#1}{\rep*<#2|}{\rep<#2|}}
\NewDocumentCommand{\rket}{sm}{\IfBooleanTF{#1}{\rep*|#2>}{\rep|#2>}}
\NewDocumentCommand{\rbraket}{smom}{
    \IfBooleanTF{#1}{
        \IfNoValueTF{#3}{\rep*<#2||#4>}{\rep*<#2|#3\rep*|#4>}
    }{
        \IfNoValueTF{#3}{\rep<#2||#4>}{\rep<#2|#3\rep|#4>}
    }
}

\NewDocumentCommand{\field}{o m e{_} e{^} o e{_} e{^}}{
\IfValueTF{#5}{\overline{
  #2\IfValueT{#3}{_#3}\IfValueT{#4}{^{\otimes #4}} %
  \otimes
  #5\IfValueT{#6}{_#6}\IfValueT{#7}{^{\otimes #7}} %
  \IfValueT{#1}{;#1}
}}{
  \IfValueTF{#4}{\overline{
     #2\IfValueT{#3}{_#3}\IfValueT{#4}{^{\otimes #4}}
     \IfValueT{#1}{;#1}
  }}
  {#2\IfValueT{#3}{_#3}}
}
}

\NewDocumentCommand{\frho}{o e{_} e{^}}{
\field[#1]{\rho}_{#2}^{#3}
}

\newcommand{\bx}{\mbf{x}}

\newcommand{\e}{a}  %

\NewDocumentCommand{\ex}{e_}{
\IfValueTF{#1}{\e_{#1}\bx_{#1}}{\e\bx}
}  %

\NewDocumentCommand{\lm}{e_}{
\IfValueTF{#1}{l_{#1}m_{#1}}{lm}
}
\NewDocumentCommand{\nlm}{e_}{
\IfValueTF{#1}{n_{#1}\lm_{#1}}{n\lm}
}
\NewDocumentCommand{\enlm}{e_}{
\IfValueTF{#1}{\e_{#1}\nlm_{#1}}{\e\nlm}
}
\NewDocumentCommand{\en}{e_}{
\IfValueTF{#1}{\e_{#1}n_{#1}}{\e n}
}
\NewDocumentCommand{\nlk}{e_}{
\IfValueTF{#1}{n_{#1}l_{#1}k_{#1}}{nlk}
}
\NewDocumentCommand{\enlk}{e_}{
\IfValueTF{#1}{\e_{#1}\nlk_{#1}}{\e\nlk}
}
\NewDocumentCommand{\enl}{e_}{
\IfValueTF{#1}{\en_{#1}l_#1}{\en l}
}

\NewDocumentCommand{\nnl}{s}{
\IfBooleanTF{#1}{n_1 n_2 l}{n_1; n_2; l}
}
\NewDocumentCommand{\ennl}{s}{
\IfBooleanTF{#1}{\en_1 \en_2 l}{\en_1; \en_2; l}
}

\NewDocumentCommand{\gslm}{s}{
\IfBooleanTF{#1}{\sigma\lambda\mu}{\sigma;\lambda\mu}
}

\fi %

\begin{document}

\title{Beyond potentials: integrated machine-learning models for materials }
\author{Michele Ceriotti}
\affiliation{Laboratory of Computational Science and Modeling, Institute of Materials, \'Ecole Polytechnique F\'ed\'erale de Lausanne, 1015 Lausanne, Switzerland}
\affiliation{National Centre for Computational Design and Discovery of Novel Materials (MARVEL), {\'E}cole Polytechnique F{\'e}d{\'e}rale de Lausanne, 1015 Lausanne, Switzerland}

\date{March 2022}

\begin{abstract}
Over the past decade inter-atomic potentials based on machine-learning (ML) techniques have become an indispensable tool in the atomic-scale modeling of materials. 
Trained on energies and forces obtained from electronic-structure calculations, they inherit their predictive accuracy, and extend greatly the length and time scales that are accessible to explicit atomistic simulations. 
Inexpensive predictions of the energetics of individual configurations have facilitated greatly the calculation of the thermodynamics of materials, including finite-temperature effects and disorder. 
More recently, machine-learning models have been closing the gap with first-principles calculations in another area: the prediction of arbitrarily complicated functional properties, from vibrational and optical spectroscopies to electronic excitations. 
The implementation of \emph{integrated} machine-learning models, that combine energetic and functional predictions with statistical and dynamical sampling of atomic-scale properties is bringing the promise of predictive, uncompromising simulations of existing and novel materials closer to its full realisation. 
\end{abstract}

\keywords{modeling, machine learning, simulation, electronic structure, statistical methods}
\setcitestyle{super}

\maketitle

\section{Predictive materials modeling}

Materials science as a discipline has benefited tremendously from the possibility of solving numerically the electronic-structure problem for a condensed phase of matter. Supported by the steady decrease of the cost of computation, the availability of high-quality open-source software\cite{sher+20jcp} and by a better understanding of the approximations that are needed to make the many-body quantum problem tractable\cite{onid+02rmp,burk12jcp}, computational discovery of materials has become a reality, pursued by high-throughput calculations\cite{cald+15cms,moun+18nn} and made available through public databases\cite{jain+13aplm,drax-sche18mrsb,tarl+20sd}.
The vast majority of these high-throughput calculations, however, describe materials in a highly-idealized fashion, as ideal crystalline structures with a static configuration of their atoms. This is a severe limitation: the properties of real materials can be strongly influenced by the presence of defects, or just the disorder that is induced by thermal fluctuations, as well as by the quantum mechanical nature of the nuclei.\cite{ceri+16cr} 
Including these effects, however, is computationally demanding: rather than evaluating the properties of a single structure, hundreds of thousands of calculations must be performed to collect representative configurations of the distorted configurations that are seen at finite temperature.
Following classical works that pioneered ``ab initio'' molecular dynamics simulations~\cite{car-parr85prl,marx+99nat}, several groups have been working to streamline, and minimize the cost, of evaluating the finite-temperature thermodynamics of materials, for metals (where the work of Neugebauer and collaborators is particularly noteworthy\cite{grab+09prb,frey+14rmp}) as well as for hydrogen-bonded systems\cite{li+11pnas} and molecular materials\cite{reil-tkat14prl,ross+16prl,ko+18prm,raim+19prm}. 

\begin{figure}[btp]
    \centering
    \includegraphics[width=1.0\linewidth]{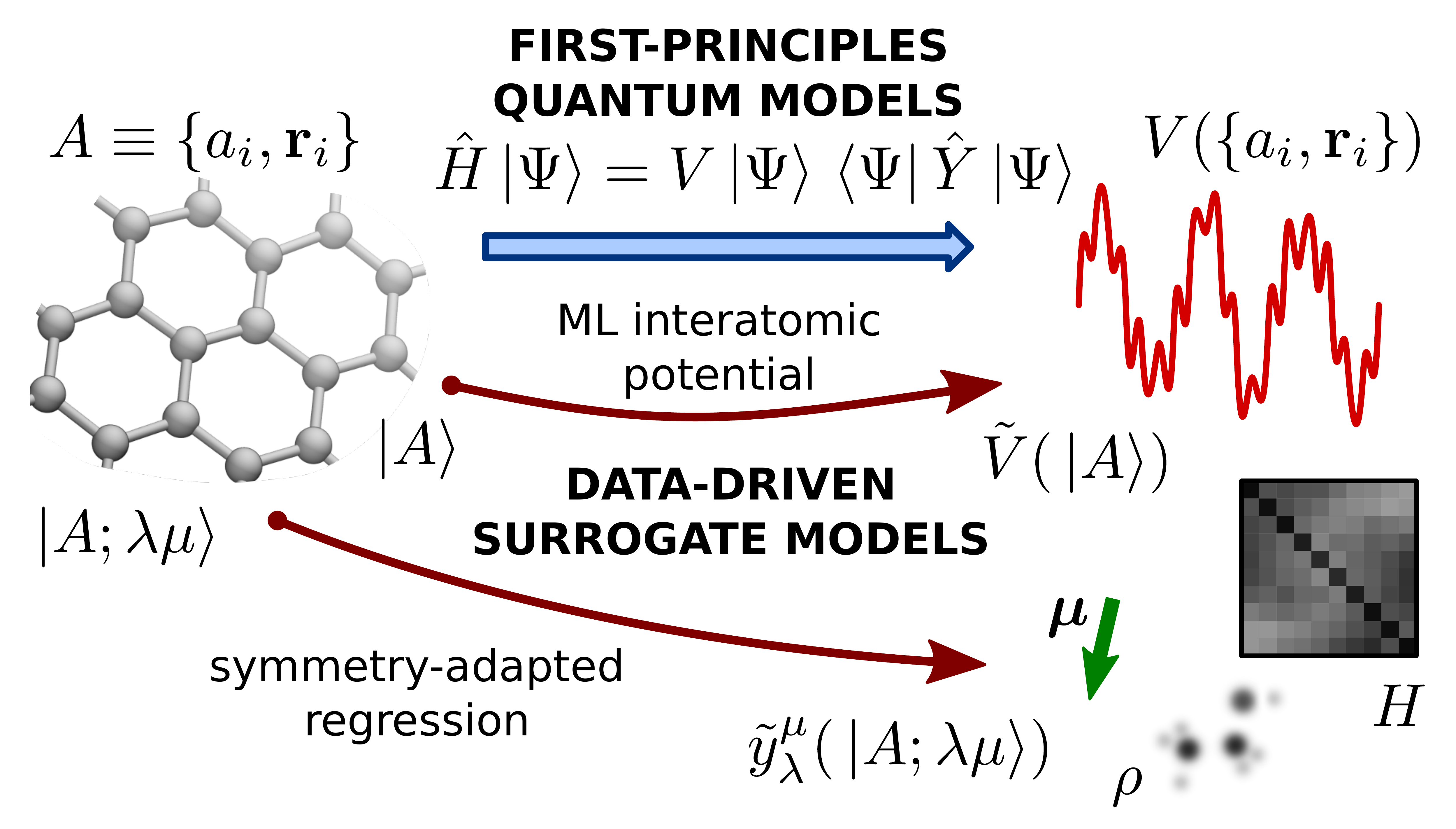}
    \caption{Predictive modeling of atomic-scale properties. Quantum calculations provide accurate, but computationally-demanding, predictions of the ground-state properties of any structure $A$, which is defined by the nature $a_i$ and coordinates $\mathbf{r}$ of each atom. 
    Data-driven surrogate models allow interpolating between a small number of reference calculations, e.g. to estimate the interatomic potential $V$. Invariant $\rep|A>$ (or equivariant $\rep|A;\lambda\mu>$, that transform like spherical harmonics $Y^\mu_\lambda$ under rotation) structural descriptors can be used to evaluate scalar (or tensorial) properties with a much reduced effort.  This includes dipole moments $\boldsymbol{\mu}$, scalar fields such as the electron density $\rho(\mathbf{x})$, or the Hamiltonian matrix $H$.}
    \label{fig:overview}
\end{figure}

The past decade has seen the rise of a different approach to the problem. Rather than solving directly a quantum mechanical problem to predict the properties $y$ of a structure $A$, the idea is to use a small number of reference calculations to fit a more or less sophisticated functional form $\tilde{y}(A)$, that can then be used to achieve predictions that are (almost) as accurate as the underlying quantum calculations, but computationally much less demanding, and with a more benign scaling with system size (Fig.~\ref{fig:overview}). 
The idea was not new: quantum chemists have been constructing potential energy functions for small molecules by interpolating between few high-end quantum calculations\cite{part-schw97jcp,huan+05jcp}, and materials scientists have for decades fitted cluster expansion expressions to compute the thermodynamics of alloys\cite{sanc+84pa}. Behler and Parrinello\cite{behl-parr07prl} - and later Bart\'ok, Cs\'anyi and collaborators\cite{bart+10prl} - demonstrated how these ideas could be extended to condensed phases and off-lattice configurations, laying the foundations for the flourishing of machine-learned interatomic potentials (MLIPs). 
The interested reader may find an historical overview, that covers also some of the latest developments, in a recent review by Behler\cite{behl21cr}. 
The essential ingredients that underlie the success of MLIPs, and that in particular are incorporated in both the structural descriptors\cite{musi+21cr,lang+22npjcm} and the models, will be familiar to many computational materials scientists: A locality ansatz that decomposes the energy of a material in a hierarchy of atom-centered, body-ordered terms -- pair, three body, many body potentials\cite{glie+18prb} that add up to give the energetic contribution from an atom and its environment; the incorporation of physical principles such as the invariance of the energy under symmetry operations\cite{behl11jcp,bart+13prb};  the construction of reference data sets that cover the range of configurations and compositions that are relevant for the problem at hand. 
Similar considerations have been incorporated in ML models of molecules\cite{rupp+12prl,manz-carr21cr}:  even though (much as for electronic structure calculations) chemical and materials modeling often use a different language, and the technical details may differ, there is substantial unity in the way machine learning has been incorporated into atomistic modeling.\cite{bart+17sa, butl+18nature} 

\begin{figure*}[tbp]
    \centering
    \includegraphics[width=1.0\linewidth]{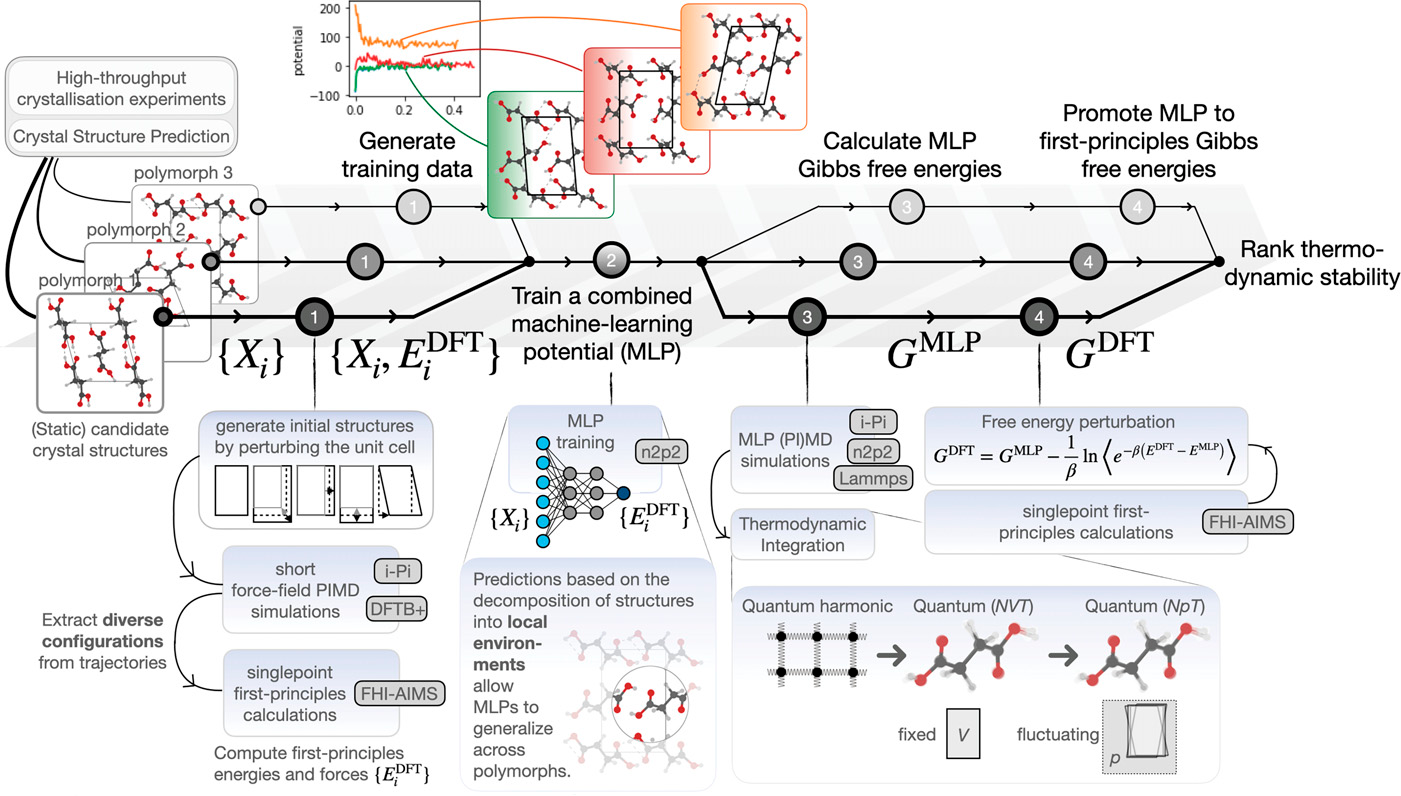}
    \caption{Schematic representation of a possible ML-based workflow to compute quantum anharmonic free-energies of different phases of molecular crystals. Original figure published in Ref.~\citenum{kapi-enge22pnas} under the CC-BY 4.0 license (\url{https://creativecommons.org/licenses/by/4.0/}). }
    \label{fig:molcryst}
\end{figure*}

\section{First-principles thermodynamics made easy}

Sampling of disorder, and thermal fluctuations, is the perfect application for data-driven potentials. Computing finite-temperature properties involves generating a large number $M$ of configurations $A$, consistent with the thermodynamic ensemble, and averaging over their properties, $\langle y \rangle =\sum_{A=1}^M y(A)/M$. 
Even for liquid configurations, the fluctuations involve comparatively small structural differences, which makes it relatively easy for interpolative techniques such as MLIPs to achieve accurate predictions from a small number of reference configurations. 
It is not by chance that many of the early applications focused on the calculation of phase diagrams or phase transitions\cite{khal+10prb,eshe+12prl} as well as finite-temperature thermodynamics of solid phases\cite{drag+18prm} and more traditional studies of extended lattice defects\cite{szla+14prb}.
From water\cite{chen+19pnas,chen+21np} to oxides\cite{wall+21prm,wall+21prr,lee+22jcp}, from metals\cite{mish21am,rose+21npjcm} and to amorphous materials\cite{nong+18jcp,caro+18cm,deri+20nc}, from molecular materials\cite{kapi-enge22pnas} to hybrid perovskites\cite{jinn+19prl}, there is virtually no class of materials for which MLIP have not been successfully used to facilitate the study of materials in disordered, or finite-temperature, conditions. 
It is worth again to note that also in molecular applications the focus has shifted from the prediction of the properties of minimum-energy geometries\cite{rupp+12prl,mont+13njp,rama+14sd} to flexible potentials\cite{smit+17cs,hoja+21sd}, the transition states of chemical reactions\cite{jack+21cs}, and the modeling of the quantum mechanical fluctuations of the nuclei\cite{chmi+18nc}.  

MLIPs have reached a considerable level of maturity, and so there is a good understanding of their mathematical foundations,\cite{musi+21cr} their efficient implementation\cite{zuo+20jpcl} and of the best practices connected with the construction of the training set to obtain a robust, transferable model\cite{rowe+20jcp}. 
What is more, several pragmatic solutions have emerged to solve their limitations, to enable applications to highly non-trivial problems. 
An obvious strategy to tackle limitations in stability and accuracy of MLIPs is not to use them alone, but in tandem with traditional materials modeling techniques. 
Rather than training a model directly to high-end quantum calculations, one could use an affordable semi-empirical method such as density-functional tight binding (DFTB)\cite{elst+98prb} as a baseline, that is corrected by a MLIP that evaluates only the difference to a higher level of theory.\cite{rama+15jctc,sun-saut19jctc} 
This can be combined with all sorts of well-established tricks, such as multiple-time stepping\cite{tuck+92jcp,kapi+16jcp}, to reduce the cost of the baseline calculations.  Ref.~\citenum{ross+20jctc}, that discusses calculations of solvation and acid-base equilibra that are relevant for regioselective homogeneous catalysis, demonstrates several of the techniques that simplify the description of a challenging, reactive system using a MLIP, including also subtle effects such as the quantum mechanical fluctuations of the nuclei. 
Another way to combine data-driven and traditional calculations involves correcting the free energies computed using a MLIP to promote them to explicit first-principles accuracy (Fig.~\ref{fig:molcryst}). This technique, that has also been used in the past to reduce the cost of direct first-principles thermodynamics\cite{grab+09prb}, involves selecting a few hundreds uncorrelated configurations $\{A\}$ from an ML-driven calculation, computing the reference and ML potentials  ($V_\text{REF}$ and $V_\text{ML}$) for them, and evaluating the free-energy perturbation correction
\begin{equation}
\Delta_{\text{ML}\rightarrow \text{REF}} = k_BT \ln M^{-1}\sum_{A=1}^M e^{-(V_\text{REF}(A)-V_\text{ML}(A))/k_B T}.
\end{equation}
This expression accounts for the possible discrepancy between $V_\text{REF}$ and $V_\text{ML}$. 
Applying the correction to each phase allows to recover relative free-energy differences which are not affected by the residual error of the ML approximation, which can be important in cases where free-energy differences are small, or dependent on energy contributions that are hard to capture by local MLIPs (e.g. proton disorder in ice\cite{chen+19pnas}). The computational cost is a tiny fraction of that which would be required for a direct first-principles evaluation of finite-temperature thermodynamics. 

Another idea that has facilitated greatly the use of ML for materials modeling is that of \emph{active learning}. \cite{li+15prl}  A preliminary version of a MLIP is used to perform molecular dynamics, or any kind of atomistic simulations that requires repeated evaluation of interatomic forces. 
Every time a configuration is encountered that departs too much from those included in the training set, it is recomputed with the reference first-principles method, and the MLIP updated (inline, or in batches) to generate a more reliable and transferable data-driven model.\cite{guba+18jcp,zhan+19prm,shap+20book,jinn+19prl}
Efficient active learning requires a way to assess the accuracy of ML predictions: some models, such as those based on Gaussian process regression,\cite{deri+21cr} provide built-in uncertainty quantification. A strategy to estimate prediction errors that is simple, inexpensive, and can be applied to any regression scheme involves constructing an ensemble of models,\cite{brei96ml} whose mean provides the best estimate of the target property, while the (calibrated) standard deviation gives a measure of uncertainty.\cite{musi+19jctc}
The ensemble of predictions can easily be used to perform uncertainty propagation and estimate the error on derived quantities, and even the error associated with the sampling of configurations performed using the mean ensemble potential\cite{imba+21jcp}. 

\section{More than energy and forces}

Even though MLIPs are widely used, and have transformed the landscape of atomic-scale materials modelling, they only replace one of the tasks that have made first-principles simulations successful. 
Having access to the ground-state (and excited state) electronic structure of a material allows evaluating all sorts of response properties\cite{baro+01rmp}, which has made it possible to compute Raman and infra-red spectra\cite{pagl+08jcp}, nuclear magnetic shieldings\cite{yate+07prb}, ferroelectricity\cite{king-vand94prb}, and much more. 
To realize the full potential of ML, all these functional properties should be as easily accessible as the interatomic potential and forces. This, however, poses additional challenges.
To begin with, there is much less empirical experience about what works and what does not work to build an effective ML model. We roughly know what range of interactions, and what parameters of the most widespread frameworks, can be used as a reliable initial guess to build a MLIP for a metal, or a molecular solid, but literature is much sparser when it comes to predicting e.g. polarizability. 
Furthermore, the algebraic structure of the target property might differ from that of the potential energy, that is a scalar and is only defined as a global, extensive property. Scalar NMR chemical shieldings,\cite{paru+18ncomm,liu+19jpcl} for instance, are atom-centered properties: even though this is may appear as a small issue, given that most MLIPs construct the total potential as a sum of atom-centered terms, the need to predict individual atomic contributions may exacerbate some of the shortcomings of the most common descriptors of local atomic structure\cite{pozd+20prl}.  
The electron density of states $g(E)$ is another global property that can be decomposed in atom-centered contributions, but each atomic environment contributes an energy-dependent term $g_i(E)$, which begs the additional question of what is the most efficient way to encode the energy dependency in the ML targets.\cite{benm+20prb}

\begin{figure}[btp]
    \centering
    \includegraphics[width=1.00\linewidth]{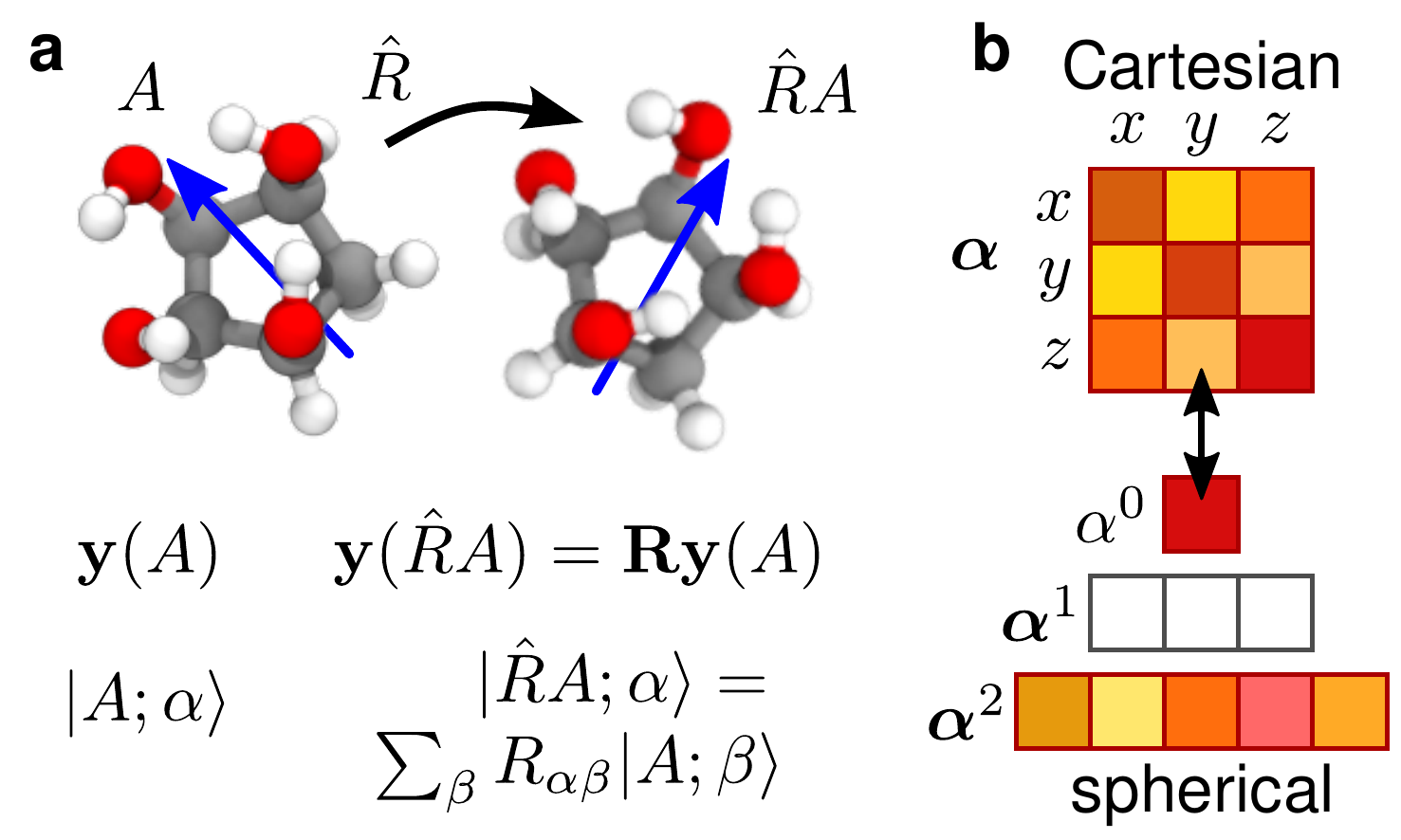}
    \caption{a) Schematic overview of the transformations of a molecular vector property under the action of a rigid rotation. The property and any symmetry-adapted model must transform in a prescribed way. Similar equivariance rules must be obeyed by symmetry-adapted descriptors.
    b) A symmetric Cartesian tensor (e.g. the polarizability $\boldsymbol{\alpha}$) can be more conveniently written in an irreducible spherical form, with a scalar component $\alpha^0$, and 5 components that transform like $l=2$ spherical harmonics. The $l=1$ component is associated with the asymmetric part of the tensor, and is therefore zero. }
    \label{fig:sa-models}
\end{figure}

A more substantial problem is associated with the prediction of properties that are not scalars, but have a  vectorial/tensorial geometric nature. To see why this is an issue, one should consider that one of the fundamental physical priors that are incorporated in the vast majority of MLIPs is the notion that energy is invariant to rigid rotations $\hat{R}$  of a structure $A$, as well as to permutation of the atom label $\hat\Pi$. 
Thus, a model $\tilde{y}$ predicting scalar properties should fulfill the invariance condition $\tilde{y}(\hat{R}\hat{\Pi} A) =\tilde{y}(A)$. This is achieved easily by ensuring that the \emph{descriptors} that are used to represent the atomic structures are themselves invariant to rotations\cite{behl11jcp}, so that any model that uses them as inputs inherits this property.\footnote{Strictly speaking, atom-centered descriptors are equivariant to permutations of atoms labels, and so are the predictions of atom-centered models of energy contributions. The \emph{sum} of these terms, however, is permutation invariant. } 
Consider instead a property that is covariant to rotations, e.g. the dipole moment of a molecule (Fig.~\ref{fig:sa-models}a). A ML model that is consistent with basic physical principles should ensure that $\tilde{y}_\alpha(\hat{R} A) = \sum_\beta R_{\alpha\beta}\tilde{y}_\beta(A)$, where $\alpha,\beta$ refer to the Cartesian components, and $R_{\alpha\beta}$ is the rotation matrix associated with $\hat{R}$.\cite{bere+15jctc,lian+17prb,glie+17prb}
The need to build symmetry-adapted models has driven the development of \emph{equivariant} machine learning schemes, that incorporate the appropriate (covariant or invariant) behavior depending on the nature of the target.\cite{gris+18prl}
Appreciating the subtleties involved requires a certain degree of mathematical sophistication. Ref.~\citenum{bron+21arxiv} provides an excellent and comparatively accessible overview from the more general perspective of geometric ML. In a nutshell, symmetry operations on tensors are most efficiently described in terms of \emph{irreducible representations} -- combinations of the Cartesian components that transform linearly into each other when the object they refer to is transformed. As a concrete example, the trace of the polarizability tensor, $\alpha_{xx}+\alpha_{yy}+\alpha_{zz}$ is invariant to rotations, and so it is easier to manipulate than the individual diagonal components (Fig.~\ref{fig:sa-models}b).  
For the case of proper rotations (corresponding to the SO(3) group), the irreducible representations correspond to spherical harmonics, and rotating a structure mixes the components $\mu$ of each angular momentum value $\lambda$ according to 
\begin{equation}
\tilde{y}^\mu_\lambda(\hat{R} A) = \sum_{\mu'} D^\lambda_{\mu\mu'}(\hat{R}) \tilde{y}^{\mu'}_\lambda(A),
\label{eq:rotation-lambdamu}
\end{equation}
where $\mathbf{D}^\lambda$ is known as a Wigner D-matrix.\cite{morri-park87ajp}

One way to obtain symmetry-adapted models is to construct them based on equivariant descriptors, which themselves transform as spherical harmonics. To understand the basic idea, consider a linear model to predict ${y}^\mu_\lambda(A)$
\begin{equation}
    \tilde{y}^\mu_\lambda(A) = \sum_q \rep<y||q>\rep<q||A; \lambda\mu>.
\end{equation}
In this notation (Ref.~\citenum{musi+21cr}, Section 3.1) $\rep<y||q>$ indicates the regression weights for the $q$-th feature, and $\rep<q||A; \lambda\mu>$ the $q$-th component of the feature vector that describes the structure $A$. 
Crucially, for each feature there are $2\lambda+1$ components, indexed by $\mu$ that describe their behavior under rotation. It is easy to see that this linear model will be consistent with the requirements of Eq.~\eqref{eq:rotation-lambdamu} as long as 
\begin{equation}
\rep<q||\hat{R}A; \lambda\mu>  = \sum_{\mu'} D^\lambda_{\mu\mu'}(\hat{R})\rep<q||A; \lambda\mu'> 
\end{equation}
holds true for each feature. 

\begin{figure}
    \centering
    \includegraphics[width=1.0 \linewidth]{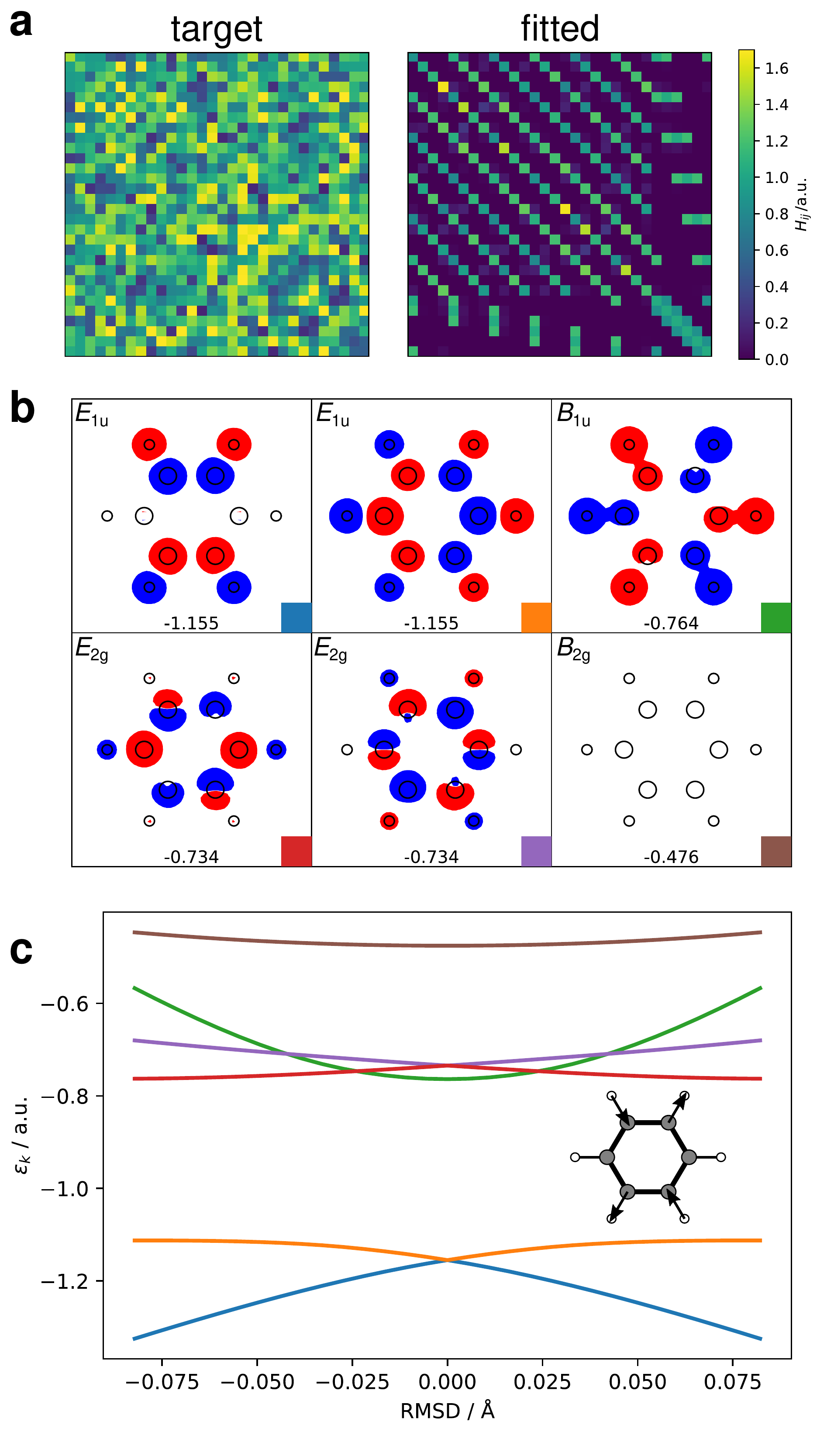}
    \caption{Symmetry-adapted modeling of the single-particle Hamiltonian of a minimal-basis model of benzene. The model enforces the correct molecular symmetries, which means that  (a) even training on a randomized Hamiltonian yields a prediction that is consistent with $D_{6h}$ point group symmetries; (b) the symmetry and degeneracies of the eigenstates are consistent with the character tables of the group; (c) the way degeneracies are split by a deformation is also consistent with molecular orbital theory. Reprinted from Ref.~\citenum{niga+22jcp}, with the permission of AIP Publishing. }
    \label{fig:benzene}
\end{figure}
These equivariant representations have been used to machine-learn force vectors\cite{glie+17prb}, molecular dipoles\cite{veit+20jcp} and polarizabilities\cite{wilk+19pnas}, an atom-based expansion of the electron density\cite{gris+19acscs,lewi+21jctc} and of the single-particle electronic Hamiltonian\cite{niga+22jcp}.
The advantage of using a symmetry-adapted model are demonstrated in Fig.~\ref{fig:benzene} for the learning of a molecular Hamiltonian: incorporating rotation and permutation equivariance ensures that point-group symmetries, when present, are automatically accounted for.  

Symmetry considerations underlie also the latest developments in geometric machine learning, with equivariant  models\cite{ande+19nips,thom+18arxiv,ande+19nips,simo+21arxiv,klic+21arxiv}  showing greatly improved performance compared to their invariant counterparts\cite{gilm+17icml}.  In fact, it is possible to treat within the same formalism shallow models based on atom-centered features and the most recent families of equivariant neural networks\cite{niga+22jcp2} - a unification that may expedite the systematic optimization of the model architecture. 
Furthermore, the use of equivariant representations or networks is not the only way to obtain a symmetry-adapted model.
For relatively rigid molecules, one can simply learn the components of the tensors in the local reference frame.\cite{bere+15jctc,lian+17prb} 
Another strategy involves predicting atomic partial charges, combining them with atomic positions to obtain a vectorial prediction for properties such as dipole moments\cite{gast+17cs,unke-meuw19jctc,veit+20jcp}. Computing dipoles by predicting the position of localized centers of charge\cite{zhan+20prb}, or constructing covariant descriptors by formally assigning fixed charges to the atoms and computing ``operator derivatives'' of atom-centered descriptors\cite{chri+19jcp} with respect to a fictitious electric field, are equivalent to atomic-dipole schemes, despite the superficial similarity with models based on the prediction of atomic charges. 
Alternative approaches are also available to predict electronic properties: the charge density can be predicted as a scalar field by using descriptors centered on grid points, rather than on atoms\cite{alre+18cst,chan+19npjcm}. The single-particle energy levels can be predicted by modeling an \emph{invariant} effective Hamiltonian\cite{west-maur21cs}.

It is also worth noting that equivariant representations are not only useful to build symmetry-adapted models of tensorial properties. The vast majority of atom-centered descriptors used to build MLIPs can be understood in terms of symmetrized correlations of the atomic neighbor density\cite{will+19jcp}, which can be constructed iteratively\cite{niga+20jcp} by projecting tensor products of the one-neighbor density onto the irreducible representations of the rotation group. 
The invariant part of these high-order descriptors provides a complete \emph{linear} basis to expand scalar atomic properties, which is used in the construction of the moment tensor potentials (MTP)\cite{shap16mms} and the atomic cluster expansion (ACE)\cite{drau19prb} -- two classes of MLIPs with an excellent accuracy/cost ratio. 
Equivariant NNs are often used to predict invariant targets, even though they obviously are ideally suited to predict tensorial properties, and recent results on single particle Hamiltonians have demonstrated remarkable performance for these tasks\cite{unke+21nips}. 

\section{Integrated machine-learning models}

\begin{figure}
    \centering
    \includegraphics[width=0.9\linewidth]{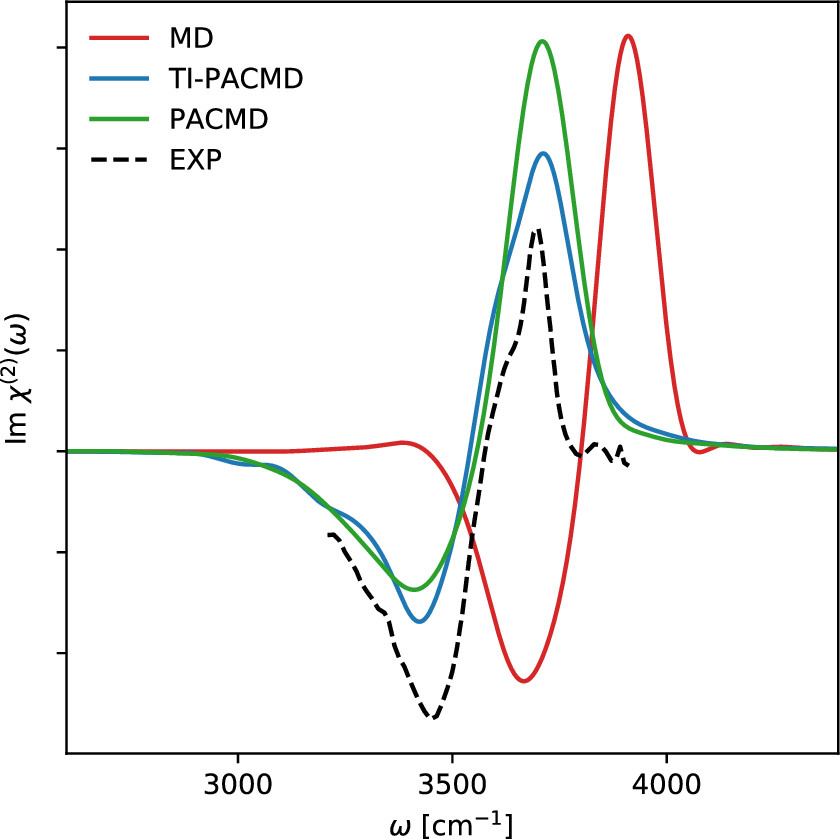}
    \caption{Sum-frequency generation spectrum for a liquid water-vacuum interface, built using an integrated model describing interatomic potentials, dipole and polarizability with ML. The curves compare experimental results\cite{adhi+15jcp} (dashed line) with two types of quantum dynamics (green, blue) and with classical MD (red) which is significantly blue-shifted. Reprinted with permission
from Ref.~\citenum{shep+21jcp}. Copyright 2021 American Chemical Society. }
    \label{fig:sfg-water}
\end{figure}

Having simultaneous access to ML predictions of energy, forces and functional properties opens the way to simulations that describe all the aspects of the electronic structure of materials, and that use state-of-the-art sampling techniques to access static and dynamic behavior at finite temperature, without compromising on the cell size or trajectory length. Advanced data analytics can then be used to interpret the outcome of increasingly complex datasets and simulations.\cite{isay+15cm,puli+17nature,glie+21cr,cers+21mlst}
For example, it is now possible to compute vibrational spectroscopies of water as the Fourier transform of the correlations between dielectric response functions, using a ML potential to drive the dynamics and to predict dipole moments and polarizability\cite{gast+17cs,kapi+20jcp,somm+20pccp,zhan+20prb}. 
Similarly, one can model with first-principles accuracy thermodynamic and dielectric properties of the archetypal ferroelectric perovskite \ce{BaTiO3}, without limitations on simulation size or duration\cite{gigli2021arxiv}. 
The fact that the models are computationally affordable means that it is possible to go beyond a purely classical description of the nuclei, including an approximate description of nuclear quantum dynamics\cite{kapi+20jcp,shep+21jcp}.
The quantum mechanical nature of the atomic nuclei, that is an important contributor to the fluctuations of light atoms such as hydrogen, has a substantial effect on many observables (Fig.~\ref{fig:sfg-water}). The availability of integrated ML models is making it easy to investigate them, as in a recent study that showed how much they contribute to the nuclear chemical shifts in molecular crystals\cite{enge+21jpcl}. 

Another area in which ML needs to go beyond potentials to describe materials involves all properties associated with the electronic degrees of freedom. This involves both predictions of the conductivity (at least through the proxy of the density of electronic states at the Fermi level\cite{deri+21nature}) as well as the estimation of the contribution to thermophysical properties arising from electronic excitations,\cite{zhan+20jcp,elli+21prb,benmahmoud2022arxiv} that is relevant for matter at extreme conditions, but also for refractory metals at high temperature\cite{grab+09prb,gong+18prb,lopa+21prm}. 
Magnetic states and magnetic excitations can also be treated with machine-learning potentials that take into account the spin degrees of freedom.\cite{novi+22npjcm}
Potentials that describe electronic excited state rather than the ground state are making it possible to perform non-adiabatic molecular dynamics\cite{dral+18jpcl,west-marq20mlst}, extending further the range of applicability of ML beyond ground-state properties. 

An important practical issue that may slow down the adoption of this new class of integrated models has to do with the need to combine different frameworks, and different types of advanced simulation strategies.  Even though general-purpose models that treat on the same footings potential energy and functional properties will become prevalent, there may still be cases in which one wants to combine models from different sources, and/or to implement complicated simulation setups that require their own dedicated software stack. 
The design of modular software that can be easily interfaced with different packages, such as the atomic simulation environment (ASE)\cite{hjor+17jpcm}, i-PI\cite{kapi+19cpc}, as well as free-energy sampling plugins PLUMED\cite{trib+14cpc} or SSAGES\cite{sidk+18jcp} provide the necessary ``glue'' to combine MLIPs, equivariant property models, and accelerated sampling with thermal and quantum fluctuations.

\section{Conclusions}

Machine-learning techniques have been incorporated, and largely normalized, as essential tools for the atomic-scale modeling of matter. The progress in the field has been incredibly fast, and shows no sign of slowing down, with new ideas being proposed to tackle the limitations of the best-established approaches, from the incompleteness of some descriptors\cite{pozd+20prl,zhan+21prl} to the lack of long-range interactions, that are now being included with point-charge and static multipole models\cite{artr+11prb,bere+18jcp}, self-consistent charge equilibration schemes \cite{ghas+15prb,nibl+21jcp,gao-rems22nc}, as well as with bespoke descriptors that reproduce the asymptotics of long-range physics\cite{gris-ceri19jcp,gris+21cs}. 
Still, some of the core concepts have begun to crystallize. The mathematical foundations of the kind of descriptors that are needed to encode atomic composition and geometry are now better understood, and clear similarities have been revealed between different approaches.\cite{will+19jcp,drau19prb,niga+22jcp2} 

Interatomic potentials, the first quantity that was directly targeted by data-driven surrogate models, have reached a certain maturity, and have made the calculation of thermodynamic properties of materials from first principles -- once requiring careful planning and large investments of computer time -- relatively routine. 
Thanks to the extension of symmetry-adapted models to vectorial and tensorial quantities, other microscopic properties that can be computed by solving the electronic-structure problem can also be predicted by ML. 
The same kind of uncompromising accuracy that has become possible for the prediction of the stability of materials may soon be available for functional properties and all sorts of experimental observables, through integrated machine-learning schemes that combine models of energy and electronic properties with advanced statistical sampling. 
Still, these developments are unlikely to mark the end of an era for first-principles simulations. On the contrary, more accurate electronic-structure methods will become necessary to bring simulations even closer to reality, and data-driven techniques will serve as an accelerator, to translate results on benchmark systems into applications to large-scale simulations.  
What is more, one can see the first signs of the emergence of hybrid modeling techniques, that combine electronic-structure calculations with data-driven approaches -- from the superficial use of semi-empirical models as a baseline\cite{rama+15jctc,ross+20jctc}, to the use of electronic-structure quantities as descriptors\cite{qiao+20jcp,fabr+22dd}, to the prediction of electronic density\cite{broc+17nc} as the first step in the calculation of standard, and soon data-driven,\cite{kali+21acr,kirk+21science} density functionals.
Bridging the divide between physics and data-driven modelling, deductive and inductive paradigms of scientific inquiry, hybrid frameworks indicate the direction to follow to complete the integration of artificial intelligence into the theory and practice of computational materials science. 

\begin{acknowledgments}
MC acknowledges funding from the European Research Council (ERC) under the European Union’s Horizon 2020 research and innovation programme (grant agreement No 101001890-FIAMMA), and the NCCR MARVEL, funded by the Swiss National Science Foundation (grant number 182892). MC would like to thank all the current and former members of the Laboratory of Computational Science and Modeling at EPFL for the substantial contributions given to the formal framework, the software implementation, and the practical application of many of the ideas discussed in this review. 

\end{acknowledgments}

\section{Conflicts of interest}

The author declares there are no conflicts of interest. 

\newcommand{\noopsort}[1]{}


\begin{thebibliography}{145}%
\makeatletter
\providecommand \@ifxundefined [1]{%
 \@ifx{#1\undefined}
}%
\providecommand \@ifnum [1]{%
 \ifnum #1\expandafter \@firstoftwo
 \else \expandafter \@secondoftwo
 \fi
}%
\providecommand \@ifx [1]{%
 \ifx #1\expandafter \@firstoftwo
 \else \expandafter \@secondoftwo
 \fi
}%
\providecommand \natexlab [1]{#1}%
\providecommand \enquote  [1]{``#1''}%
\providecommand \bibnamefont  [1]{#1}%
\providecommand \bibfnamefont [1]{#1}%
\providecommand \citenamefont [1]{#1}%
\providecommand \href@noop [0]{\@secondoftwo}%
\providecommand \href [0]{\begingroup \@sanitize@url \@href}%
\providecommand \@href[1]{\@@startlink{#1}\@@href}%
\providecommand \@@href[1]{\endgroup#1\@@endlink}%
\providecommand \@sanitize@url [0]{\catcode `\\12\catcode `\$12\catcode
  `\&12\catcode `\#12\catcode `\^12\catcode `\_12\catcode `\%12\relax}%
\providecommand \@@startlink[1]{}%
\providecommand \@@endlink[0]{}%
\providecommand \url  [0]{\begingroup\@sanitize@url \@url }%
\providecommand \@url [1]{\endgroup\@href {#1}{\urlprefix }}%
\providecommand \urlprefix  [0]{URL }%
\providecommand \Eprint [0]{\href }%
\providecommand \doibase [0]{http://dx.doi.org/}%
\providecommand \selectlanguage [0]{\@gobble}%
\providecommand \bibinfo  [0]{\@secondoftwo}%
\providecommand \bibfield  [0]{\@secondoftwo}%
\providecommand \translation [1]{[#1]}%
\providecommand \BibitemOpen [0]{}%
\providecommand \bibitemStop [0]{}%
\providecommand \bibitemNoStop [0]{.\EOS\space}%
\providecommand \EOS [0]{\spacefactor3000\relax}%
\providecommand \BibitemShut  [1]{\csname bibitem#1\endcsname}%
\let\auto@bib@innerbib\@empty
\bibitem [{\citenamefont {Sherrill}\ \emph {et~al.}(2020)\citenamefont
  {Sherrill}, \citenamefont {Manolopoulos}, \citenamefont {Mart{\'i}nez},\ and\
  \citenamefont {Michaelides}}]{sher+20jcp}%
  \BibitemOpen
  \bibfield  {author} {\bibinfo {author} {\bibfnamefont {C.~D.}\ \bibnamefont
  {Sherrill}}, \bibinfo {author} {\bibfnamefont {D.~E.}\ \bibnamefont
  {Manolopoulos}}, \bibinfo {author} {\bibfnamefont {T.~J.}\ \bibnamefont
  {Mart{\'i}nez}}, \ and\ \bibinfo {author} {\bibfnamefont {A.}~\bibnamefont
  {Michaelides}},\ }\href {\doibase 10.1063/5.0023185} {\bibfield  {journal}
  {\bibinfo  {journal} {J. Chem. Phys.}\ }\textbf {\bibinfo {volume} {153}},\
  \bibinfo {pages} {070401} (\bibinfo {year} {2020})}\BibitemShut {NoStop}%
\bibitem [{\citenamefont {Onida}\ \emph {et~al.}(2002)\citenamefont {Onida},
  \citenamefont {Reining},\ and\ \citenamefont {Rubio}}]{onid+02rmp}%
  \BibitemOpen
  \bibfield  {author} {\bibinfo {author} {\bibfnamefont {G.}~\bibnamefont
  {Onida}}, \bibinfo {author} {\bibfnamefont {L.}~\bibnamefont {Reining}}, \
  and\ \bibinfo {author} {\bibfnamefont {A.}~\bibnamefont {Rubio}},\ }\href
  {\doibase 10.1103/RevModPhys.74.601} {\bibfield  {journal} {\bibinfo
  {journal} {Rev. Mod. Phys.}\ }\textbf {\bibinfo {volume} {74}},\ \bibinfo
  {pages} {601} (\bibinfo {year} {2002})}\BibitemShut {NoStop}%
\bibitem [{\citenamefont {Burke}(2012)}]{burk12jcp}%
  \BibitemOpen
  \bibfield  {author} {\bibinfo {author} {\bibfnamefont {K.}~\bibnamefont
  {Burke}},\ }\href {\doibase 10.1063/1.4704546} {\bibfield  {journal}
  {\bibinfo  {journal} {J. Chem. Phys.}\ }\textbf {\bibinfo {volume} {136}},\
  \bibinfo {pages} {150901} (\bibinfo {year} {2012})}\BibitemShut {NoStop}%
\bibitem [{\citenamefont {Calderon}\ \emph {et~al.}(2015)\citenamefont
  {Calderon}, \citenamefont {Plata}, \citenamefont {Toher}, \citenamefont
  {Oses}, \citenamefont {Levy}, \citenamefont {Fornari}, \citenamefont {Natan},
  \citenamefont {Mehl}, \citenamefont {Hart}, \citenamefont
  {Buongiorno~Nardelli},\ and\ \citenamefont {Curtarolo}}]{cald+15cms}%
  \BibitemOpen
  \bibfield  {author} {\bibinfo {author} {\bibfnamefont {C.~E.}\ \bibnamefont
  {Calderon}}, \bibinfo {author} {\bibfnamefont {J.~J.}\ \bibnamefont {Plata}},
  \bibinfo {author} {\bibfnamefont {C.}~\bibnamefont {Toher}}, \bibinfo
  {author} {\bibfnamefont {C.}~\bibnamefont {Oses}}, \bibinfo {author}
  {\bibfnamefont {O.}~\bibnamefont {Levy}}, \bibinfo {author} {\bibfnamefont
  {M.}~\bibnamefont {Fornari}}, \bibinfo {author} {\bibfnamefont
  {A.}~\bibnamefont {Natan}}, \bibinfo {author} {\bibfnamefont {M.~J.}\
  \bibnamefont {Mehl}}, \bibinfo {author} {\bibfnamefont {G.}~\bibnamefont
  {Hart}}, \bibinfo {author} {\bibfnamefont {M.}~\bibnamefont
  {Buongiorno~Nardelli}}, \ and\ \bibinfo {author} {\bibfnamefont
  {S.}~\bibnamefont {Curtarolo}},\ }\href {\doibase
  10.1016/j.commatsci.2015.07.019} {\bibfield  {journal} {\bibinfo  {journal}
  {Computational Materials Science}\ }\textbf {\bibinfo {volume} {108}},\
  \bibinfo {pages} {233} (\bibinfo {year} {2015})}\BibitemShut {NoStop}%
\bibitem [{\citenamefont {Mounet}\ \emph {et~al.}(2018)\citenamefont {Mounet},
  \citenamefont {Gibertini}, \citenamefont {Schwaller}, \citenamefont {Campi},
  \citenamefont {Merkys}, \citenamefont {Marrazzo}, \citenamefont {Sohier},
  \citenamefont {Castelli}, \citenamefont {Cepellotti}, \citenamefont {Pizzi},\
  and\ \citenamefont {Marzari}}]{moun+18nn}%
  \BibitemOpen
  \bibfield  {author} {\bibinfo {author} {\bibfnamefont {N.}~\bibnamefont
  {Mounet}}, \bibinfo {author} {\bibfnamefont {M.}~\bibnamefont {Gibertini}},
  \bibinfo {author} {\bibfnamefont {P.}~\bibnamefont {Schwaller}}, \bibinfo
  {author} {\bibfnamefont {D.}~\bibnamefont {Campi}}, \bibinfo {author}
  {\bibfnamefont {A.}~\bibnamefont {Merkys}}, \bibinfo {author} {\bibfnamefont
  {A.}~\bibnamefont {Marrazzo}}, \bibinfo {author} {\bibfnamefont
  {T.}~\bibnamefont {Sohier}}, \bibinfo {author} {\bibfnamefont {I.~E.}\
  \bibnamefont {Castelli}}, \bibinfo {author} {\bibfnamefont {A.}~\bibnamefont
  {Cepellotti}}, \bibinfo {author} {\bibfnamefont {G.}~\bibnamefont {Pizzi}}, \
  and\ \bibinfo {author} {\bibfnamefont {N.}~\bibnamefont {Marzari}},\ }\href
  {\doibase 10.1038/s41565-017-0035-5} {\bibfield  {journal} {\bibinfo
  {journal} {Nat. Nanotechnol.}\ }\textbf {\bibinfo {volume} {13}},\ \bibinfo
  {pages} {246} (\bibinfo {year} {2018})}\BibitemShut {NoStop}%
\bibitem [{\citenamefont {Jain}\ \emph {et~al.}(2013)\citenamefont {Jain},
  \citenamefont {Ong}, \citenamefont {Hautier}, \citenamefont {Chen},
  \citenamefont {Richards}, \citenamefont {Dacek}, \citenamefont {Cholia},
  \citenamefont {Gunter}, \citenamefont {Skinner}, \citenamefont {Ceder},\ and\
  \citenamefont {Persson}}]{jain+13aplm}%
  \BibitemOpen
  \bibfield  {author} {\bibinfo {author} {\bibfnamefont {A.}~\bibnamefont
  {Jain}}, \bibinfo {author} {\bibfnamefont {S.~P.}\ \bibnamefont {Ong}},
  \bibinfo {author} {\bibfnamefont {G.}~\bibnamefont {Hautier}}, \bibinfo
  {author} {\bibfnamefont {W.}~\bibnamefont {Chen}}, \bibinfo {author}
  {\bibfnamefont {W.~D.}\ \bibnamefont {Richards}}, \bibinfo {author}
  {\bibfnamefont {S.}~\bibnamefont {Dacek}}, \bibinfo {author} {\bibfnamefont
  {S.}~\bibnamefont {Cholia}}, \bibinfo {author} {\bibfnamefont
  {D.}~\bibnamefont {Gunter}}, \bibinfo {author} {\bibfnamefont
  {D.}~\bibnamefont {Skinner}}, \bibinfo {author} {\bibfnamefont
  {G.}~\bibnamefont {Ceder}}, \ and\ \bibinfo {author} {\bibfnamefont {K.~A.}\
  \bibnamefont {Persson}},\ }\href {\doibase 10.1063/1.4812323} {\bibfield
  {journal} {\bibinfo  {journal} {APL Materials}\ }\textbf {\bibinfo {volume}
  {1}},\ \bibinfo {pages} {011002} (\bibinfo {year} {2013})}\BibitemShut
  {NoStop}%
\bibitem [{\citenamefont {Draxl}\ and\ \citenamefont
  {Scheffler}(2018)}]{drax-sche18mrsb}%
  \BibitemOpen
  \bibfield  {author} {\bibinfo {author} {\bibfnamefont {C.}~\bibnamefont
  {Draxl}}\ and\ \bibinfo {author} {\bibfnamefont {M.}~\bibnamefont
  {Scheffler}},\ }\href {\doibase 10.1557/mrs.2018.208} {\bibfield  {journal}
  {\bibinfo  {journal} {MRS Bull.}\ }\textbf {\bibinfo {volume} {43}},\
  \bibinfo {pages} {676} (\bibinfo {year} {2018})}\BibitemShut {NoStop}%
\bibitem [{\citenamefont {Talirz}\ \emph {et~al.}(2020)\citenamefont {Talirz},
  \citenamefont {Kumbhar}, \citenamefont {Passaro}, \citenamefont {Yakutovich},
  \citenamefont {Granata}, \citenamefont {Gargiulo}, \citenamefont {Borelli},
  \citenamefont {Uhrin}, \citenamefont {Huber}, \citenamefont {Zoupanos},
  \citenamefont {Adorf}, \citenamefont {Andersen}, \citenamefont {Sch{\"u}tt},
  \citenamefont {Pignedoli}, \citenamefont {Passerone}, \citenamefont
  {VandeVondele}, \citenamefont {Schulthess}, \citenamefont {Smit},
  \citenamefont {Pizzi},\ and\ \citenamefont {Marzari}}]{tarl+20sd}%
  \BibitemOpen
  \bibfield  {author} {\bibinfo {author} {\bibfnamefont {L.}~\bibnamefont
  {Talirz}}, \bibinfo {author} {\bibfnamefont {S.}~\bibnamefont {Kumbhar}},
  \bibinfo {author} {\bibfnamefont {E.}~\bibnamefont {Passaro}}, \bibinfo
  {author} {\bibfnamefont {A.~V.}\ \bibnamefont {Yakutovich}}, \bibinfo
  {author} {\bibfnamefont {V.}~\bibnamefont {Granata}}, \bibinfo {author}
  {\bibfnamefont {F.}~\bibnamefont {Gargiulo}}, \bibinfo {author}
  {\bibfnamefont {M.}~\bibnamefont {Borelli}}, \bibinfo {author} {\bibfnamefont
  {M.}~\bibnamefont {Uhrin}}, \bibinfo {author} {\bibfnamefont {S.~P.}\
  \bibnamefont {Huber}}, \bibinfo {author} {\bibfnamefont {S.}~\bibnamefont
  {Zoupanos}}, \bibinfo {author} {\bibfnamefont {C.~S.}\ \bibnamefont {Adorf}},
  \bibinfo {author} {\bibfnamefont {C.~W.}\ \bibnamefont {Andersen}}, \bibinfo
  {author} {\bibfnamefont {O.}~\bibnamefont {Sch{\"u}tt}}, \bibinfo {author}
  {\bibfnamefont {C.~A.}\ \bibnamefont {Pignedoli}}, \bibinfo {author}
  {\bibfnamefont {D.}~\bibnamefont {Passerone}}, \bibinfo {author}
  {\bibfnamefont {J.}~\bibnamefont {VandeVondele}}, \bibinfo {author}
  {\bibfnamefont {T.~C.}\ \bibnamefont {Schulthess}}, \bibinfo {author}
  {\bibfnamefont {B.}~\bibnamefont {Smit}}, \bibinfo {author} {\bibfnamefont
  {G.}~\bibnamefont {Pizzi}}, \ and\ \bibinfo {author} {\bibfnamefont
  {N.}~\bibnamefont {Marzari}},\ }\href {\doibase 10.1038/s41597-020-00637-5}
  {\bibfield  {journal} {\bibinfo  {journal} {Sci Data}\ }\textbf {\bibinfo
  {volume} {7}},\ \bibinfo {pages} {299} (\bibinfo {year} {2020})}\BibitemShut
  {NoStop}%
\bibitem [{\citenamefont {Ceriotti}\ \emph {et~al.}(2016)\citenamefont
  {Ceriotti}, \citenamefont {Fang}, \citenamefont {Kusalik}, \citenamefont
  {McKenzie}, \citenamefont {Michaelides}, \citenamefont {Morales},\ and\
  \citenamefont {Markland}}]{ceri+16cr}%
  \BibitemOpen
  \bibfield  {author} {\bibinfo {author} {\bibfnamefont {M.}~\bibnamefont
  {Ceriotti}}, \bibinfo {author} {\bibfnamefont {W.}~\bibnamefont {Fang}},
  \bibinfo {author} {\bibfnamefont {P.~G.}\ \bibnamefont {Kusalik}}, \bibinfo
  {author} {\bibfnamefont {R.~H.}\ \bibnamefont {McKenzie}}, \bibinfo {author}
  {\bibfnamefont {A.}~\bibnamefont {Michaelides}}, \bibinfo {author}
  {\bibfnamefont {M.~A.}\ \bibnamefont {Morales}}, \ and\ \bibinfo {author}
  {\bibfnamefont {T.~E.}\ \bibnamefont {Markland}},\ }\href {\doibase
  10.1021/acs.chemrev.5b00674} {\bibfield  {journal} {\bibinfo  {journal}
  {Chem. Rev.}\ }\textbf {\bibinfo {volume} {116}},\ \bibinfo {pages} {7529}
  (\bibinfo {year} {2016})}\BibitemShut {NoStop}%
\bibitem [{\citenamefont {Car}\ and\ \citenamefont
  {Parrinello}(1985)}]{car-parr85prl}%
  \BibitemOpen
  \bibfield  {author} {\bibinfo {author} {\bibfnamefont {R.}~\bibnamefont
  {Car}}\ and\ \bibinfo {author} {\bibfnamefont {M.}~\bibnamefont
  {Parrinello}},\ }\href@noop {} {\bibfield  {journal} {\bibinfo  {journal}
  {Phys. Rev. Lett.}\ }\textbf {\bibinfo {volume} {55}},\ \bibinfo {pages}
  {2471} (\bibinfo {year} {1985})}\BibitemShut {NoStop}%
\bibitem [{\citenamefont {Marx}\ \emph {et~al.}(1999)\citenamefont {Marx},
  \citenamefont {Tuckerman}, \citenamefont {Hutter},\ and\ \citenamefont
  {Parrinello}}]{marx+99nat}%
  \BibitemOpen
  \bibfield  {author} {\bibinfo {author} {\bibfnamefont {D.}~\bibnamefont
  {Marx}}, \bibinfo {author} {\bibfnamefont {M.~E.}\ \bibnamefont {Tuckerman}},
  \bibinfo {author} {\bibfnamefont {J.}~\bibnamefont {Hutter}}, \ and\ \bibinfo
  {author} {\bibfnamefont {M.}~\bibnamefont {Parrinello}},\ }\href@noop {}
  {\bibfield  {journal} {\bibinfo  {journal} {Nature}\ }\textbf {\bibinfo
  {volume} {397}},\ \bibinfo {pages} {601} (\bibinfo {year}
  {1999})}\BibitemShut {NoStop}%
\bibitem [{\citenamefont {Grabowski}\ \emph {et~al.}(2009)\citenamefont
  {Grabowski}, \citenamefont {Ismer}, \citenamefont {Hickel},\ and\
  \citenamefont {Neugebauer}}]{grab+09prb}%
  \BibitemOpen
  \bibfield  {author} {\bibinfo {author} {\bibfnamefont {B.}~\bibnamefont
  {Grabowski}}, \bibinfo {author} {\bibfnamefont {L.}~\bibnamefont {Ismer}},
  \bibinfo {author} {\bibfnamefont {T.}~\bibnamefont {Hickel}}, \ and\ \bibinfo
  {author} {\bibfnamefont {J.}~\bibnamefont {Neugebauer}},\ }\href {\doibase
  10.1103/physrevb.79.134106} {\bibfield  {journal} {\bibinfo  {journal} {Phys.
  Rev. B}\ }\textbf {\bibinfo {volume} {79}} (\bibinfo {year} {2009}),\
  10.1103/physrevb.79.134106}\BibitemShut {NoStop}%
\bibitem [{\citenamefont {Freysoldt}\ \emph {et~al.}(2014)\citenamefont
  {Freysoldt}, \citenamefont {Grabowski}, \citenamefont {Hickel}, \citenamefont
  {Neugebauer}, \citenamefont {Kresse}, \citenamefont {Janotti},\ and\
  \citenamefont {{Van de Walle}}}]{frey+14rmp}%
  \BibitemOpen
  \bibfield  {author} {\bibinfo {author} {\bibfnamefont {C.}~\bibnamefont
  {Freysoldt}}, \bibinfo {author} {\bibfnamefont {B.}~\bibnamefont
  {Grabowski}}, \bibinfo {author} {\bibfnamefont {T.}~\bibnamefont {Hickel}},
  \bibinfo {author} {\bibfnamefont {J.}~\bibnamefont {Neugebauer}}, \bibinfo
  {author} {\bibfnamefont {G.}~\bibnamefont {Kresse}}, \bibinfo {author}
  {\bibfnamefont {A.}~\bibnamefont {Janotti}}, \ and\ \bibinfo {author}
  {\bibfnamefont {C.~G.}\ \bibnamefont {{Van de Walle}}},\ }\href {\doibase
  10.1103/RevModPhys.86.253} {\bibfield  {journal} {\bibinfo  {journal} {Rev.
  Mod. Phys.}\ }\textbf {\bibinfo {volume} {86}},\ \bibinfo {pages} {253}
  (\bibinfo {year} {2014})}\BibitemShut {NoStop}%
\bibitem [{\citenamefont {Li}\ \emph {et~al.}(2011)\citenamefont {Li},
  \citenamefont {Walker},\ and\ \citenamefont {Michaelides}}]{li+11pnas}%
  \BibitemOpen
  \bibfield  {author} {\bibinfo {author} {\bibfnamefont {X.-Z.}\ \bibnamefont
  {Li}}, \bibinfo {author} {\bibfnamefont {B.}~\bibnamefont {Walker}}, \ and\
  \bibinfo {author} {\bibfnamefont {A.}~\bibnamefont {Michaelides}},\ }\href
  {\doibase 10.1073/pnas.1016653108} {\bibfield  {journal} {\bibinfo  {journal}
  {Proc. Natl. Acad. Sci.}\ }\textbf {\bibinfo {volume} {108}},\ \bibinfo
  {pages} {6369} (\bibinfo {year} {2011})}\BibitemShut {NoStop}%
\bibitem [{\citenamefont {Reilly}\ and\ \citenamefont
  {Tkatchenko}(2014)}]{reil-tkat14prl}%
  \BibitemOpen
  \bibfield  {author} {\bibinfo {author} {\bibfnamefont {A.~M.}\ \bibnamefont
  {Reilly}}\ and\ \bibinfo {author} {\bibfnamefont {A.}~\bibnamefont
  {Tkatchenko}},\ }\href {\doibase 10.1103/PhysRevLett.113.055701} {\bibfield
  {journal} {\bibinfo  {journal} {Phys. Rev. Lett.}\ }\textbf {\bibinfo
  {volume} {113}},\ \bibinfo {pages} {055701} (\bibinfo {year}
  {2014})}\BibitemShut {NoStop}%
\bibitem [{\citenamefont {Rossi}\ \emph {et~al.}(2016)\citenamefont {Rossi},
  \citenamefont {Gasparotto},\ and\ \citenamefont {Ceriotti}}]{ross+16prl}%
  \BibitemOpen
  \bibfield  {author} {\bibinfo {author} {\bibfnamefont {M.}~\bibnamefont
  {Rossi}}, \bibinfo {author} {\bibfnamefont {P.}~\bibnamefont {Gasparotto}}, \
  and\ \bibinfo {author} {\bibfnamefont {M.}~\bibnamefont {Ceriotti}},\ }\href
  {\doibase 10.1103/PhysRevLett.117.115702} {\bibfield  {journal} {\bibinfo
  {journal} {Phys. Rev. Lett.}\ }\textbf {\bibinfo {volume} {117}},\ \bibinfo
  {pages} {115702} (\bibinfo {year} {2016})}\BibitemShut {NoStop}%
\bibitem [{\citenamefont {Ko}\ \emph {et~al.}(2018)\citenamefont {Ko},
  \citenamefont {DiStasio}, \citenamefont {Santra},\ and\ \citenamefont
  {Car}}]{ko+18prm}%
  \BibitemOpen
  \bibfield  {author} {\bibinfo {author} {\bibfnamefont {H.-Y.}\ \bibnamefont
  {Ko}}, \bibinfo {author} {\bibfnamefont {R.~A.}\ \bibnamefont {DiStasio}},
  \bibinfo {author} {\bibfnamefont {B.}~\bibnamefont {Santra}}, \ and\ \bibinfo
  {author} {\bibfnamefont {R.}~\bibnamefont {Car}},\ }\href {\doibase
  10.1103/PhysRevMaterials.2.055603} {\bibfield  {journal} {\bibinfo  {journal}
  {Phys. Rev. Materials}\ }\textbf {\bibinfo {volume} {2}},\ \bibinfo {pages}
  {055603} (\bibinfo {year} {2018})}\BibitemShut {NoStop}%
\bibitem [{\citenamefont {Raimbault}\ \emph {et~al.}(2019)\citenamefont
  {Raimbault}, \citenamefont {Athavale},\ and\ \citenamefont
  {Rossi}}]{raim+19prm}%
  \BibitemOpen
  \bibfield  {author} {\bibinfo {author} {\bibfnamefont {N.}~\bibnamefont
  {Raimbault}}, \bibinfo {author} {\bibfnamefont {V.}~\bibnamefont {Athavale}},
  \ and\ \bibinfo {author} {\bibfnamefont {M.}~\bibnamefont {Rossi}},\ }\href
  {\doibase 10.1103/PhysRevMaterials.3.053605} {\bibfield  {journal} {\bibinfo
  {journal} {Phys. Rev. Materials}\ }\textbf {\bibinfo {volume} {3}},\ \bibinfo
  {pages} {053605} (\bibinfo {year} {2019})}\BibitemShut {NoStop}%
\bibitem [{\citenamefont {Partridge}\ and\ \citenamefont
  {Schwenke}(1997)}]{part-schw97jcp}%
  \BibitemOpen
  \bibfield  {author} {\bibinfo {author} {\bibfnamefont {H.}~\bibnamefont
  {Partridge}}\ and\ \bibinfo {author} {\bibfnamefont {D.~W.}\ \bibnamefont
  {Schwenke}},\ }\href {\doibase 10.1063/1.473987} {\bibfield  {journal}
  {\bibinfo  {journal} {J. Chem. Phys.}\ }\textbf {\bibinfo {volume} {106}},\
  \bibinfo {pages} {4618} (\bibinfo {year} {1997})}\BibitemShut {NoStop}%
\bibitem [{\citenamefont {Huang}\ \emph {et~al.}(2005)\citenamefont {Huang},
  \citenamefont {Braams},\ and\ \citenamefont {Bowman}}]{huan+05jcp}%
  \BibitemOpen
  \bibfield  {author} {\bibinfo {author} {\bibfnamefont {X.}~\bibnamefont
  {Huang}}, \bibinfo {author} {\bibfnamefont {B.~J.}\ \bibnamefont {Braams}}, \
  and\ \bibinfo {author} {\bibfnamefont {J.~M.}\ \bibnamefont {Bowman}},\
  }\href {\doibase 10.1063/1.1834500} {\bibfield  {journal} {\bibinfo
  {journal} {J. Chem. Phys.}\ }\textbf {\bibinfo {volume} {122}},\ \bibinfo
  {pages} {44308} (\bibinfo {year} {2005})}\BibitemShut {NoStop}%
\bibitem [{\citenamefont {Sanchez}\ \emph {et~al.}(1984)\citenamefont
  {Sanchez}, \citenamefont {Ducastelle},\ and\ \citenamefont
  {Gratias}}]{sanc+84pa}%
  \BibitemOpen
  \bibfield  {author} {\bibinfo {author} {\bibfnamefont {J.}~\bibnamefont
  {Sanchez}}, \bibinfo {author} {\bibfnamefont {F.}~\bibnamefont {Ducastelle}},
  \ and\ \bibinfo {author} {\bibfnamefont {D.}~\bibnamefont {Gratias}},\ }\href
  {\doibase 10.1016/0378-4371(84)90096-7} {\bibfield  {journal} {\bibinfo
  {journal} {Physica A: Statistical Mechanics and its Applications}\ }\textbf
  {\bibinfo {volume} {128}},\ \bibinfo {pages} {334} (\bibinfo {year}
  {1984})}\BibitemShut {NoStop}%
\bibitem [{\citenamefont {Behler}\ and\ \citenamefont
  {Parrinello}(2007)}]{behl-parr07prl}%
  \BibitemOpen
  \bibfield  {author} {\bibinfo {author} {\bibfnamefont {J.}~\bibnamefont
  {Behler}}\ and\ \bibinfo {author} {\bibfnamefont {M.}~\bibnamefont
  {Parrinello}},\ }\href {\doibase 10.1103/PhysRevLett.98.146401} {\bibfield
  {journal} {\bibinfo  {journal} {Phys. Rev. Lett.}\ }\textbf {\bibinfo
  {volume} {98}},\ \bibinfo {pages} {146401} (\bibinfo {year}
  {2007})}\BibitemShut {NoStop}%
\bibitem [{\citenamefont {Bart{\'o}k}\ \emph {et~al.}(2010)\citenamefont
  {Bart{\'o}k}, \citenamefont {Payne}, \citenamefont {Kondor},\ and\
  \citenamefont {Cs{\'a}nyi}}]{bart+10prl}%
  \BibitemOpen
  \bibfield  {author} {\bibinfo {author} {\bibfnamefont {A.~P.}\ \bibnamefont
  {Bart{\'o}k}}, \bibinfo {author} {\bibfnamefont {M.~C.}\ \bibnamefont
  {Payne}}, \bibinfo {author} {\bibfnamefont {R.}~\bibnamefont {Kondor}}, \
  and\ \bibinfo {author} {\bibfnamefont {G.}~\bibnamefont {Cs{\'a}nyi}},\
  }\href {\doibase 10.1103/PhysRevLett.104.136403} {\bibfield  {journal}
  {\bibinfo  {journal} {Phys. Rev. Lett.}\ }\textbf {\bibinfo {volume} {104}},\
  \bibinfo {pages} {136403} (\bibinfo {year} {2010})}\BibitemShut {NoStop}%
\bibitem [{\citenamefont {Behler}(2021)}]{behl21cr}%
  \BibitemOpen
  \bibfield  {author} {\bibinfo {author} {\bibfnamefont {J.}~\bibnamefont
  {Behler}},\ }\href {\doibase 10.1021/acs.chemrev.0c00868} {\bibfield
  {journal} {\bibinfo  {journal} {Chem. Rev.}\ }\textbf {\bibinfo {volume}
  {121}},\ \bibinfo {pages} {10037} (\bibinfo {year} {2021})}\BibitemShut
  {NoStop}%
\bibitem [{\citenamefont {Musil}\ \emph {et~al.}(2021)\citenamefont {Musil},
  \citenamefont {Grisafi}, \citenamefont {Bart{\'o}k}, \citenamefont {Ortner},
  \citenamefont {Cs{\'a}nyi},\ and\ \citenamefont {Ceriotti}}]{musi+21cr}%
  \BibitemOpen
  \bibfield  {author} {\bibinfo {author} {\bibfnamefont {F.}~\bibnamefont
  {Musil}}, \bibinfo {author} {\bibfnamefont {A.}~\bibnamefont {Grisafi}},
  \bibinfo {author} {\bibfnamefont {A.~P.}\ \bibnamefont {Bart{\'o}k}},
  \bibinfo {author} {\bibfnamefont {C.}~\bibnamefont {Ortner}}, \bibinfo
  {author} {\bibfnamefont {G.}~\bibnamefont {Cs{\'a}nyi}}, \ and\ \bibinfo
  {author} {\bibfnamefont {M.}~\bibnamefont {Ceriotti}},\ }\href {\doibase
  10.1021/acs.chemrev.1c00021} {\bibfield  {journal} {\bibinfo  {journal}
  {Chem. Rev.}\ }\textbf {\bibinfo {volume} {121}},\ \bibinfo {pages} {9759}
  (\bibinfo {year} {2021})}\BibitemShut {NoStop}%
\bibitem [{\citenamefont {Langer}\ \emph {et~al.}(2022)\citenamefont {Langer},
  \citenamefont {Goe{\ss}mann},\ and\ \citenamefont {Rupp}}]{lang+22npjcm}%
  \BibitemOpen
  \bibfield  {author} {\bibinfo {author} {\bibfnamefont {M.~F.}\ \bibnamefont
  {Langer}}, \bibinfo {author} {\bibfnamefont {A.}~\bibnamefont
  {Goe{\ss}mann}}, \ and\ \bibinfo {author} {\bibfnamefont {M.}~\bibnamefont
  {Rupp}},\ }\href {\doibase 10.1038/s41524-022-00721-x} {\bibfield  {journal}
  {\bibinfo  {journal} {npj Comput Mater}\ }\textbf {\bibinfo {volume} {8}},\
  \bibinfo {pages} {41} (\bibinfo {year} {2022})}\BibitemShut {NoStop}%
\bibitem [{\citenamefont {Glielmo}\ \emph {et~al.}(2018)\citenamefont
  {Glielmo}, \citenamefont {Zeni},\ and\ \citenamefont {De~Vita}}]{glie+18prb}%
  \BibitemOpen
  \bibfield  {author} {\bibinfo {author} {\bibfnamefont {A.}~\bibnamefont
  {Glielmo}}, \bibinfo {author} {\bibfnamefont {C.}~\bibnamefont {Zeni}}, \
  and\ \bibinfo {author} {\bibfnamefont {A.}~\bibnamefont {De~Vita}},\ }\href
  {\doibase 10.1103/PhysRevB.97.184307} {\bibfield  {journal} {\bibinfo
  {journal} {Phys. Rev. B}\ }\textbf {\bibinfo {volume} {97}},\ \bibinfo
  {pages} {184307} (\bibinfo {year} {2018})}\BibitemShut {NoStop}%
\bibitem [{\citenamefont {Behler}(2011)}]{behl11jcp}%
  \BibitemOpen
  \bibfield  {author} {\bibinfo {author} {\bibfnamefont {J.}~\bibnamefont
  {Behler}},\ }\href {\doibase 10.1063/1.3553717} {\bibfield  {journal}
  {\bibinfo  {journal} {The Journal of Chemical Physics}\ }\textbf {\bibinfo
  {volume} {134}},\ \bibinfo {pages} {074106} (\bibinfo {year}
  {2011})}\BibitemShut {NoStop}%
\bibitem [{\citenamefont {Bart{\'o}k}\ \emph {et~al.}(2013)\citenamefont
  {Bart{\'o}k}, \citenamefont {Kondor},\ and\ \citenamefont
  {Cs{\'a}nyi}}]{bart+13prb}%
  \BibitemOpen
  \bibfield  {author} {\bibinfo {author} {\bibfnamefont {A.~P.}\ \bibnamefont
  {Bart{\'o}k}}, \bibinfo {author} {\bibfnamefont {R.}~\bibnamefont {Kondor}},
  \ and\ \bibinfo {author} {\bibfnamefont {G.}~\bibnamefont {Cs{\'a}nyi}},\
  }\href {\doibase 10.1103/PhysRevB.87.184115} {\bibfield  {journal} {\bibinfo
  {journal} {Phys. Rev. B}\ }\textbf {\bibinfo {volume} {87}},\ \bibinfo
  {pages} {184115} (\bibinfo {year} {2013})}\BibitemShut {NoStop}%
\bibitem [{\citenamefont {Rupp}\ \emph {et~al.}(2012)\citenamefont {Rupp},
  \citenamefont {Tkatchenko}, \citenamefont {M{\"u}ller},\ and\ \citenamefont
  {{\noopsort{lilienfeld}}{von Lilienfeld}}}]{rupp+12prl}%
  \BibitemOpen
  \bibfield  {author} {\bibinfo {author} {\bibfnamefont {M.}~\bibnamefont
  {Rupp}}, \bibinfo {author} {\bibfnamefont {A.}~\bibnamefont {Tkatchenko}},
  \bibinfo {author} {\bibfnamefont {K.-R.}\ \bibnamefont {M{\"u}ller}}, \ and\
  \bibinfo {author} {\bibfnamefont {O.~A.}\ \bibnamefont
  {{\noopsort{lilienfeld}}{von Lilienfeld}}},\ }\href {\doibase
  10.1103/PhysRevLett.108.058301} {\bibfield  {journal} {\bibinfo  {journal}
  {Phys. Rev. Lett.}\ }\textbf {\bibinfo {volume} {108}},\ \bibinfo {pages}
  {058301} (\bibinfo {year} {2012})}\BibitemShut {NoStop}%
\bibitem [{\citenamefont {Manzhos}\ and\ \citenamefont
  {Carrington}(2021)}]{manz-carr21cr}%
  \BibitemOpen
  \bibfield  {author} {\bibinfo {author} {\bibfnamefont {S.}~\bibnamefont
  {Manzhos}}\ and\ \bibinfo {author} {\bibfnamefont {T.}~\bibnamefont
  {Carrington}},\ }\href {\doibase 10.1021/acs.chemrev.0c00665} {\bibfield
  {journal} {\bibinfo  {journal} {Chem. Rev.}\ }\textbf {\bibinfo {volume}
  {121}},\ \bibinfo {pages} {10187} (\bibinfo {year} {2021})}\BibitemShut
  {NoStop}%
\bibitem [{\citenamefont {Bart{\'o}k}\ \emph {et~al.}(2017)\citenamefont
  {Bart{\'o}k}, \citenamefont {De}, \citenamefont {Poelking}, \citenamefont
  {Bernstein}, \citenamefont {Kermode}, \citenamefont {Cs{\'a}nyi},\ and\
  \citenamefont {Ceriotti}}]{bart+17sa}%
  \BibitemOpen
  \bibfield  {author} {\bibinfo {author} {\bibfnamefont {A.~P.}\ \bibnamefont
  {Bart{\'o}k}}, \bibinfo {author} {\bibfnamefont {S.}~\bibnamefont {De}},
  \bibinfo {author} {\bibfnamefont {C.}~\bibnamefont {Poelking}}, \bibinfo
  {author} {\bibfnamefont {N.}~\bibnamefont {Bernstein}}, \bibinfo {author}
  {\bibfnamefont {J.~R.}\ \bibnamefont {Kermode}}, \bibinfo {author}
  {\bibfnamefont {G.}~\bibnamefont {Cs{\'a}nyi}}, \ and\ \bibinfo {author}
  {\bibfnamefont {M.}~\bibnamefont {Ceriotti}},\ }\href {\doibase
  10.1126/sciadv.1701816} {\bibfield  {journal} {\bibinfo  {journal} {Sci.
  Adv.}\ }\textbf {\bibinfo {volume} {3}},\ \bibinfo {pages} {e1701816}
  (\bibinfo {year} {2017})}\BibitemShut {NoStop}%
\bibitem [{\citenamefont {Butler}\ \emph {et~al.}(2018)\citenamefont {Butler},
  \citenamefont {Davies}, \citenamefont {Cartwright}, \citenamefont {Isayev},\
  and\ \citenamefont {Walsh}}]{butl+18nature}%
  \BibitemOpen
  \bibfield  {author} {\bibinfo {author} {\bibfnamefont {K.~T.}\ \bibnamefont
  {Butler}}, \bibinfo {author} {\bibfnamefont {D.~W.}\ \bibnamefont {Davies}},
  \bibinfo {author} {\bibfnamefont {H.}~\bibnamefont {Cartwright}}, \bibinfo
  {author} {\bibfnamefont {O.}~\bibnamefont {Isayev}}, \ and\ \bibinfo {author}
  {\bibfnamefont {A.}~\bibnamefont {Walsh}},\ }\href {\doibase
  10.1038/s41586-018-0337-2} {\bibfield  {journal} {\bibinfo  {journal}
  {Nature}\ }\textbf {\bibinfo {volume} {559}},\ \bibinfo {pages} {547}
  (\bibinfo {year} {2018})}\BibitemShut {NoStop}%
\bibitem [{\citenamefont {Kapil}\ and\ \citenamefont
  {Engel}(2022)}]{kapi-enge22pnas}%
  \BibitemOpen
  \bibfield  {author} {\bibinfo {author} {\bibfnamefont {V.}~\bibnamefont
  {Kapil}}\ and\ \bibinfo {author} {\bibfnamefont {E.~A.}\ \bibnamefont
  {Engel}},\ }\href {\doibase 10.1073/pnas.2111769119} {\bibfield  {journal}
  {\bibinfo  {journal} {Proc. Natl. Acad. Sci. U.S.A.}\ }\textbf {\bibinfo
  {volume} {119}},\ \bibinfo {pages} {e2111769119} (\bibinfo {year}
  {2022})}\BibitemShut {NoStop}%
\bibitem [{\citenamefont {Khaliullin}\ \emph {et~al.}(2010)\citenamefont
  {Khaliullin}, \citenamefont {Eshet}, \citenamefont {K{\"u}hne}, \citenamefont
  {Behler},\ and\ \citenamefont {Parrinello}}]{khal+10prb}%
  \BibitemOpen
  \bibfield  {author} {\bibinfo {author} {\bibfnamefont {R.~Z.}\ \bibnamefont
  {Khaliullin}}, \bibinfo {author} {\bibfnamefont {H.}~\bibnamefont {Eshet}},
  \bibinfo {author} {\bibfnamefont {T.~D.}\ \bibnamefont {K{\"u}hne}}, \bibinfo
  {author} {\bibfnamefont {J.}~\bibnamefont {Behler}}, \ and\ \bibinfo {author}
  {\bibfnamefont {M.}~\bibnamefont {Parrinello}},\ }\href@noop {} {\bibfield
  {journal} {\bibinfo  {journal} {Phys. Rev. B}\ }\textbf {\bibinfo {volume}
  {81}},\ \bibinfo {pages} {100103} (\bibinfo {year} {2010})}\BibitemShut
  {NoStop}%
\bibitem [{\citenamefont {Eshet}\ \emph {et~al.}(2012)\citenamefont {Eshet},
  \citenamefont {Khaliullin}, \citenamefont {K{\"u}hne}, \citenamefont
  {Behler},\ and\ \citenamefont {Parrinello}}]{eshe+12prl}%
  \BibitemOpen
  \bibfield  {author} {\bibinfo {author} {\bibfnamefont {H.}~\bibnamefont
  {Eshet}}, \bibinfo {author} {\bibfnamefont {R.~Z.}\ \bibnamefont
  {Khaliullin}}, \bibinfo {author} {\bibfnamefont {T.~D.}\ \bibnamefont
  {K{\"u}hne}}, \bibinfo {author} {\bibfnamefont {J.}~\bibnamefont {Behler}}, \
  and\ \bibinfo {author} {\bibfnamefont {M.}~\bibnamefont {Parrinello}},\
  }\href {\doibase 10.1103/PhysRevLett.108.115701} {\bibfield  {journal}
  {\bibinfo  {journal} {Phys. Rev. Lett.}\ }\textbf {\bibinfo {volume} {108}},\
  \bibinfo {pages} {115701} (\bibinfo {year} {2012})}\BibitemShut {NoStop}%
\bibitem [{\citenamefont {Dragoni}\ \emph {et~al.}(2018)\citenamefont
  {Dragoni}, \citenamefont {Daff}, \citenamefont {Cs{\'a}nyi},\ and\
  \citenamefont {Marzari}}]{drag+18prm}%
  \BibitemOpen
  \bibfield  {author} {\bibinfo {author} {\bibfnamefont {D.}~\bibnamefont
  {Dragoni}}, \bibinfo {author} {\bibfnamefont {T.~D.}\ \bibnamefont {Daff}},
  \bibinfo {author} {\bibfnamefont {G.}~\bibnamefont {Cs{\'a}nyi}}, \ and\
  \bibinfo {author} {\bibfnamefont {N.}~\bibnamefont {Marzari}},\ }\href
  {\doibase 10.1103/PhysRevMaterials.2.013808} {\bibfield  {journal} {\bibinfo
  {journal} {Phys. Rev. Materials}\ }\textbf {\bibinfo {volume} {2}},\ \bibinfo
  {pages} {013808} (\bibinfo {year} {2018})}\BibitemShut {NoStop}%
\bibitem [{\citenamefont {Szlachta}\ \emph {et~al.}(2014)\citenamefont
  {Szlachta}, \citenamefont {Bart{\'o}k},\ and\ \citenamefont
  {Cs{\'a}nyi}}]{szla+14prb}%
  \BibitemOpen
  \bibfield  {author} {\bibinfo {author} {\bibfnamefont {W.~J.}\ \bibnamefont
  {Szlachta}}, \bibinfo {author} {\bibfnamefont {A.~P.}\ \bibnamefont
  {Bart{\'o}k}}, \ and\ \bibinfo {author} {\bibfnamefont {G.}~\bibnamefont
  {Cs{\'a}nyi}},\ }\href {\doibase 10.1103/PhysRevB.90.104108} {\bibfield
  {journal} {\bibinfo  {journal} {Phys. Rev. B}\ }\textbf {\bibinfo {volume}
  {90}},\ \bibinfo {pages} {104108} (\bibinfo {year} {2014})}\BibitemShut
  {NoStop}%
\bibitem [{\citenamefont {Cheng}\ \emph {et~al.}(2019)\citenamefont {Cheng},
  \citenamefont {Engel}, \citenamefont {Behler}, \citenamefont {Dellago},\ and\
  \citenamefont {Ceriotti}}]{chen+19pnas}%
  \BibitemOpen
  \bibfield  {author} {\bibinfo {author} {\bibfnamefont {B.}~\bibnamefont
  {Cheng}}, \bibinfo {author} {\bibfnamefont {E.~A.}\ \bibnamefont {Engel}},
  \bibinfo {author} {\bibfnamefont {J.}~\bibnamefont {Behler}}, \bibinfo
  {author} {\bibfnamefont {C.}~\bibnamefont {Dellago}}, \ and\ \bibinfo
  {author} {\bibfnamefont {M.}~\bibnamefont {Ceriotti}},\ }\href {\doibase
  10.1073/pnas.1815117116} {\bibfield  {journal} {\bibinfo  {journal} {Proc.
  Natl. Acad. Sci. U. S. A.}\ }\textbf {\bibinfo {volume} {116}},\ \bibinfo
  {pages} {1110} (\bibinfo {year} {2019})}\BibitemShut {NoStop}%
\bibitem [{\citenamefont {Cheng}\ \emph {et~al.}(2021)\citenamefont {Cheng},
  \citenamefont {Bethkenhagen}, \citenamefont {Pickard},\ and\ \citenamefont
  {Hamel}}]{chen+21np}%
  \BibitemOpen
  \bibfield  {author} {\bibinfo {author} {\bibfnamefont {B.}~\bibnamefont
  {Cheng}}, \bibinfo {author} {\bibfnamefont {M.}~\bibnamefont {Bethkenhagen}},
  \bibinfo {author} {\bibfnamefont {C.~J.}\ \bibnamefont {Pickard}}, \ and\
  \bibinfo {author} {\bibfnamefont {S.}~\bibnamefont {Hamel}},\ }\href
  {\doibase 10.1038/s41567-021-01334-9} {\bibfield  {journal} {\bibinfo
  {journal} {Nat. Phys.}\ }\textbf {\bibinfo {volume} {17}},\ \bibinfo {pages}
  {1228} (\bibinfo {year} {2021})}\BibitemShut {NoStop}%
\bibitem [{\citenamefont {Wallace}\ \emph
  {et~al.}(2021{\natexlab{a}})\citenamefont {Wallace}, \citenamefont
  {Bochkarev}, \citenamefont {{\noopsort{roekeghem}}{van Roekeghem}},
  \citenamefont {Carrasco}, \citenamefont {Shapeev},\ and\ \citenamefont
  {Mingo}}]{wall+21prm}%
  \BibitemOpen
  \bibfield  {author} {\bibinfo {author} {\bibfnamefont {S.~K.}\ \bibnamefont
  {Wallace}}, \bibinfo {author} {\bibfnamefont {A.~S.}\ \bibnamefont
  {Bochkarev}}, \bibinfo {author} {\bibfnamefont {A.}~\bibnamefont
  {{\noopsort{roekeghem}}{van Roekeghem}}}, \bibinfo {author} {\bibfnamefont
  {J.}~\bibnamefont {Carrasco}}, \bibinfo {author} {\bibfnamefont
  {A.}~\bibnamefont {Shapeev}}, \ and\ \bibinfo {author} {\bibfnamefont
  {N.}~\bibnamefont {Mingo}},\ }\href {\doibase
  10.1103/PhysRevMaterials.5.035402} {\bibfield  {journal} {\bibinfo  {journal}
  {Phys. Rev. Materials}\ }\textbf {\bibinfo {volume} {5}},\ \bibinfo {pages}
  {035402} (\bibinfo {year} {2021}{\natexlab{a}})}\BibitemShut {NoStop}%
\bibitem [{\citenamefont {Wallace}\ \emph
  {et~al.}(2021{\natexlab{b}})\citenamefont {Wallace}, \citenamefont
  {{\noopsort{roekeghem}}{van Roekeghem}}, \citenamefont {Bochkarev},
  \citenamefont {Carrasco}, \citenamefont {Shapeev},\ and\ \citenamefont
  {Mingo}}]{wall+21prr}%
  \BibitemOpen
  \bibfield  {author} {\bibinfo {author} {\bibfnamefont {S.~K.}\ \bibnamefont
  {Wallace}}, \bibinfo {author} {\bibfnamefont {A.}~\bibnamefont
  {{\noopsort{roekeghem}}{van Roekeghem}}}, \bibinfo {author} {\bibfnamefont
  {A.~S.}\ \bibnamefont {Bochkarev}}, \bibinfo {author} {\bibfnamefont
  {J.}~\bibnamefont {Carrasco}}, \bibinfo {author} {\bibfnamefont
  {A.}~\bibnamefont {Shapeev}}, \ and\ \bibinfo {author} {\bibfnamefont
  {N.}~\bibnamefont {Mingo}},\ }\href {\doibase
  10.1103/PhysRevResearch.3.013139} {\bibfield  {journal} {\bibinfo  {journal}
  {Phys. Rev. Research}\ }\textbf {\bibinfo {volume} {3}},\ \bibinfo {pages}
  {013139} (\bibinfo {year} {2021}{\natexlab{b}})}\BibitemShut {NoStop}%
\bibitem [{\citenamefont {Lee}\ \emph {et~al.}(2022)\citenamefont {Lee},
  \citenamefont {Pickard},\ and\ \citenamefont {Cheng}}]{lee+22jcp}%
  \BibitemOpen
  \bibfield  {author} {\bibinfo {author} {\bibfnamefont {J.~G.}\ \bibnamefont
  {Lee}}, \bibinfo {author} {\bibfnamefont {C.~J.}\ \bibnamefont {Pickard}}, \
  and\ \bibinfo {author} {\bibfnamefont {B.}~\bibnamefont {Cheng}},\ }\href
  {\doibase 10.1063/5.0079844} {\bibfield  {journal} {\bibinfo  {journal} {J.
  Chem. Phys.}\ }\textbf {\bibinfo {volume} {156}},\ \bibinfo {pages} {074106}
  (\bibinfo {year} {2022})}\BibitemShut {NoStop}%
\bibitem [{\citenamefont {Mishin}(2021)}]{mish21am}%
  \BibitemOpen
  \bibfield  {author} {\bibinfo {author} {\bibfnamefont {Y.}~\bibnamefont
  {Mishin}},\ }\href {\doibase 10.1016/j.actamat.2021.116980} {\bibfield
  {journal} {\bibinfo  {journal} {Acta Materialia}\ }\textbf {\bibinfo {volume}
  {214}},\ \bibinfo {pages} {116980} (\bibinfo {year} {2021})}\BibitemShut
  {NoStop}%
\bibitem [{\citenamefont {Rosenbrock}\ \emph {et~al.}(2021)\citenamefont
  {Rosenbrock}, \citenamefont {Gubaev}, \citenamefont {Shapeev}, \citenamefont
  {P{\'a}rtay}, \citenamefont {Bernstein}, \citenamefont {Cs{\'a}nyi},\ and\
  \citenamefont {Hart}}]{rose+21npjcm}%
  \BibitemOpen
  \bibfield  {author} {\bibinfo {author} {\bibfnamefont {C.~W.}\ \bibnamefont
  {Rosenbrock}}, \bibinfo {author} {\bibfnamefont {K.}~\bibnamefont {Gubaev}},
  \bibinfo {author} {\bibfnamefont {A.~V.}\ \bibnamefont {Shapeev}}, \bibinfo
  {author} {\bibfnamefont {L.~B.}\ \bibnamefont {P{\'a}rtay}}, \bibinfo
  {author} {\bibfnamefont {N.}~\bibnamefont {Bernstein}}, \bibinfo {author}
  {\bibfnamefont {G.}~\bibnamefont {Cs{\'a}nyi}}, \ and\ \bibinfo {author}
  {\bibfnamefont {G.~L.~W.}\ \bibnamefont {Hart}},\ }\href {\doibase
  10.1038/s41524-020-00477-2} {\bibfield  {journal} {\bibinfo  {journal} {npj
  Comput Mater}\ }\textbf {\bibinfo {volume} {7}},\ \bibinfo {pages} {24}
  (\bibinfo {year} {2021})}\BibitemShut {NoStop}%
\bibitem [{\citenamefont {Artrith}\ \emph {et~al.}(2018)\citenamefont
  {Artrith}, \citenamefont {Urban},\ and\ \citenamefont {Ceder}}]{nong+18jcp}%
  \BibitemOpen
  \bibfield  {author} {\bibinfo {author} {\bibfnamefont {N.}~\bibnamefont
  {Artrith}}, \bibinfo {author} {\bibfnamefont {A.}~\bibnamefont {Urban}}, \
  and\ \bibinfo {author} {\bibfnamefont {G.}~\bibnamefont {Ceder}},\ }\href
  {\doibase 10.1063/1.5017661} {\bibfield  {journal} {\bibinfo  {journal} {The
  Journal of Chemical Physics}\ }\textbf {\bibinfo {volume} {148}},\ \bibinfo
  {pages} {241711} (\bibinfo {year} {2018})}\BibitemShut {NoStop}%
\bibitem [{\citenamefont {Caro}\ \emph {et~al.}(2018)\citenamefont {Caro},
  \citenamefont {Aarva}, \citenamefont {Deringer}, \citenamefont {Cs{\'a}nyi},\
  and\ \citenamefont {Laurila}}]{caro+18cm}%
  \BibitemOpen
  \bibfield  {author} {\bibinfo {author} {\bibfnamefont {M.~A.}\ \bibnamefont
  {Caro}}, \bibinfo {author} {\bibfnamefont {A.}~\bibnamefont {Aarva}},
  \bibinfo {author} {\bibfnamefont {V.~L.}\ \bibnamefont {Deringer}}, \bibinfo
  {author} {\bibfnamefont {G.}~\bibnamefont {Cs{\'a}nyi}}, \ and\ \bibinfo
  {author} {\bibfnamefont {T.}~\bibnamefont {Laurila}},\ }\href {\doibase
  10.1021/acs.chemmater.8b03353} {\bibfield  {journal} {\bibinfo  {journal}
  {Chem. Mater.}\ }\textbf {\bibinfo {volume} {30}},\ \bibinfo {pages} {7446}
  (\bibinfo {year} {2018})}\BibitemShut {NoStop}%
\bibitem [{\citenamefont {Deringer}\ \emph {et~al.}(2020)\citenamefont
  {Deringer}, \citenamefont {Caro},\ and\ \citenamefont
  {Cs{\'a}nyi}}]{deri+20nc}%
  \BibitemOpen
  \bibfield  {author} {\bibinfo {author} {\bibfnamefont {V.~L.}\ \bibnamefont
  {Deringer}}, \bibinfo {author} {\bibfnamefont {M.~A.}\ \bibnamefont {Caro}},
  \ and\ \bibinfo {author} {\bibfnamefont {G.}~\bibnamefont {Cs{\'a}nyi}},\
  }\href {\doibase 10.1038/s41467-020-19168-z} {\bibfield  {journal} {\bibinfo
  {journal} {Nat Commun}\ }\textbf {\bibinfo {volume} {11}},\ \bibinfo {pages}
  {5461} (\bibinfo {year} {2020})}\BibitemShut {NoStop}%
\bibitem [{\citenamefont {Jinnouchi}\ \emph {et~al.}(2019)\citenamefont
  {Jinnouchi}, \citenamefont {Lahnsteiner}, \citenamefont {Karsai},
  \citenamefont {Kresse},\ and\ \citenamefont {Bokdam}}]{jinn+19prl}%
  \BibitemOpen
  \bibfield  {author} {\bibinfo {author} {\bibfnamefont {R.}~\bibnamefont
  {Jinnouchi}}, \bibinfo {author} {\bibfnamefont {J.}~\bibnamefont
  {Lahnsteiner}}, \bibinfo {author} {\bibfnamefont {F.}~\bibnamefont {Karsai}},
  \bibinfo {author} {\bibfnamefont {G.}~\bibnamefont {Kresse}}, \ and\ \bibinfo
  {author} {\bibfnamefont {M.}~\bibnamefont {Bokdam}},\ }\href {\doibase
  10.1103/PhysRevLett.122.225701} {\bibfield  {journal} {\bibinfo  {journal}
  {Phys. Rev. Lett.}\ }\textbf {\bibinfo {volume} {122}},\ \bibinfo {pages}
  {225701} (\bibinfo {year} {2019})}\BibitemShut {NoStop}%
\bibitem [{\citenamefont {Montavon}\ \emph {et~al.}(2013)\citenamefont
  {Montavon}, \citenamefont {Rupp}, \citenamefont {Gobre}, \citenamefont
  {{Vazquez-Mayagoitia}}, \citenamefont {Hansen}, \citenamefont {Tkatchenko},
  \citenamefont {M{\"u}ller},\ and\ \citenamefont {Anatole
  Von~Lilienfeld}}]{mont+13njp}%
  \BibitemOpen
  \bibfield  {author} {\bibinfo {author} {\bibfnamefont {G.}~\bibnamefont
  {Montavon}}, \bibinfo {author} {\bibfnamefont {M.}~\bibnamefont {Rupp}},
  \bibinfo {author} {\bibfnamefont {V.}~\bibnamefont {Gobre}}, \bibinfo
  {author} {\bibfnamefont {A.}~\bibnamefont {{Vazquez-Mayagoitia}}}, \bibinfo
  {author} {\bibfnamefont {K.}~\bibnamefont {Hansen}}, \bibinfo {author}
  {\bibfnamefont {A.}~\bibnamefont {Tkatchenko}}, \bibinfo {author}
  {\bibfnamefont {K.~R.}\ \bibnamefont {M{\"u}ller}}, \ and\ \bibinfo {author}
  {\bibfnamefont {O.}~\bibnamefont {Anatole Von~Lilienfeld}},\ }\href {\doibase
  10.1088/1367-2630/15/9/095003} {\bibfield  {journal} {\bibinfo  {journal}
  {New J. Phys.}\ }\textbf {\bibinfo {volume} {15}},\ \bibinfo {pages} {095003}
  (\bibinfo {year} {2013})}\BibitemShut {NoStop}%
\bibitem [{\citenamefont {Ramakrishnan}\ \emph {et~al.}(2014)\citenamefont
  {Ramakrishnan}, \citenamefont {Dral}, \citenamefont {Rupp},\ and\
  \citenamefont {Von~Lilienfeld}}]{rama+14sd}%
  \BibitemOpen
  \bibfield  {author} {\bibinfo {author} {\bibfnamefont {R.}~\bibnamefont
  {Ramakrishnan}}, \bibinfo {author} {\bibfnamefont {P.~O.}\ \bibnamefont
  {Dral}}, \bibinfo {author} {\bibfnamefont {M.}~\bibnamefont {Rupp}}, \ and\
  \bibinfo {author} {\bibfnamefont {O.~A.}\ \bibnamefont {Von~Lilienfeld}},\
  }\href {\doibase 10.1038/sdata.2014.22} {\bibfield  {journal} {\bibinfo
  {journal} {Sci. Data}\ }\textbf {\bibinfo {volume} {1}},\ \bibinfo {pages}
  {1} (\bibinfo {year} {2014})}\BibitemShut {NoStop}%
\bibitem [{\citenamefont {Smith}\ \emph {et~al.}(2017)\citenamefont {Smith},
  \citenamefont {Isayev},\ and\ \citenamefont {Roitberg}}]{smit+17cs}%
  \BibitemOpen
  \bibfield  {author} {\bibinfo {author} {\bibfnamefont {J.~S.}\ \bibnamefont
  {Smith}}, \bibinfo {author} {\bibfnamefont {O.}~\bibnamefont {Isayev}}, \
  and\ \bibinfo {author} {\bibfnamefont {A.~E.}\ \bibnamefont {Roitberg}},\
  }\href {\doibase 10.1039/C6SC05720A} {\bibfield  {journal} {\bibinfo
  {journal} {Chem. Sci.}\ }\textbf {\bibinfo {volume} {8}},\ \bibinfo {pages}
  {3192} (\bibinfo {year} {2017})}\BibitemShut {NoStop}%
\bibitem [{\citenamefont {Hoja}\ \emph {et~al.}(2021)\citenamefont {Hoja},
  \citenamefont {Medrano~Sandonas}, \citenamefont {Ernst}, \citenamefont
  {{Vazquez-Mayagoitia}}, \citenamefont {DiStasio},\ and\ \citenamefont
  {Tkatchenko}}]{hoja+21sd}%
  \BibitemOpen
  \bibfield  {author} {\bibinfo {author} {\bibfnamefont {J.}~\bibnamefont
  {Hoja}}, \bibinfo {author} {\bibfnamefont {L.}~\bibnamefont
  {Medrano~Sandonas}}, \bibinfo {author} {\bibfnamefont {B.~G.}\ \bibnamefont
  {Ernst}}, \bibinfo {author} {\bibfnamefont {A.}~\bibnamefont
  {{Vazquez-Mayagoitia}}}, \bibinfo {author} {\bibfnamefont {R.~A.}\
  \bibnamefont {DiStasio}}, \ and\ \bibinfo {author} {\bibfnamefont
  {A.}~\bibnamefont {Tkatchenko}},\ }\href {\doibase
  10.1038/s41597-021-00812-2} {\bibfield  {journal} {\bibinfo  {journal} {Sci
  Data}\ }\textbf {\bibinfo {volume} {8}},\ \bibinfo {pages} {43} (\bibinfo
  {year} {2021})}\BibitemShut {NoStop}%
\bibitem [{\citenamefont {Jackson}\ \emph {et~al.}(2021)\citenamefont
  {Jackson}, \citenamefont {Zhang},\ and\ \citenamefont {Pearson}}]{jack+21cs}%
  \BibitemOpen
  \bibfield  {author} {\bibinfo {author} {\bibfnamefont {R.}~\bibnamefont
  {Jackson}}, \bibinfo {author} {\bibfnamefont {W.}~\bibnamefont {Zhang}}, \
  and\ \bibinfo {author} {\bibfnamefont {J.}~\bibnamefont {Pearson}},\ }\href
  {\doibase 10.1039/D1SC01206A} {\bibfield  {journal} {\bibinfo  {journal}
  {Chem. Sci.}\ }\textbf {\bibinfo {volume} {12}},\ \bibinfo {pages} {10022}
  (\bibinfo {year} {2021})}\BibitemShut {NoStop}%
\bibitem [{\citenamefont {Chmiela}\ \emph {et~al.}(2018)\citenamefont
  {Chmiela}, \citenamefont {Sauceda}, \citenamefont {M{\"u}ller},\ and\
  \citenamefont {Tkatchenko}}]{chmi+18nc}%
  \BibitemOpen
  \bibfield  {author} {\bibinfo {author} {\bibfnamefont {S.}~\bibnamefont
  {Chmiela}}, \bibinfo {author} {\bibfnamefont {H.~E.}\ \bibnamefont
  {Sauceda}}, \bibinfo {author} {\bibfnamefont {K.-R.}\ \bibnamefont
  {M{\"u}ller}}, \ and\ \bibinfo {author} {\bibfnamefont {A.}~\bibnamefont
  {Tkatchenko}},\ }\href {\doibase 10.1038/s41467-018-06169-2} {\bibfield
  {journal} {\bibinfo  {journal} {Nat Commun}\ }\textbf {\bibinfo {volume}
  {9}},\ \bibinfo {pages} {3887} (\bibinfo {year} {2018})}\BibitemShut
  {NoStop}%
\bibitem [{\citenamefont {Zuo}\ \emph {et~al.}(2020)\citenamefont {Zuo},
  \citenamefont {Chen}, \citenamefont {Li}, \citenamefont {Deng}, \citenamefont
  {Chen}, \citenamefont {Behler}, \citenamefont {Cs{\'a}nyi}, \citenamefont
  {Shapeev}, \citenamefont {Thompson}, \citenamefont {Wood},\ and\
  \citenamefont {Ong}}]{zuo+20jpcl}%
  \BibitemOpen
  \bibfield  {author} {\bibinfo {author} {\bibfnamefont {Y.}~\bibnamefont
  {Zuo}}, \bibinfo {author} {\bibfnamefont {C.}~\bibnamefont {Chen}}, \bibinfo
  {author} {\bibfnamefont {X.}~\bibnamefont {Li}}, \bibinfo {author}
  {\bibfnamefont {Z.}~\bibnamefont {Deng}}, \bibinfo {author} {\bibfnamefont
  {Y.}~\bibnamefont {Chen}}, \bibinfo {author} {\bibfnamefont {J.}~\bibnamefont
  {Behler}}, \bibinfo {author} {\bibfnamefont {G.}~\bibnamefont {Cs{\'a}nyi}},
  \bibinfo {author} {\bibfnamefont {A.~V.}\ \bibnamefont {Shapeev}}, \bibinfo
  {author} {\bibfnamefont {A.~P.}\ \bibnamefont {Thompson}}, \bibinfo {author}
  {\bibfnamefont {M.~A.}\ \bibnamefont {Wood}}, \ and\ \bibinfo {author}
  {\bibfnamefont {S.~P.}\ \bibnamefont {Ong}},\ }\href {\doibase
  10.1021/acs.jpca.9b08723} {\bibfield  {journal} {\bibinfo  {journal} {J.
  Phys. Chem. A}\ ,\ \bibinfo {pages} {acs.jpca.9b08723}} (\bibinfo {year}
  {2020})}\BibitemShut {NoStop}%
\bibitem [{\citenamefont {Rowe}\ \emph {et~al.}(2020)\citenamefont {Rowe},
  \citenamefont {Deringer}, \citenamefont {Gasparotto}, \citenamefont
  {Cs{\'a}nyi},\ and\ \citenamefont {Michaelides}}]{rowe+20jcp}%
  \BibitemOpen
  \bibfield  {author} {\bibinfo {author} {\bibfnamefont {P.}~\bibnamefont
  {Rowe}}, \bibinfo {author} {\bibfnamefont {V.~L.}\ \bibnamefont {Deringer}},
  \bibinfo {author} {\bibfnamefont {P.}~\bibnamefont {Gasparotto}}, \bibinfo
  {author} {\bibfnamefont {G.}~\bibnamefont {Cs{\'a}nyi}}, \ and\ \bibinfo
  {author} {\bibfnamefont {A.}~\bibnamefont {Michaelides}},\ }\href {\doibase
  10.1063/5.0005084} {\bibfield  {journal} {\bibinfo  {journal} {J. Chem.
  Phys.}\ }\textbf {\bibinfo {volume} {153}},\ \bibinfo {pages} {034702}
  (\bibinfo {year} {2020})}\BibitemShut {NoStop}%
\bibitem [{\citenamefont {Elstner}\ \emph {et~al.}(1998)\citenamefont
  {Elstner}, \citenamefont {Porezag}, \citenamefont {Jungnickel}, \citenamefont
  {Elsner}, \citenamefont {Haugk}, \citenamefont {Frauenheim}, \citenamefont
  {Suhai},\ and\ \citenamefont {Seifert}}]{elst+98prb}%
  \BibitemOpen
  \bibfield  {author} {\bibinfo {author} {\bibfnamefont {M.}~\bibnamefont
  {Elstner}}, \bibinfo {author} {\bibfnamefont {D.}~\bibnamefont {Porezag}},
  \bibinfo {author} {\bibfnamefont {G.}~\bibnamefont {Jungnickel}}, \bibinfo
  {author} {\bibfnamefont {J.}~\bibnamefont {Elsner}}, \bibinfo {author}
  {\bibfnamefont {M.}~\bibnamefont {Haugk}}, \bibinfo {author} {\bibfnamefont
  {T.}~\bibnamefont {Frauenheim}}, \bibinfo {author} {\bibfnamefont
  {S.}~\bibnamefont {Suhai}}, \ and\ \bibinfo {author} {\bibfnamefont
  {G.}~\bibnamefont {Seifert}},\ }\href {\doibase 10.1103/PhysRevB.58.7260}
  {\bibfield  {journal} {\bibinfo  {journal} {Phys. Rev. B}\ }\textbf {\bibinfo
  {volume} {58}},\ \bibinfo {pages} {7260} (\bibinfo {year}
  {1998})}\BibitemShut {NoStop}%
\bibitem [{\citenamefont {Ramakrishnan}\ \emph {et~al.}(2015)\citenamefont
  {Ramakrishnan}, \citenamefont {Dral}, \citenamefont {Rupp},\ and\
  \citenamefont {Von~Lilienfeld}}]{rama+15jctc}%
  \BibitemOpen
  \bibfield  {author} {\bibinfo {author} {\bibfnamefont {R.}~\bibnamefont
  {Ramakrishnan}}, \bibinfo {author} {\bibfnamefont {P.~O.}\ \bibnamefont
  {Dral}}, \bibinfo {author} {\bibfnamefont {M.}~\bibnamefont {Rupp}}, \ and\
  \bibinfo {author} {\bibfnamefont {O.~A.}\ \bibnamefont {Von~Lilienfeld}},\
  }\href {\doibase 10.1021/acs.jctc.5b00099} {\bibfield  {journal} {\bibinfo
  {journal} {J. Chem. Theory Comput.}\ }\textbf {\bibinfo {volume} {11}},\
  \bibinfo {pages} {2087} (\bibinfo {year} {2015})}\BibitemShut {NoStop}%
\bibitem [{\citenamefont {Sun}\ and\ \citenamefont
  {Sautet}(2019)}]{sun-saut19jctc}%
  \BibitemOpen
  \bibfield  {author} {\bibinfo {author} {\bibfnamefont {G.}~\bibnamefont
  {Sun}}\ and\ \bibinfo {author} {\bibfnamefont {P.}~\bibnamefont {Sautet}},\
  }\href {\doibase 10.1021/acs.jctc.9b00465} {\bibfield  {journal} {\bibinfo
  {journal} {J. Chem. Theory Comput.}\ }\textbf {\bibinfo {volume} {15}},\
  \bibinfo {pages} {5614} (\bibinfo {year} {2019})}\BibitemShut {NoStop}%
\bibitem [{\citenamefont {Tuckerman}\ \emph {et~al.}(1992)\citenamefont
  {Tuckerman}, \citenamefont {Berne},\ and\ \citenamefont
  {Martyna}}]{tuck+92jcp}%
  \BibitemOpen
  \bibfield  {author} {\bibinfo {author} {\bibfnamefont {M.}~\bibnamefont
  {Tuckerman}}, \bibinfo {author} {\bibfnamefont {B.~J.}\ \bibnamefont
  {Berne}}, \ and\ \bibinfo {author} {\bibfnamefont {G.~J.}\ \bibnamefont
  {Martyna}},\ }\href {\doibase 10.1063/1.463137} {\bibfield  {journal}
  {\bibinfo  {journal} {J. Chem. Phys.}\ }\textbf {\bibinfo {volume} {97}},\
  \bibinfo {pages} {1990} (\bibinfo {year} {1992})}\BibitemShut {NoStop}%
\bibitem [{\citenamefont {Kapil}\ \emph {et~al.}(2016)\citenamefont {Kapil},
  \citenamefont {VandeVondele},\ and\ \citenamefont {Ceriotti}}]{kapi+16jcp}%
  \BibitemOpen
  \bibfield  {author} {\bibinfo {author} {\bibfnamefont {V.}~\bibnamefont
  {Kapil}}, \bibinfo {author} {\bibfnamefont {J.}~\bibnamefont {VandeVondele}},
  \ and\ \bibinfo {author} {\bibfnamefont {M.}~\bibnamefont {Ceriotti}},\
  }\href {\doibase 10.1063/1.4941091} {\bibfield  {journal} {\bibinfo
  {journal} {J. Chem. Phys.}\ }\textbf {\bibinfo {volume} {144}},\ \bibinfo
  {pages} {054111} (\bibinfo {year} {2016})}\BibitemShut {NoStop}%
\bibitem [{\citenamefont {Rossi}\ \emph {et~al.}(2020)\citenamefont {Rossi},
  \citenamefont {Jur{\'a}skov{\'a}}, \citenamefont {Wischert}, \citenamefont
  {Garel}, \citenamefont {Corminboeuf},\ and\ \citenamefont
  {Ceriotti}}]{ross+20jctc}%
  \BibitemOpen
  \bibfield  {author} {\bibinfo {author} {\bibfnamefont {K.}~\bibnamefont
  {Rossi}}, \bibinfo {author} {\bibfnamefont {V.}~\bibnamefont
  {Jur{\'a}skov{\'a}}}, \bibinfo {author} {\bibfnamefont {R.}~\bibnamefont
  {Wischert}}, \bibinfo {author} {\bibfnamefont {L.}~\bibnamefont {Garel}},
  \bibinfo {author} {\bibfnamefont {C.}~\bibnamefont {Corminboeuf}}, \ and\
  \bibinfo {author} {\bibfnamefont {M.}~\bibnamefont {Ceriotti}},\ }\href
  {\doibase 10.1021/acs.jctc.0c00362} {\bibfield  {journal} {\bibinfo
  {journal} {J. Chem. Theory Comput.}\ }\textbf {\bibinfo {volume} {16}},\
  \bibinfo {pages} {5139} (\bibinfo {year} {2020})}\BibitemShut {NoStop}%
\bibitem [{\citenamefont {Li}\ \emph {et~al.}(2015)\citenamefont {Li},
  \citenamefont {Kermode},\ and\ \citenamefont {De~Vita}}]{li+15prl}%
  \BibitemOpen
  \bibfield  {author} {\bibinfo {author} {\bibfnamefont {Z.}~\bibnamefont
  {Li}}, \bibinfo {author} {\bibfnamefont {J.~R.}\ \bibnamefont {Kermode}}, \
  and\ \bibinfo {author} {\bibfnamefont {A.}~\bibnamefont {De~Vita}},\ }\href
  {\doibase 10.1103/PhysRevLett.114.096405} {\bibfield  {journal} {\bibinfo
  {journal} {Phys. Rev. Lett.}\ }\textbf {\bibinfo {volume} {114}},\ \bibinfo
  {pages} {096405} (\bibinfo {year} {2015})}\BibitemShut {NoStop}%
\bibitem [{\citenamefont {Gubaev}\ \emph {et~al.}(2018)\citenamefont {Gubaev},
  \citenamefont {Podryabinkin},\ and\ \citenamefont {Shapeev}}]{guba+18jcp}%
  \BibitemOpen
  \bibfield  {author} {\bibinfo {author} {\bibfnamefont {K.}~\bibnamefont
  {Gubaev}}, \bibinfo {author} {\bibfnamefont {E.~V.}\ \bibnamefont
  {Podryabinkin}}, \ and\ \bibinfo {author} {\bibfnamefont {A.~V.}\
  \bibnamefont {Shapeev}},\ }\href {\doibase 10.1063/1.5005095} {\bibfield
  {journal} {\bibinfo  {journal} {The Journal of Chemical Physics}\ }\textbf
  {\bibinfo {volume} {148}},\ \bibinfo {pages} {241727} (\bibinfo {year}
  {2018})}\BibitemShut {NoStop}%
\bibitem [{\citenamefont {Zhang}\ \emph {et~al.}(2019)\citenamefont {Zhang},
  \citenamefont {Lin}, \citenamefont {Wang}, \citenamefont {Car},\ and\
  \citenamefont {E}}]{zhan+19prm}%
  \BibitemOpen
  \bibfield  {author} {\bibinfo {author} {\bibfnamefont {L.}~\bibnamefont
  {Zhang}}, \bibinfo {author} {\bibfnamefont {D.-Y.}\ \bibnamefont {Lin}},
  \bibinfo {author} {\bibfnamefont {H.}~\bibnamefont {Wang}}, \bibinfo {author}
  {\bibfnamefont {R.}~\bibnamefont {Car}}, \ and\ \bibinfo {author}
  {\bibfnamefont {W.}~\bibnamefont {E}},\ }\href {\doibase
  10.1103/PhysRevMaterials.3.023804} {\bibfield  {journal} {\bibinfo  {journal}
  {Phys. Rev. Materials}\ }\textbf {\bibinfo {volume} {3}},\ \bibinfo {pages}
  {023804} (\bibinfo {year} {2019})}\BibitemShut {NoStop}%
\bibitem [{\citenamefont {Shapeev}\ \emph {et~al.}(2020)\citenamefont
  {Shapeev}, \citenamefont {Gubaev}, \citenamefont {Tsymbalov},\ and\
  \citenamefont {Podryabinkin}}]{shap+20book}%
  \BibitemOpen
  \bibfield  {author} {\bibinfo {author} {\bibfnamefont {A.}~\bibnamefont
  {Shapeev}}, \bibinfo {author} {\bibfnamefont {K.}~\bibnamefont {Gubaev}},
  \bibinfo {author} {\bibfnamefont {E.}~\bibnamefont {Tsymbalov}}, \ and\
  \bibinfo {author} {\bibfnamefont {E.}~\bibnamefont {Podryabinkin}},\ }in\
  \href {\doibase 10.1007/978-3-030-40245-7_15} {\emph {\bibinfo {booktitle}
  {Machine {{Learning Meets Quantum Physics}}}}},\ Vol.\ \bibinfo {volume}
  {968},\ \bibinfo {editor} {edited by\ \bibinfo {editor} {\bibfnamefont
  {K.~T.}\ \bibnamefont {Sch{\"u}tt}}, \bibinfo {editor} {\bibfnamefont
  {S.}~\bibnamefont {Chmiela}}, \bibinfo {editor} {\bibfnamefont {O.~A.}\
  \bibnamefont {{\noopsort{lilienfeld}}{von Lilienfeld}}}, \bibinfo {editor}
  {\bibfnamefont {A.}~\bibnamefont {Tkatchenko}}, \bibinfo {editor}
  {\bibfnamefont {K.}~\bibnamefont {Tsuda}}, \ and\ \bibinfo {editor}
  {\bibfnamefont {K.-R.}\ \bibnamefont {M{\"u}ller}}}\ (\bibinfo  {publisher}
  {{Springer International Publishing}},\ \bibinfo {address} {{Cham}},\
  \bibinfo {year} {2020})\ pp.\ \bibinfo {pages} {309--329}\BibitemShut
  {NoStop}%
\bibitem [{\citenamefont {Deringer}\ \emph
  {et~al.}(2021{\natexlab{a}})\citenamefont {Deringer}, \citenamefont
  {Bart{\'o}k}, \citenamefont {Bernstein}, \citenamefont {Wilkins},
  \citenamefont {Ceriotti},\ and\ \citenamefont {Cs{\'a}nyi}}]{deri+21cr}%
  \BibitemOpen
  \bibfield  {author} {\bibinfo {author} {\bibfnamefont {V.~L.}\ \bibnamefont
  {Deringer}}, \bibinfo {author} {\bibfnamefont {A.~P.}\ \bibnamefont
  {Bart{\'o}k}}, \bibinfo {author} {\bibfnamefont {N.}~\bibnamefont
  {Bernstein}}, \bibinfo {author} {\bibfnamefont {D.~M.}\ \bibnamefont
  {Wilkins}}, \bibinfo {author} {\bibfnamefont {M.}~\bibnamefont {Ceriotti}}, \
  and\ \bibinfo {author} {\bibfnamefont {G.}~\bibnamefont {Cs{\'a}nyi}},\
  }\href {\doibase 10.1021/acs.chemrev.1c00022} {\bibfield  {journal} {\bibinfo
   {journal} {Chem. Rev.}\ }\textbf {\bibinfo {volume} {121}},\ \bibinfo
  {pages} {10073} (\bibinfo {year} {2021}{\natexlab{a}})}\BibitemShut {NoStop}%
\bibitem [{\citenamefont {Breiman}(1996)}]{brei96ml}%
  \BibitemOpen
  \bibfield  {author} {\bibinfo {author} {\bibfnamefont {L.}~\bibnamefont
  {Breiman}},\ }\href {\doibase 10.1007/BF00058655} {\bibfield  {journal}
  {\bibinfo  {journal} {Mach Learn}\ }\textbf {\bibinfo {volume} {24}},\
  \bibinfo {pages} {123} (\bibinfo {year} {1996})}\BibitemShut {NoStop}%
\bibitem [{\citenamefont {Musil}\ \emph {et~al.}(2019)\citenamefont {Musil},
  \citenamefont {Willatt}, \citenamefont {Langovoy},\ and\ \citenamefont
  {Ceriotti}}]{musi+19jctc}%
  \BibitemOpen
  \bibfield  {author} {\bibinfo {author} {\bibfnamefont {F.}~\bibnamefont
  {Musil}}, \bibinfo {author} {\bibfnamefont {M.~J.}\ \bibnamefont {Willatt}},
  \bibinfo {author} {\bibfnamefont {M.~A.}\ \bibnamefont {Langovoy}}, \ and\
  \bibinfo {author} {\bibfnamefont {M.}~\bibnamefont {Ceriotti}},\ }\href
  {\doibase 10.1021/acs.jctc.8b00959} {\bibfield  {journal} {\bibinfo
  {journal} {J. Chem. Theory Comput.}\ }\textbf {\bibinfo {volume} {15}},\
  \bibinfo {pages} {906} (\bibinfo {year} {2019})}\BibitemShut {NoStop}%
\bibitem [{\citenamefont {Imbalzano}\ \emph {et~al.}(2021)\citenamefont
  {Imbalzano}, \citenamefont {Zhuang}, \citenamefont {Kapil}, \citenamefont
  {Rossi}, \citenamefont {Engel}, \citenamefont {Grasselli},\ and\
  \citenamefont {Ceriotti}}]{imba+21jcp}%
  \BibitemOpen
  \bibfield  {author} {\bibinfo {author} {\bibfnamefont {G.}~\bibnamefont
  {Imbalzano}}, \bibinfo {author} {\bibfnamefont {Y.}~\bibnamefont {Zhuang}},
  \bibinfo {author} {\bibfnamefont {V.}~\bibnamefont {Kapil}}, \bibinfo
  {author} {\bibfnamefont {K.}~\bibnamefont {Rossi}}, \bibinfo {author}
  {\bibfnamefont {E.~A.}\ \bibnamefont {Engel}}, \bibinfo {author}
  {\bibfnamefont {F.}~\bibnamefont {Grasselli}}, \ and\ \bibinfo {author}
  {\bibfnamefont {M.}~\bibnamefont {Ceriotti}},\ }\href {\doibase
  10.1063/5.0036522} {\bibfield  {journal} {\bibinfo  {journal} {J. Chem.
  Phys.}\ }\textbf {\bibinfo {volume} {154}},\ \bibinfo {pages} {074102}
  (\bibinfo {year} {2021})}\BibitemShut {NoStop}%
\bibitem [{\citenamefont {Baroni}\ \emph {et~al.}(2001)\citenamefont {Baroni},
  \citenamefont {De~Gironcoli}, \citenamefont {Dal~Corso},\ and\ \citenamefont
  {Giannozzi}}]{baro+01rmp}%
  \BibitemOpen
  \bibfield  {author} {\bibinfo {author} {\bibfnamefont {S.}~\bibnamefont
  {Baroni}}, \bibinfo {author} {\bibfnamefont {S.}~\bibnamefont
  {De~Gironcoli}}, \bibinfo {author} {\bibfnamefont {A.}~\bibnamefont
  {Dal~Corso}}, \ and\ \bibinfo {author} {\bibfnamefont {P.}~\bibnamefont
  {Giannozzi}},\ }\href@noop {} {\bibfield  {journal} {\bibinfo  {journal}
  {Rev. Mod. Phys.}\ }\textbf {\bibinfo {volume} {73}},\ \bibinfo {pages} {515}
  (\bibinfo {year} {2001})}\BibitemShut {NoStop}%
\bibitem [{\citenamefont {Pagliai}\ \emph {et~al.}(2008)\citenamefont
  {Pagliai}, \citenamefont {Cavazzoni}, \citenamefont {Cardini}, \citenamefont
  {Erbacci}, \citenamefont {Parrinello},\ and\ \citenamefont
  {Schettino}}]{pagl+08jcp}%
  \BibitemOpen
  \bibfield  {author} {\bibinfo {author} {\bibfnamefont {M.}~\bibnamefont
  {Pagliai}}, \bibinfo {author} {\bibfnamefont {C.}~\bibnamefont {Cavazzoni}},
  \bibinfo {author} {\bibfnamefont {G.}~\bibnamefont {Cardini}}, \bibinfo
  {author} {\bibfnamefont {G.}~\bibnamefont {Erbacci}}, \bibinfo {author}
  {\bibfnamefont {M.}~\bibnamefont {Parrinello}}, \ and\ \bibinfo {author}
  {\bibfnamefont {V.}~\bibnamefont {Schettino}},\ }\href {\doibase
  10.1063/1.2936988} {\bibfield  {journal} {\bibinfo  {journal} {The Journal of
  Chemical Physics}\ }\textbf {\bibinfo {volume} {128}},\ \bibinfo {pages}
  {224514} (\bibinfo {year} {2008})}\BibitemShut {NoStop}%
\bibitem [{\citenamefont {Yates}\ \emph {et~al.}(2007)\citenamefont {Yates},
  \citenamefont {Pickard},\ and\ \citenamefont {Mauri}}]{yate+07prb}%
  \BibitemOpen
  \bibfield  {author} {\bibinfo {author} {\bibfnamefont {J.~R.}\ \bibnamefont
  {Yates}}, \bibinfo {author} {\bibfnamefont {C.~J.}\ \bibnamefont {Pickard}},
  \ and\ \bibinfo {author} {\bibfnamefont {F.}~\bibnamefont {Mauri}},\ }\href
  {\doibase 10.1103/PhysRevB.76.024401} {\bibfield  {journal} {\bibinfo
  {journal} {Phys. Rev. B}\ }\textbf {\bibinfo {volume} {76}},\ \bibinfo
  {pages} {024401} (\bibinfo {year} {2007})}\BibitemShut {NoStop}%
\bibitem [{\citenamefont {{King-Smith}}\ and\ \citenamefont
  {Vanderbilt}(1994)}]{king-vand94prb}%
  \BibitemOpen
  \bibfield  {author} {\bibinfo {author} {\bibfnamefont {R.~D.}\ \bibnamefont
  {{King-Smith}}}\ and\ \bibinfo {author} {\bibfnamefont {D.}~\bibnamefont
  {Vanderbilt}},\ }\href@noop {} {\bibfield  {journal} {\bibinfo  {journal}
  {Phys. Rev. B}\ }\textbf {\bibinfo {volume} {49}},\ \bibinfo {pages} {5828}
  (\bibinfo {year} {1994})}\BibitemShut {NoStop}%
\bibitem [{\citenamefont {Paruzzo}\ \emph {et~al.}(2018)\citenamefont
  {Paruzzo}, \citenamefont {Hofstetter}, \citenamefont {Musil}, \citenamefont
  {De}, \citenamefont {Ceriotti},\ and\ \citenamefont {Emsley}}]{paru+18ncomm}%
  \BibitemOpen
  \bibfield  {author} {\bibinfo {author} {\bibfnamefont {F.~M.}\ \bibnamefont
  {Paruzzo}}, \bibinfo {author} {\bibfnamefont {A.}~\bibnamefont {Hofstetter}},
  \bibinfo {author} {\bibfnamefont {F.}~\bibnamefont {Musil}}, \bibinfo
  {author} {\bibfnamefont {S.}~\bibnamefont {De}}, \bibinfo {author}
  {\bibfnamefont {M.}~\bibnamefont {Ceriotti}}, \ and\ \bibinfo {author}
  {\bibfnamefont {L.}~\bibnamefont {Emsley}},\ }\href {\doibase
  10.1038/s41467-018-06972-x} {\bibfield  {journal} {\bibinfo  {journal} {Nat.
  Commun.}\ }\textbf {\bibinfo {volume} {9}},\ \bibinfo {pages} {4501}
  (\bibinfo {year} {2018})}\BibitemShut {NoStop}%
\bibitem [{\citenamefont {Liu}\ \emph {et~al.}(2019)\citenamefont {Liu},
  \citenamefont {Li}, \citenamefont {Bennett}, \citenamefont {Ganoe},
  \citenamefont {Stauch}, \citenamefont {{Head-Gordon}}, \citenamefont
  {Hexemer}, \citenamefont {Ushizima},\ and\ \citenamefont
  {{Head-Gordon}}}]{liu+19jpcl}%
  \BibitemOpen
  \bibfield  {author} {\bibinfo {author} {\bibfnamefont {S.}~\bibnamefont
  {Liu}}, \bibinfo {author} {\bibfnamefont {J.}~\bibnamefont {Li}}, \bibinfo
  {author} {\bibfnamefont {K.~C.}\ \bibnamefont {Bennett}}, \bibinfo {author}
  {\bibfnamefont {B.}~\bibnamefont {Ganoe}}, \bibinfo {author} {\bibfnamefont
  {T.}~\bibnamefont {Stauch}}, \bibinfo {author} {\bibfnamefont
  {M.}~\bibnamefont {{Head-Gordon}}}, \bibinfo {author} {\bibfnamefont
  {A.}~\bibnamefont {Hexemer}}, \bibinfo {author} {\bibfnamefont
  {D.}~\bibnamefont {Ushizima}}, \ and\ \bibinfo {author} {\bibfnamefont
  {T.}~\bibnamefont {{Head-Gordon}}},\ }\href {\doibase
  10.1021/acs.jpclett.9b01570} {\bibfield  {journal} {\bibinfo  {journal} {J.
  Phys. Chem. Lett.}\ }\textbf {\bibinfo {volume} {10}},\ \bibinfo {pages}
  {4558} (\bibinfo {year} {2019})}\BibitemShut {NoStop}%
\bibitem [{\citenamefont {Pozdnyakov}\ \emph {et~al.}(2020)\citenamefont
  {Pozdnyakov}, \citenamefont {Willatt}, \citenamefont {Bart{\'o}k},
  \citenamefont {Ortner}, \citenamefont {Cs{\'a}nyi},\ and\ \citenamefont
  {Ceriotti}}]{pozd+20prl}%
  \BibitemOpen
  \bibfield  {author} {\bibinfo {author} {\bibfnamefont {S.~N.}\ \bibnamefont
  {Pozdnyakov}}, \bibinfo {author} {\bibfnamefont {M.~J.}\ \bibnamefont
  {Willatt}}, \bibinfo {author} {\bibfnamefont {A.~P.}\ \bibnamefont
  {Bart{\'o}k}}, \bibinfo {author} {\bibfnamefont {C.}~\bibnamefont {Ortner}},
  \bibinfo {author} {\bibfnamefont {G.}~\bibnamefont {Cs{\'a}nyi}}, \ and\
  \bibinfo {author} {\bibfnamefont {M.}~\bibnamefont {Ceriotti}},\ }\href
  {\doibase 10.1103/PhysRevLett.125.166001} {\bibfield  {journal} {\bibinfo
  {journal} {Phys. Rev. Lett.}\ }\textbf {\bibinfo {volume} {125}},\ \bibinfo
  {pages} {166001} (\bibinfo {year} {2020})}\BibitemShut {NoStop}%
\bibitem [{\citenamefont {Ben~Mahmoud}\ \emph {et~al.}(2020)\citenamefont
  {Ben~Mahmoud}, \citenamefont {Anelli}, \citenamefont {Cs{\'a}nyi},\ and\
  \citenamefont {Ceriotti}}]{benm+20prb}%
  \BibitemOpen
  \bibfield  {author} {\bibinfo {author} {\bibfnamefont {C.}~\bibnamefont
  {Ben~Mahmoud}}, \bibinfo {author} {\bibfnamefont {A.}~\bibnamefont {Anelli}},
  \bibinfo {author} {\bibfnamefont {G.}~\bibnamefont {Cs{\'a}nyi}}, \ and\
  \bibinfo {author} {\bibfnamefont {M.}~\bibnamefont {Ceriotti}},\ }\href
  {\doibase 10.1103/PhysRevB.102.235130} {\bibfield  {journal} {\bibinfo
  {journal} {Phys. Rev. B}\ }\textbf {\bibinfo {volume} {102}},\ \bibinfo
  {pages} {235130} (\bibinfo {year} {2020})}\BibitemShut {NoStop}%
\bibitem [{Note1()}]{Note1}%
  \BibitemOpen
  \bibinfo {note} {Strictly speaking, atom-centered descriptors are equivariant
  to permutations of atoms labels, and so are the predictions of atom-centered
  models of energy contributions. The \protect \emph {sum} of these terms,
  however, is permutation invariant.}\BibitemShut {Stop}%
\bibitem [{\citenamefont {Bereau}\ \emph {et~al.}(2015)\citenamefont {Bereau},
  \citenamefont {Andrienko},\ and\ \citenamefont
  {Von~Lilienfeld}}]{bere+15jctc}%
  \BibitemOpen
  \bibfield  {author} {\bibinfo {author} {\bibfnamefont {T.}~\bibnamefont
  {Bereau}}, \bibinfo {author} {\bibfnamefont {D.}~\bibnamefont {Andrienko}}, \
  and\ \bibinfo {author} {\bibfnamefont {O.~A.}\ \bibnamefont
  {Von~Lilienfeld}},\ }\href {\doibase 10.1021/acs.jctc.5b00301} {\bibfield
  {journal} {\bibinfo  {journal} {J. Chem. Theory Comput.}\ }\textbf {\bibinfo
  {volume} {11}},\ \bibinfo {pages} {3225} (\bibinfo {year}
  {2015})}\BibitemShut {NoStop}%
\bibitem [{\citenamefont {Liang}\ \emph {et~al.}(2017)\citenamefont {Liang},
  \citenamefont {Tocci}, \citenamefont {Wilkins}, \citenamefont {Grisafi},
  \citenamefont {Roke},\ and\ \citenamefont {Ceriotti}}]{lian+17prb}%
  \BibitemOpen
  \bibfield  {author} {\bibinfo {author} {\bibfnamefont {C.}~\bibnamefont
  {Liang}}, \bibinfo {author} {\bibfnamefont {G.}~\bibnamefont {Tocci}},
  \bibinfo {author} {\bibfnamefont {D.~M.}\ \bibnamefont {Wilkins}}, \bibinfo
  {author} {\bibfnamefont {A.}~\bibnamefont {Grisafi}}, \bibinfo {author}
  {\bibfnamefont {S.}~\bibnamefont {Roke}}, \ and\ \bibinfo {author}
  {\bibfnamefont {M.}~\bibnamefont {Ceriotti}},\ }\href {\doibase
  10.1103/PhysRevB.96.041407} {\bibfield  {journal} {\bibinfo  {journal} {Phys.
  Rev. B}\ }\textbf {\bibinfo {volume} {96}},\ \bibinfo {pages} {041407}
  (\bibinfo {year} {2017})}\BibitemShut {NoStop}%
\bibitem [{\citenamefont {Glielmo}\ \emph {et~al.}(2017)\citenamefont
  {Glielmo}, \citenamefont {Sollich},\ and\ \citenamefont
  {De~Vita}}]{glie+17prb}%
  \BibitemOpen
  \bibfield  {author} {\bibinfo {author} {\bibfnamefont {A.}~\bibnamefont
  {Glielmo}}, \bibinfo {author} {\bibfnamefont {P.}~\bibnamefont {Sollich}}, \
  and\ \bibinfo {author} {\bibfnamefont {A.}~\bibnamefont {De~Vita}},\ }\href
  {\doibase 10.1103/PhysRevB.95.214302} {\bibfield  {journal} {\bibinfo
  {journal} {Phys. Rev. B}\ }\textbf {\bibinfo {volume} {95}},\ \bibinfo
  {pages} {214302} (\bibinfo {year} {2017})}\BibitemShut {NoStop}%
\bibitem [{\citenamefont {Grisafi}\ \emph {et~al.}(2018)\citenamefont
  {Grisafi}, \citenamefont {Wilkins}, \citenamefont {Cs{\'a}nyi},\ and\
  \citenamefont {Ceriotti}}]{gris+18prl}%
  \BibitemOpen
  \bibfield  {author} {\bibinfo {author} {\bibfnamefont {A.}~\bibnamefont
  {Grisafi}}, \bibinfo {author} {\bibfnamefont {D.~M.}\ \bibnamefont
  {Wilkins}}, \bibinfo {author} {\bibfnamefont {G.}~\bibnamefont {Cs{\'a}nyi}},
  \ and\ \bibinfo {author} {\bibfnamefont {M.}~\bibnamefont {Ceriotti}},\
  }\href {\doibase 10.1103/PhysRevLett.120.036002} {\bibfield  {journal}
  {\bibinfo  {journal} {Phys. Rev. Lett.}\ }\textbf {\bibinfo {volume} {120}},\
  \bibinfo {pages} {036002} (\bibinfo {year} {2018})}\BibitemShut {NoStop}%
\bibitem [{\citenamefont {Bronstein}\ \emph {et~al.}(2021)\citenamefont
  {Bronstein}, \citenamefont {Bruna}, \citenamefont {Cohen},\ and\
  \citenamefont {Veli{\v c}kovi{\'c}}}]{bron+21arxiv}%
  \BibitemOpen
  \bibfield  {author} {\bibinfo {author} {\bibfnamefont {M.~M.}\ \bibnamefont
  {Bronstein}}, \bibinfo {author} {\bibfnamefont {J.}~\bibnamefont {Bruna}},
  \bibinfo {author} {\bibfnamefont {T.}~\bibnamefont {Cohen}}, \ and\ \bibinfo
  {author} {\bibfnamefont {P.}~\bibnamefont {Veli{\v c}kovi{\'c}}},\ }\href
  {http://arxiv.org/abs/2104.13478v2} {\bibfield  {journal} {\bibinfo
  {journal} {arxiv:2104.13478}\ } (\bibinfo {year} {2021})}\BibitemShut
  {NoStop}%
\bibitem [{\citenamefont {A~Morrison}\ and\ \citenamefont
  {A~Parker}(1987)}]{morri-park87ajp}%
  \BibitemOpen
  \bibfield  {author} {\bibinfo {author} {\bibfnamefont {M.}~\bibnamefont
  {A~Morrison}}\ and\ \bibinfo {author} {\bibfnamefont {G.}~\bibnamefont
  {A~Parker}},\ }\href {\doibase 10.1071/ph870465} {\bibfield  {journal}
  {\bibinfo  {journal} {Aust. J. Phys.}\ }\textbf {\bibinfo {volume} {40}},\
  \bibinfo {pages} {465} (\bibinfo {year} {1987})}\BibitemShut {NoStop}%
\bibitem [{\citenamefont {Nigam}\ \emph
  {et~al.}(2022{\natexlab{a}})\citenamefont {Nigam}, \citenamefont {Willatt},\
  and\ \citenamefont {Ceriotti}}]{niga+22jcp}%
  \BibitemOpen
  \bibfield  {author} {\bibinfo {author} {\bibfnamefont {J.}~\bibnamefont
  {Nigam}}, \bibinfo {author} {\bibfnamefont {M.~J.}\ \bibnamefont {Willatt}},
  \ and\ \bibinfo {author} {\bibfnamefont {M.}~\bibnamefont {Ceriotti}},\
  }\href {\doibase 10.1063/5.0072784} {\bibfield  {journal} {\bibinfo
  {journal} {J. Chem. Phys.}\ }\textbf {\bibinfo {volume} {156}},\ \bibinfo
  {pages} {014115} (\bibinfo {year} {2022}{\natexlab{a}})}\BibitemShut
  {NoStop}%
\bibitem [{\citenamefont {Veit}\ \emph {et~al.}(2020)\citenamefont {Veit},
  \citenamefont {Wilkins}, \citenamefont {Yang}, \citenamefont {DiStasio},\
  and\ \citenamefont {Ceriotti}}]{veit+20jcp}%
  \BibitemOpen
  \bibfield  {author} {\bibinfo {author} {\bibfnamefont {M.}~\bibnamefont
  {Veit}}, \bibinfo {author} {\bibfnamefont {D.~M.}\ \bibnamefont {Wilkins}},
  \bibinfo {author} {\bibfnamefont {Y.}~\bibnamefont {Yang}}, \bibinfo {author}
  {\bibfnamefont {R.~A.}\ \bibnamefont {DiStasio}}, \ and\ \bibinfo {author}
  {\bibfnamefont {M.}~\bibnamefont {Ceriotti}},\ }\href {\doibase
  10.1063/5.0009106} {\bibfield  {journal} {\bibinfo  {journal} {J. Chem.
  Phys.}\ }\textbf {\bibinfo {volume} {153}},\ \bibinfo {pages} {024113}
  (\bibinfo {year} {2020})}\BibitemShut {NoStop}%
\bibitem [{\citenamefont {Wilkins}\ \emph {et~al.}(2019)\citenamefont
  {Wilkins}, \citenamefont {Grisafi}, \citenamefont {Yang}, \citenamefont
  {Lao}, \citenamefont {DiStasio},\ and\ \citenamefont
  {Ceriotti}}]{wilk+19pnas}%
  \BibitemOpen
  \bibfield  {author} {\bibinfo {author} {\bibfnamefont {D.~M.}\ \bibnamefont
  {Wilkins}}, \bibinfo {author} {\bibfnamefont {A.}~\bibnamefont {Grisafi}},
  \bibinfo {author} {\bibfnamefont {Y.}~\bibnamefont {Yang}}, \bibinfo {author}
  {\bibfnamefont {K.~U.}\ \bibnamefont {Lao}}, \bibinfo {author} {\bibfnamefont
  {R.~A.}\ \bibnamefont {DiStasio}}, \ and\ \bibinfo {author} {\bibfnamefont
  {M.}~\bibnamefont {Ceriotti}},\ }\href {\doibase 10.1073/pnas.1816132116}
  {\bibfield  {journal} {\bibinfo  {journal} {Proc. Natl. Acad. Sci. U. S. A.}\
  }\textbf {\bibinfo {volume} {116}},\ \bibinfo {pages} {3401} (\bibinfo {year}
  {2019})}\BibitemShut {NoStop}%
\bibitem [{\citenamefont {Grisafi}\ \emph {et~al.}(2019)\citenamefont
  {Grisafi}, \citenamefont {Fabrizio}, \citenamefont {Meyer}, \citenamefont
  {Wilkins}, \citenamefont {Corminboeuf},\ and\ \citenamefont
  {Ceriotti}}]{gris+19acscs}%
  \BibitemOpen
  \bibfield  {author} {\bibinfo {author} {\bibfnamefont {A.}~\bibnamefont
  {Grisafi}}, \bibinfo {author} {\bibfnamefont {A.}~\bibnamefont {Fabrizio}},
  \bibinfo {author} {\bibfnamefont {B.}~\bibnamefont {Meyer}}, \bibinfo
  {author} {\bibfnamefont {D.~M.}\ \bibnamefont {Wilkins}}, \bibinfo {author}
  {\bibfnamefont {C.}~\bibnamefont {Corminboeuf}}, \ and\ \bibinfo {author}
  {\bibfnamefont {M.}~\bibnamefont {Ceriotti}},\ }\href {\doibase
  10.1021/acscentsci.8b00551} {\bibfield  {journal} {\bibinfo  {journal} {ACS
  Cent. Sci.}\ }\textbf {\bibinfo {volume} {5}},\ \bibinfo {pages} {57}
  (\bibinfo {year} {2019})}\BibitemShut {NoStop}%
\bibitem [{\citenamefont {Lewis}\ \emph {et~al.}(2021)\citenamefont {Lewis},
  \citenamefont {Grisafi}, \citenamefont {Ceriotti},\ and\ \citenamefont
  {Rossi}}]{lewi+21jctc}%
  \BibitemOpen
  \bibfield  {author} {\bibinfo {author} {\bibfnamefont {A.~M.}\ \bibnamefont
  {Lewis}}, \bibinfo {author} {\bibfnamefont {A.}~\bibnamefont {Grisafi}},
  \bibinfo {author} {\bibfnamefont {M.}~\bibnamefont {Ceriotti}}, \ and\
  \bibinfo {author} {\bibfnamefont {M.}~\bibnamefont {Rossi}},\ }\href
  {\doibase 10.1021/acs.jctc.1c00576} {\bibfield  {journal} {\bibinfo
  {journal} {J. Chem. Theory Comput.}\ }\textbf {\bibinfo {volume} {17}},\
  \bibinfo {pages} {7203} (\bibinfo {year} {2021})}\BibitemShut {NoStop}%
\bibitem [{\citenamefont {Anderson}\ \emph {et~al.}(2019)\citenamefont
  {Anderson}, \citenamefont {Hy},\ and\ \citenamefont {Kondor}}]{ande+19nips}%
  \BibitemOpen
  \bibfield  {author} {\bibinfo {author} {\bibfnamefont {B.}~\bibnamefont
  {Anderson}}, \bibinfo {author} {\bibfnamefont {T.~S.}\ \bibnamefont {Hy}}, \
  and\ \bibinfo {author} {\bibfnamefont {R.}~\bibnamefont {Kondor}},\ }in\
  \href@noop {} {\emph {\bibinfo {booktitle} {{{NeurIPS}}}}}\ (\bibinfo {year}
  {2019})\ p.~\bibinfo {pages} {10}\BibitemShut {NoStop}%
\bibitem [{\citenamefont {Thomas}\ \emph {et~al.}(2018)\citenamefont {Thomas},
  \citenamefont {Smidt}, \citenamefont {Kearnes}, \citenamefont {Yang},
  \citenamefont {Li}, \citenamefont {Kohlhoff},\ and\ \citenamefont
  {Riley}}]{thom+18arxiv}%
  \BibitemOpen
  \bibfield  {author} {\bibinfo {author} {\bibfnamefont {N.}~\bibnamefont
  {Thomas}}, \bibinfo {author} {\bibfnamefont {T.}~\bibnamefont {Smidt}},
  \bibinfo {author} {\bibfnamefont {S.}~\bibnamefont {Kearnes}}, \bibinfo
  {author} {\bibfnamefont {L.}~\bibnamefont {Yang}}, \bibinfo {author}
  {\bibfnamefont {L.}~\bibnamefont {Li}}, \bibinfo {author} {\bibfnamefont
  {K.}~\bibnamefont {Kohlhoff}}, \ and\ \bibinfo {author} {\bibfnamefont
  {P.}~\bibnamefont {Riley}},\ }\href {http://arxiv.org/abs/1802.08219v3}
  {\bibfield  {journal} {\bibinfo  {journal} {arxiv:1802.08219}\ } (\bibinfo
  {year} {2018})}\BibitemShut {NoStop}%
\bibitem [{\citenamefont {Batzner}\ \emph {et~al.}(2021)\citenamefont
  {Batzner}, \citenamefont {Musaelian}, \citenamefont {Sun}, \citenamefont
  {Geiger}, \citenamefont {Mailoa}, \citenamefont {Kornbluth}, \citenamefont
  {Molinari}, \citenamefont {Smidt},\ and\ \citenamefont
  {Kozinsky}}]{simo+21arxiv}%
  \BibitemOpen
  \bibfield  {author} {\bibinfo {author} {\bibfnamefont {S.}~\bibnamefont
  {Batzner}}, \bibinfo {author} {\bibfnamefont {A.}~\bibnamefont {Musaelian}},
  \bibinfo {author} {\bibfnamefont {L.}~\bibnamefont {Sun}}, \bibinfo {author}
  {\bibfnamefont {M.}~\bibnamefont {Geiger}}, \bibinfo {author} {\bibfnamefont
  {J.~P.}\ \bibnamefont {Mailoa}}, \bibinfo {author} {\bibfnamefont
  {M.}~\bibnamefont {Kornbluth}}, \bibinfo {author} {\bibfnamefont
  {N.}~\bibnamefont {Molinari}}, \bibinfo {author} {\bibfnamefont {T.~E.}\
  \bibnamefont {Smidt}}, \ and\ \bibinfo {author} {\bibfnamefont
  {B.}~\bibnamefont {Kozinsky}},\ }\href {http://arxiv.org/abs/2101.03164v3}
  {\bibfield  {journal} {\bibinfo  {journal} {arxiv:2101.03164}\ } (\bibinfo
  {year} {2021})}\BibitemShut {NoStop}%
\bibitem [{\citenamefont {Klicpera}\ \emph {et~al.}(2021)\citenamefont
  {Klicpera}, \citenamefont {Becker},\ and\ \citenamefont
  {G{\"u}nnemann}}]{klic+21arxiv}%
  \BibitemOpen
  \bibfield  {author} {\bibinfo {author} {\bibfnamefont {J.}~\bibnamefont
  {Klicpera}}, \bibinfo {author} {\bibfnamefont {F.}~\bibnamefont {Becker}}, \
  and\ \bibinfo {author} {\bibfnamefont {S.}~\bibnamefont {G{\"u}nnemann}},\
  }\href {http://arxiv.org/abs/2106.08903v8} {\bibfield  {journal} {\bibinfo
  {journal} {arxiv:2106.08903}\ } (\bibinfo {year} {2021})}\BibitemShut
  {NoStop}%
\bibitem [{\citenamefont {Gilmer}\ \emph {et~al.}(2017)\citenamefont {Gilmer},
  \citenamefont {Schoenholz}, \citenamefont {Riley}, \citenamefont {Vinyals},\
  and\ \citenamefont {Dahl}}]{gilm+17icml}%
  \BibitemOpen
  \bibfield  {author} {\bibinfo {author} {\bibfnamefont {J.}~\bibnamefont
  {Gilmer}}, \bibinfo {author} {\bibfnamefont {S.~S.}\ \bibnamefont
  {Schoenholz}}, \bibinfo {author} {\bibfnamefont {P.~F.}\ \bibnamefont
  {Riley}}, \bibinfo {author} {\bibfnamefont {O.}~\bibnamefont {Vinyals}}, \
  and\ \bibinfo {author} {\bibfnamefont {G.~E.}\ \bibnamefont {Dahl}},\ }in\
  \href@noop {} {\emph {\bibinfo {booktitle} {International Conference on
  Machine Learning}}}\ (\bibinfo {year} {2017})\ pp.\ \bibinfo {pages}
  {1263--1272}\BibitemShut {NoStop}%
\bibitem [{\citenamefont {Nigam}\ \emph
  {et~al.}(2022{\natexlab{b}})\citenamefont {Nigam}, \citenamefont
  {Pozdnyakov}, \citenamefont {Fraux},\ and\ \citenamefont
  {Ceriotti}}]{niga+22jcp2}%
  \BibitemOpen
  \bibfield  {author} {\bibinfo {author} {\bibfnamefont {J.}~\bibnamefont
  {Nigam}}, \bibinfo {author} {\bibfnamefont {S.}~\bibnamefont {Pozdnyakov}},
  \bibinfo {author} {\bibfnamefont {G.}~\bibnamefont {Fraux}}, \ and\ \bibinfo
  {author} {\bibfnamefont {M.}~\bibnamefont {Ceriotti}},\ }\href {\doibase
  10.1063/5.0087042} {\bibfield  {journal} {\bibinfo  {journal} {J. Chem.
  Phys.}\ } (\bibinfo {year} {2022}{\natexlab{b}}),\
  10.1063/5.0087042}\BibitemShut {NoStop}%
\bibitem [{\citenamefont {Gastegger}\ \emph {et~al.}(2017)\citenamefont
  {Gastegger}, \citenamefont {Behler},\ and\ \citenamefont
  {Marquetand}}]{gast+17cs}%
  \BibitemOpen
  \bibfield  {author} {\bibinfo {author} {\bibfnamefont {M.}~\bibnamefont
  {Gastegger}}, \bibinfo {author} {\bibfnamefont {J.}~\bibnamefont {Behler}}, \
  and\ \bibinfo {author} {\bibfnamefont {P.}~\bibnamefont {Marquetand}},\
  }\href {\doibase 10.1039/C7SC02267K} {\bibfield  {journal} {\bibinfo
  {journal} {Chem. Sci.}\ }\textbf {\bibinfo {volume} {8}},\ \bibinfo {pages}
  {6924} (\bibinfo {year} {2017})}\BibitemShut {NoStop}%
\bibitem [{\citenamefont {Unke}\ and\ \citenamefont
  {Meuwly}(2019)}]{unke-meuw19jctc}%
  \BibitemOpen
  \bibfield  {author} {\bibinfo {author} {\bibfnamefont {O.~T.}\ \bibnamefont
  {Unke}}\ and\ \bibinfo {author} {\bibfnamefont {M.}~\bibnamefont {Meuwly}},\
  }\href {\doibase 10.1021/acs.jctc.9b00181} {\bibfield  {journal} {\bibinfo
  {journal} {J. Chem. Theory Comput.}\ }\textbf {\bibinfo {volume} {15}},\
  \bibinfo {pages} {3678} (\bibinfo {year} {2019})}\BibitemShut {NoStop}%
\bibitem [{\citenamefont {Zhang}\ \emph
  {et~al.}(2020{\natexlab{a}})\citenamefont {Zhang}, \citenamefont {Chen},
  \citenamefont {Wu}, \citenamefont {Wang}, \citenamefont {E},\ and\
  \citenamefont {Car}}]{zhan+20prb}%
  \BibitemOpen
  \bibfield  {author} {\bibinfo {author} {\bibfnamefont {L.}~\bibnamefont
  {Zhang}}, \bibinfo {author} {\bibfnamefont {M.}~\bibnamefont {Chen}},
  \bibinfo {author} {\bibfnamefont {X.}~\bibnamefont {Wu}}, \bibinfo {author}
  {\bibfnamefont {H.}~\bibnamefont {Wang}}, \bibinfo {author} {\bibfnamefont
  {W.}~\bibnamefont {E}}, \ and\ \bibinfo {author} {\bibfnamefont
  {R.}~\bibnamefont {Car}},\ }\href {\doibase 10.1103/PhysRevB.102.041121}
  {\bibfield  {journal} {\bibinfo  {journal} {Phys. Rev. B}\ }\textbf {\bibinfo
  {volume} {102}},\ \bibinfo {pages} {041121} (\bibinfo {year}
  {2020}{\natexlab{a}})}\BibitemShut {NoStop}%
\bibitem [{\citenamefont {Christensen}\ \emph {et~al.}(2019)\citenamefont
  {Christensen}, \citenamefont {Faber},\ and\ \citenamefont
  {{\noopsort{lilienfeld}}{von Lilienfeld}}}]{chri+19jcp}%
  \BibitemOpen
  \bibfield  {author} {\bibinfo {author} {\bibfnamefont {A.~S.}\ \bibnamefont
  {Christensen}}, \bibinfo {author} {\bibfnamefont {F.~A.}\ \bibnamefont
  {Faber}}, \ and\ \bibinfo {author} {\bibfnamefont {O.~A.}\ \bibnamefont
  {{\noopsort{lilienfeld}}{von Lilienfeld}}},\ }\href {\doibase
  10.1063/1.5053562} {\bibfield  {journal} {\bibinfo  {journal} {J. Chem.
  Phys.}\ }\textbf {\bibinfo {volume} {150}},\ \bibinfo {pages} {064105}
  (\bibinfo {year} {2019})}\BibitemShut {NoStop}%
\bibitem [{\citenamefont {Alred}\ \emph {et~al.}(2018)\citenamefont {Alred},
  \citenamefont {Bets}, \citenamefont {Xie},\ and\ \citenamefont
  {Yakobson}}]{alre+18cst}%
  \BibitemOpen
  \bibfield  {author} {\bibinfo {author} {\bibfnamefont {J.~M.}\ \bibnamefont
  {Alred}}, \bibinfo {author} {\bibfnamefont {K.~V.}\ \bibnamefont {Bets}},
  \bibinfo {author} {\bibfnamefont {Y.}~\bibnamefont {Xie}}, \ and\ \bibinfo
  {author} {\bibfnamefont {B.~I.}\ \bibnamefont {Yakobson}},\ }\href {\doibase
  10.1016/j.compscitech.2018.03.035} {\bibfield  {journal} {\bibinfo  {journal}
  {Composites Science and Technology}\ }\textbf {\bibinfo {volume} {166}},\
  \bibinfo {pages} {3} (\bibinfo {year} {2018})}\BibitemShut {NoStop}%
\bibitem [{\citenamefont {Chandrasekaran}\ \emph {et~al.}(2019)\citenamefont
  {Chandrasekaran}, \citenamefont {Kamal}, \citenamefont {Batra}, \citenamefont
  {Kim}, \citenamefont {Chen},\ and\ \citenamefont {Ramprasad}}]{chan+19npjcm}%
  \BibitemOpen
  \bibfield  {author} {\bibinfo {author} {\bibfnamefont {A.}~\bibnamefont
  {Chandrasekaran}}, \bibinfo {author} {\bibfnamefont {D.}~\bibnamefont
  {Kamal}}, \bibinfo {author} {\bibfnamefont {R.}~\bibnamefont {Batra}},
  \bibinfo {author} {\bibfnamefont {C.}~\bibnamefont {Kim}}, \bibinfo {author}
  {\bibfnamefont {L.}~\bibnamefont {Chen}}, \ and\ \bibinfo {author}
  {\bibfnamefont {R.}~\bibnamefont {Ramprasad}},\ }\href {\doibase
  10.1038/s41524-019-0162-7} {\bibfield  {journal} {\bibinfo  {journal} {npj
  Comput Mater}\ }\textbf {\bibinfo {volume} {5}},\ \bibinfo {pages} {22}
  (\bibinfo {year} {2019})}\BibitemShut {NoStop}%
\bibitem [{\citenamefont {Westermayr}\ and\ \citenamefont
  {Maurer}(2021)}]{west-maur21cs}%
  \BibitemOpen
  \bibfield  {author} {\bibinfo {author} {\bibfnamefont {J.}~\bibnamefont
  {Westermayr}}\ and\ \bibinfo {author} {\bibfnamefont {R.~J.}\ \bibnamefont
  {Maurer}},\ }\href {\doibase 10.1039/D1SC01542G} {\bibfield  {journal}
  {\bibinfo  {journal} {Chem. Sci.}\ }\textbf {\bibinfo {volume} {12}},\
  \bibinfo {pages} {10755} (\bibinfo {year} {2021})}\BibitemShut {NoStop}%
\bibitem [{\citenamefont {Willatt}\ \emph {et~al.}(2019)\citenamefont
  {Willatt}, \citenamefont {Musil},\ and\ \citenamefont
  {Ceriotti}}]{will+19jcp}%
  \BibitemOpen
  \bibfield  {author} {\bibinfo {author} {\bibfnamefont {M.~J.}\ \bibnamefont
  {Willatt}}, \bibinfo {author} {\bibfnamefont {F.}~\bibnamefont {Musil}}, \
  and\ \bibinfo {author} {\bibfnamefont {M.}~\bibnamefont {Ceriotti}},\ }\href
  {\doibase 10.1063/1.5090481} {\bibfield  {journal} {\bibinfo  {journal} {J.
  Chem. Phys.}\ }\textbf {\bibinfo {volume} {150}},\ \bibinfo {pages} {154110}
  (\bibinfo {year} {2019})}\BibitemShut {NoStop}%
\bibitem [{\citenamefont {Nigam}\ \emph {et~al.}(2020)\citenamefont {Nigam},
  \citenamefont {Pozdnyakov},\ and\ \citenamefont {Ceriotti}}]{niga+20jcp}%
  \BibitemOpen
  \bibfield  {author} {\bibinfo {author} {\bibfnamefont {J.}~\bibnamefont
  {Nigam}}, \bibinfo {author} {\bibfnamefont {S.}~\bibnamefont {Pozdnyakov}}, \
  and\ \bibinfo {author} {\bibfnamefont {M.}~\bibnamefont {Ceriotti}},\ }\href
  {\doibase 10.1063/5.0021116} {\bibfield  {journal} {\bibinfo  {journal} {J.
  Chem. Phys.}\ }\textbf {\bibinfo {volume} {153}},\ \bibinfo {pages} {121101}
  (\bibinfo {year} {2020})}\BibitemShut {NoStop}%
\bibitem [{\citenamefont {Shapeev}(2016)}]{shap16mms}%
  \BibitemOpen
  \bibfield  {author} {\bibinfo {author} {\bibfnamefont {A.~V.}\ \bibnamefont
  {Shapeev}},\ }\href {\doibase 10.1137/15M1054183} {\bibfield  {journal}
  {\bibinfo  {journal} {Multiscale Model. Simul.}\ }\textbf {\bibinfo {volume}
  {14}},\ \bibinfo {pages} {1153} (\bibinfo {year} {2016})}\BibitemShut
  {NoStop}%
\bibitem [{\citenamefont {Drautz}(2019)}]{drau19prb}%
  \BibitemOpen
  \bibfield  {author} {\bibinfo {author} {\bibfnamefont {R.}~\bibnamefont
  {Drautz}},\ }\href {\doibase 10.1103/PhysRevB.99.014104} {\bibfield
  {journal} {\bibinfo  {journal} {Phys. Rev. B}\ }\textbf {\bibinfo {volume}
  {99}},\ \bibinfo {pages} {014104} (\bibinfo {year} {2019})}\BibitemShut
  {NoStop}%
\bibitem [{\citenamefont {Unke}\ \emph {et~al.}(2021)\citenamefont {Unke},
  \citenamefont {Bogojeski}, \citenamefont {Gastegger}, \citenamefont {Geiger},
  \citenamefont {Smidt},\ and\ \citenamefont {M{\"u}ller}}]{unke+21nips}%
  \BibitemOpen
  \bibfield  {author} {\bibinfo {author} {\bibfnamefont {O.}~\bibnamefont
  {Unke}}, \bibinfo {author} {\bibfnamefont {M.}~\bibnamefont {Bogojeski}},
  \bibinfo {author} {\bibfnamefont {M.}~\bibnamefont {Gastegger}}, \bibinfo
  {author} {\bibfnamefont {M.}~\bibnamefont {Geiger}}, \bibinfo {author}
  {\bibfnamefont {T.}~\bibnamefont {Smidt}}, \ and\ \bibinfo {author}
  {\bibfnamefont {K.-R.}\ \bibnamefont {M{\"u}ller}},\ }\href@noop {}
  {\bibfield  {journal} {\bibinfo  {journal} {Adv. Neural Inf. Process. Syst.}\
  }\textbf {\bibinfo {volume} {34}} (\bibinfo {year} {2021})}\BibitemShut
  {NoStop}%
\bibitem [{\citenamefont {Isayev}\ \emph {et~al.}(2015)\citenamefont {Isayev},
  \citenamefont {Fourches}, \citenamefont {Muratov}, \citenamefont {Oses},
  \citenamefont {Rasch}, \citenamefont {Tropsha},\ and\ \citenamefont
  {Curtarolo}}]{isay+15cm}%
  \BibitemOpen
  \bibfield  {author} {\bibinfo {author} {\bibfnamefont {O.}~\bibnamefont
  {Isayev}}, \bibinfo {author} {\bibfnamefont {D.}~\bibnamefont {Fourches}},
  \bibinfo {author} {\bibfnamefont {E.~N.}\ \bibnamefont {Muratov}}, \bibinfo
  {author} {\bibfnamefont {C.}~\bibnamefont {Oses}}, \bibinfo {author}
  {\bibfnamefont {K.}~\bibnamefont {Rasch}}, \bibinfo {author} {\bibfnamefont
  {A.}~\bibnamefont {Tropsha}}, \ and\ \bibinfo {author} {\bibfnamefont
  {S.}~\bibnamefont {Curtarolo}},\ }\href {\doibase 10.1021/cm503507h}
  {\bibfield  {journal} {\bibinfo  {journal} {Chem. Mater.}\ }\textbf {\bibinfo
  {volume} {27}},\ \bibinfo {pages} {735} (\bibinfo {year} {2015})}\BibitemShut
  {NoStop}%
\bibitem [{\citenamefont {Pulido}\ \emph {et~al.}(2017)\citenamefont {Pulido},
  \citenamefont {Chen}, \citenamefont {Kaczorowski}, \citenamefont {Holden},
  \citenamefont {Little}, \citenamefont {Chong}, \citenamefont {Slater},
  \citenamefont {McMahon}, \citenamefont {Bonillo}, \citenamefont {Stackhouse},
  \citenamefont {Stephenson}, \citenamefont {Kane}, \citenamefont {Clowes},
  \citenamefont {Hasell}, \citenamefont {Cooper},\ and\ \citenamefont
  {Day}}]{puli+17nature}%
  \BibitemOpen
  \bibfield  {author} {\bibinfo {author} {\bibfnamefont {A.}~\bibnamefont
  {Pulido}}, \bibinfo {author} {\bibfnamefont {L.}~\bibnamefont {Chen}},
  \bibinfo {author} {\bibfnamefont {T.}~\bibnamefont {Kaczorowski}}, \bibinfo
  {author} {\bibfnamefont {D.}~\bibnamefont {Holden}}, \bibinfo {author}
  {\bibfnamefont {M.~A.}\ \bibnamefont {Little}}, \bibinfo {author}
  {\bibfnamefont {S.~Y.}\ \bibnamefont {Chong}}, \bibinfo {author}
  {\bibfnamefont {B.~J.}\ \bibnamefont {Slater}}, \bibinfo {author}
  {\bibfnamefont {D.~P.}\ \bibnamefont {McMahon}}, \bibinfo {author}
  {\bibfnamefont {B.}~\bibnamefont {Bonillo}}, \bibinfo {author} {\bibfnamefont
  {C.~J.}\ \bibnamefont {Stackhouse}}, \bibinfo {author} {\bibfnamefont
  {A.}~\bibnamefont {Stephenson}}, \bibinfo {author} {\bibfnamefont {C.~M.}\
  \bibnamefont {Kane}}, \bibinfo {author} {\bibfnamefont {R.}~\bibnamefont
  {Clowes}}, \bibinfo {author} {\bibfnamefont {T.}~\bibnamefont {Hasell}},
  \bibinfo {author} {\bibfnamefont {A.~I.}\ \bibnamefont {Cooper}}, \ and\
  \bibinfo {author} {\bibfnamefont {G.~M.}\ \bibnamefont {Day}},\ }\href
  {\doibase 10.1038/nature21419} {\bibfield  {journal} {\bibinfo  {journal}
  {Nature}\ }\textbf {\bibinfo {volume} {543}},\ \bibinfo {pages} {657}
  (\bibinfo {year} {2017})}\BibitemShut {NoStop}%
\bibitem [{\citenamefont {Glielmo}\ \emph {et~al.}(2021)\citenamefont
  {Glielmo}, \citenamefont {Husic}, \citenamefont {Rodriguez}, \citenamefont
  {Clementi}, \citenamefont {No{\'e}},\ and\ \citenamefont {Laio}}]{glie+21cr}%
  \BibitemOpen
  \bibfield  {author} {\bibinfo {author} {\bibfnamefont {A.}~\bibnamefont
  {Glielmo}}, \bibinfo {author} {\bibfnamefont {B.~E.}\ \bibnamefont {Husic}},
  \bibinfo {author} {\bibfnamefont {A.}~\bibnamefont {Rodriguez}}, \bibinfo
  {author} {\bibfnamefont {C.}~\bibnamefont {Clementi}}, \bibinfo {author}
  {\bibfnamefont {F.}~\bibnamefont {No{\'e}}}, \ and\ \bibinfo {author}
  {\bibfnamefont {A.}~\bibnamefont {Laio}},\ }\href {\doibase
  10.1021/acs.chemrev.0c01195} {\bibfield  {journal} {\bibinfo  {journal}
  {Chem. Rev.}\ }\textbf {\bibinfo {volume} {121}},\ \bibinfo {pages} {9722}
  (\bibinfo {year} {2021})}\BibitemShut {NoStop}%
\bibitem [{\citenamefont {Cersonsky}\ \emph {et~al.}(2021)\citenamefont
  {Cersonsky}, \citenamefont {Helfrecht}, \citenamefont {Engel}, \citenamefont
  {Kliavinek},\ and\ \citenamefont {Ceriotti}}]{cers+21mlst}%
  \BibitemOpen
  \bibfield  {author} {\bibinfo {author} {\bibfnamefont {R.~K.}\ \bibnamefont
  {Cersonsky}}, \bibinfo {author} {\bibfnamefont {B.~A.}\ \bibnamefont
  {Helfrecht}}, \bibinfo {author} {\bibfnamefont {E.~A.}\ \bibnamefont
  {Engel}}, \bibinfo {author} {\bibfnamefont {S.}~\bibnamefont {Kliavinek}}, \
  and\ \bibinfo {author} {\bibfnamefont {M.}~\bibnamefont {Ceriotti}},\ }\href
  {\doibase 10.1088/2632-2153/abfe7c} {\bibfield  {journal} {\bibinfo
  {journal} {Mach. Learn.: Sci. Technol.}\ }\textbf {\bibinfo {volume} {2}},\
  \bibinfo {pages} {035038} (\bibinfo {year} {2021})}\BibitemShut {NoStop}%
\bibitem [{\citenamefont {Kapil}\ \emph {et~al.}(2020)\citenamefont {Kapil},
  \citenamefont {Wilkins}, \citenamefont {Lan},\ and\ \citenamefont
  {Ceriotti}}]{kapi+20jcp}%
  \BibitemOpen
  \bibfield  {author} {\bibinfo {author} {\bibfnamefont {V.}~\bibnamefont
  {Kapil}}, \bibinfo {author} {\bibfnamefont {D.~M.}\ \bibnamefont {Wilkins}},
  \bibinfo {author} {\bibfnamefont {J.}~\bibnamefont {Lan}}, \ and\ \bibinfo
  {author} {\bibfnamefont {M.}~\bibnamefont {Ceriotti}},\ }\href {\doibase
  10.1063/1.5141950} {\bibfield  {journal} {\bibinfo  {journal} {J. Chem.
  Phys.}\ }\textbf {\bibinfo {volume} {152}},\ \bibinfo {pages} {124104}
  (\bibinfo {year} {2020})}\BibitemShut {NoStop}%
\bibitem [{\citenamefont {Sommers}\ \emph {et~al.}(2020)\citenamefont
  {Sommers}, \citenamefont {Calegari~Andrade}, \citenamefont {Zhang},
  \citenamefont {Wang},\ and\ \citenamefont {Car}}]{somm+20pccp}%
  \BibitemOpen
  \bibfield  {author} {\bibinfo {author} {\bibfnamefont {G.~M.}\ \bibnamefont
  {Sommers}}, \bibinfo {author} {\bibfnamefont {M.~F.}\ \bibnamefont
  {Calegari~Andrade}}, \bibinfo {author} {\bibfnamefont {L.}~\bibnamefont
  {Zhang}}, \bibinfo {author} {\bibfnamefont {H.}~\bibnamefont {Wang}}, \ and\
  \bibinfo {author} {\bibfnamefont {R.}~\bibnamefont {Car}},\ }\href {\doibase
  10.1039/D0CP01893G} {\bibfield  {journal} {\bibinfo  {journal} {Phys. Chem.
  Chem. Phys.}\ }\textbf {\bibinfo {volume} {22}},\ \bibinfo {pages} {10592}
  (\bibinfo {year} {2020})}\BibitemShut {NoStop}%
\bibitem [{\citenamefont {Gigli}\ \emph {et~al.}(2021)\citenamefont {Gigli},
  \citenamefont {Veit}, \citenamefont {Kotiuga}, \citenamefont {Pizzi},
  \citenamefont {Marzari},\ and\ \citenamefont {Ceriotti}}]{gigli2021arxiv}%
  \BibitemOpen
  \bibfield  {author} {\bibinfo {author} {\bibfnamefont {L.}~\bibnamefont
  {Gigli}}, \bibinfo {author} {\bibfnamefont {M.}~\bibnamefont {Veit}},
  \bibinfo {author} {\bibfnamefont {M.}~\bibnamefont {Kotiuga}}, \bibinfo
  {author} {\bibfnamefont {G.}~\bibnamefont {Pizzi}}, \bibinfo {author}
  {\bibfnamefont {N.}~\bibnamefont {Marzari}}, \ and\ \bibinfo {author}
  {\bibfnamefont {M.}~\bibnamefont {Ceriotti}},\ }\href
  {http://arxiv.org/abs/2111.05129v1} {\bibfield  {journal} {\bibinfo
  {journal} {arxiv:2111.05129}\ } (\bibinfo {year} {2021})}\BibitemShut
  {NoStop}%
\bibitem [{\citenamefont {Shepherd}\ \emph {et~al.}(2021)\citenamefont
  {Shepherd}, \citenamefont {Lan}, \citenamefont {Wilkins},\ and\ \citenamefont
  {Kapil}}]{shep+21jcp}%
  \BibitemOpen
  \bibfield  {author} {\bibinfo {author} {\bibfnamefont {S.}~\bibnamefont
  {Shepherd}}, \bibinfo {author} {\bibfnamefont {J.}~\bibnamefont {Lan}},
  \bibinfo {author} {\bibfnamefont {D.~M.}\ \bibnamefont {Wilkins}}, \ and\
  \bibinfo {author} {\bibfnamefont {V.}~\bibnamefont {Kapil}},\ }\href
  {\doibase 10.1021/acs.jpclett.1c02574} {\bibfield  {journal} {\bibinfo
  {journal} {J. Phys. Chem. Lett.}\ }\textbf {\bibinfo {volume} {12}},\
  \bibinfo {pages} {9108} (\bibinfo {year} {2021})}\BibitemShut {NoStop}%
\bibitem [{\citenamefont {Engel}\ \emph {et~al.}(2021)\citenamefont {Engel},
  \citenamefont {Kapil},\ and\ \citenamefont {Ceriotti}}]{enge+21jpcl}%
  \BibitemOpen
  \bibfield  {author} {\bibinfo {author} {\bibfnamefont {E.~A.}\ \bibnamefont
  {Engel}}, \bibinfo {author} {\bibfnamefont {V.}~\bibnamefont {Kapil}}, \ and\
  \bibinfo {author} {\bibfnamefont {M.}~\bibnamefont {Ceriotti}},\ }\href
  {\doibase 10.1021/acs.jpclett.1c01987} {\bibfield  {journal} {\bibinfo
  {journal} {J. Phys. Chem. Lett.}\ }\textbf {\bibinfo {volume} {12}},\
  \bibinfo {pages} {7701} (\bibinfo {year} {2021})}\BibitemShut {NoStop}%
\bibitem [{\citenamefont {Adhikari}(2015)}]{adhi+15jcp}%
  \BibitemOpen
  \bibfield  {author} {\bibinfo {author} {\bibfnamefont {A.}~\bibnamefont
  {Adhikari}},\ }\href {\doibase 10.1063/1.4931485} {\bibfield  {journal}
  {\bibinfo  {journal} {The Journal of Chemical Physics}\ }\textbf {\bibinfo
  {volume} {143}},\ \bibinfo {pages} {124707} (\bibinfo {year}
  {2015})}\BibitemShut {NoStop}%
\bibitem [{\citenamefont {Deringer}\ \emph
  {et~al.}(2021{\natexlab{b}})\citenamefont {Deringer}, \citenamefont
  {Bernstein}, \citenamefont {Cs{\'a}nyi}, \citenamefont {Ben~Mahmoud},
  \citenamefont {Ceriotti}, \citenamefont {Wilson}, \citenamefont {Drabold},\
  and\ \citenamefont {Elliott}}]{deri+21nature}%
  \BibitemOpen
  \bibfield  {author} {\bibinfo {author} {\bibfnamefont {V.~L.}\ \bibnamefont
  {Deringer}}, \bibinfo {author} {\bibfnamefont {N.}~\bibnamefont {Bernstein}},
  \bibinfo {author} {\bibfnamefont {G.}~\bibnamefont {Cs{\'a}nyi}}, \bibinfo
  {author} {\bibfnamefont {C.}~\bibnamefont {Ben~Mahmoud}}, \bibinfo {author}
  {\bibfnamefont {M.}~\bibnamefont {Ceriotti}}, \bibinfo {author}
  {\bibfnamefont {M.}~\bibnamefont {Wilson}}, \bibinfo {author} {\bibfnamefont
  {D.~A.}\ \bibnamefont {Drabold}}, \ and\ \bibinfo {author} {\bibfnamefont
  {S.~R.}\ \bibnamefont {Elliott}},\ }\href {\doibase
  10.1038/s41586-020-03072-z} {\bibfield  {journal} {\bibinfo  {journal}
  {Nature}\ }\textbf {\bibinfo {volume} {589}},\ \bibinfo {pages} {59}
  (\bibinfo {year} {2021}{\natexlab{b}})}\BibitemShut {NoStop}%
\bibitem [{\citenamefont {Zhang}\ \emph
  {et~al.}(2020{\natexlab{b}})\citenamefont {Zhang}, \citenamefont {Gao},
  \citenamefont {Liu}, \citenamefont {Zhang}, \citenamefont {Wang},\ and\
  \citenamefont {Chen}}]{zhan+20jcp}%
  \BibitemOpen
  \bibfield  {author} {\bibinfo {author} {\bibfnamefont {Y.}~\bibnamefont
  {Zhang}}, \bibinfo {author} {\bibfnamefont {C.}~\bibnamefont {Gao}}, \bibinfo
  {author} {\bibfnamefont {Q.}~\bibnamefont {Liu}}, \bibinfo {author}
  {\bibfnamefont {L.}~\bibnamefont {Zhang}}, \bibinfo {author} {\bibfnamefont
  {H.}~\bibnamefont {Wang}}, \ and\ \bibinfo {author} {\bibfnamefont
  {M.}~\bibnamefont {Chen}},\ }\href {\doibase 10.1063/5.0023265} {\bibfield
  {journal} {\bibinfo  {journal} {Physics of Plasmas}\ }\textbf {\bibinfo
  {volume} {27}},\ \bibinfo {pages} {122704} (\bibinfo {year}
  {2020}{\natexlab{b}})}\BibitemShut {NoStop}%
\bibitem [{\citenamefont {Ellis}\ \emph {et~al.}(2021)\citenamefont {Ellis},
  \citenamefont {Fiedler}, \citenamefont {Popoola}, \citenamefont {Modine},
  \citenamefont {Stephens}, \citenamefont {Thompson}, \citenamefont {Cangi},\
  and\ \citenamefont {Rajamanickam}}]{elli+21prb}%
  \BibitemOpen
  \bibfield  {author} {\bibinfo {author} {\bibfnamefont {J.~A.}\ \bibnamefont
  {Ellis}}, \bibinfo {author} {\bibfnamefont {L.}~\bibnamefont {Fiedler}},
  \bibinfo {author} {\bibfnamefont {G.~A.}\ \bibnamefont {Popoola}}, \bibinfo
  {author} {\bibfnamefont {N.~A.}\ \bibnamefont {Modine}}, \bibinfo {author}
  {\bibfnamefont {J.~A.}\ \bibnamefont {Stephens}}, \bibinfo {author}
  {\bibfnamefont {A.~P.}\ \bibnamefont {Thompson}}, \bibinfo {author}
  {\bibfnamefont {A.}~\bibnamefont {Cangi}}, \ and\ \bibinfo {author}
  {\bibfnamefont {S.}~\bibnamefont {Rajamanickam}},\ }\href {\doibase
  10.1103/PhysRevB.104.035120} {\bibfield  {journal} {\bibinfo  {journal}
  {Phys. Rev. B}\ }\textbf {\bibinfo {volume} {104}},\ \bibinfo {pages}
  {035120} (\bibinfo {year} {2021})}\BibitemShut {NoStop}%
\bibitem [{\citenamefont {Mahmoud}\ \emph {et~al.}(2022)\citenamefont
  {Mahmoud}, \citenamefont {Grasselli},\ and\ \citenamefont
  {Ceriotti}}]{benmahmoud2022arxiv}%
  \BibitemOpen
  \bibfield  {author} {\bibinfo {author} {\bibfnamefont {C.~B.}\ \bibnamefont
  {Mahmoud}}, \bibinfo {author} {\bibfnamefont {F.}~\bibnamefont {Grasselli}},
  \ and\ \bibinfo {author} {\bibfnamefont {M.}~\bibnamefont {Ceriotti}},\
  }\href {http://arxiv.org/abs/2205.05591v1} {\bibfield  {journal} {\bibinfo
  {journal} {arxiv:2205.05591}\ } (\bibinfo {year} {2022})}\BibitemShut
  {NoStop}%
\bibitem [{\citenamefont {Gong}\ \emph {et~al.}(2018)\citenamefont {Gong},
  \citenamefont {Grabowski}, \citenamefont {Glensk}, \citenamefont
  {K{\"o}rmann}, \citenamefont {Neugebauer},\ and\ \citenamefont
  {Reed}}]{gong+18prb}%
  \BibitemOpen
  \bibfield  {author} {\bibinfo {author} {\bibfnamefont {Y.}~\bibnamefont
  {Gong}}, \bibinfo {author} {\bibfnamefont {B.}~\bibnamefont {Grabowski}},
  \bibinfo {author} {\bibfnamefont {A.}~\bibnamefont {Glensk}}, \bibinfo
  {author} {\bibfnamefont {F.}~\bibnamefont {K{\"o}rmann}}, \bibinfo {author}
  {\bibfnamefont {J.}~\bibnamefont {Neugebauer}}, \ and\ \bibinfo {author}
  {\bibfnamefont {R.~C.}\ \bibnamefont {Reed}},\ }\href {\doibase
  10.1103/PhysRevB.97.214106} {\bibfield  {journal} {\bibinfo  {journal} {Phys.
  Rev. B}\ }\textbf {\bibinfo {volume} {97}},\ \bibinfo {pages} {214106}
  (\bibinfo {year} {2018})}\BibitemShut {NoStop}%
\bibitem [{\citenamefont {Lopanitsyna}\ \emph {et~al.}(2021)\citenamefont
  {Lopanitsyna}, \citenamefont {Ben~Mahmoud},\ and\ \citenamefont
  {Ceriotti}}]{lopa+21prm}%
  \BibitemOpen
  \bibfield  {author} {\bibinfo {author} {\bibfnamefont {N.}~\bibnamefont
  {Lopanitsyna}}, \bibinfo {author} {\bibfnamefont {C.}~\bibnamefont
  {Ben~Mahmoud}}, \ and\ \bibinfo {author} {\bibfnamefont {M.}~\bibnamefont
  {Ceriotti}},\ }\href {\doibase 10.1103/PhysRevMaterials.5.043802} {\bibfield
  {journal} {\bibinfo  {journal} {Phys. Rev. Materials}\ }\textbf {\bibinfo
  {volume} {5}},\ \bibinfo {pages} {043802} (\bibinfo {year}
  {2021})}\BibitemShut {NoStop}%
\bibitem [{\citenamefont {Novikov}\ \emph {et~al.}(2022)\citenamefont
  {Novikov}, \citenamefont {Grabowski}, \citenamefont {K{\"o}rmann},\ and\
  \citenamefont {Shapeev}}]{novi+22npjcm}%
  \BibitemOpen
  \bibfield  {author} {\bibinfo {author} {\bibfnamefont {I.}~\bibnamefont
  {Novikov}}, \bibinfo {author} {\bibfnamefont {B.}~\bibnamefont {Grabowski}},
  \bibinfo {author} {\bibfnamefont {F.}~\bibnamefont {K{\"o}rmann}}, \ and\
  \bibinfo {author} {\bibfnamefont {A.}~\bibnamefont {Shapeev}},\ }\href
  {\doibase 10.1038/s41524-022-00696-9} {\bibfield  {journal} {\bibinfo
  {journal} {npj Comput Mater}\ }\textbf {\bibinfo {volume} {8}},\ \bibinfo
  {pages} {13} (\bibinfo {year} {2022})}\BibitemShut {NoStop}%
\bibitem [{\citenamefont {Dral}\ \emph {et~al.}(2018)\citenamefont {Dral},
  \citenamefont {Barbatti},\ and\ \citenamefont {Thiel}}]{dral+18jpcl}%
  \BibitemOpen
  \bibfield  {author} {\bibinfo {author} {\bibfnamefont {P.~O.}\ \bibnamefont
  {Dral}}, \bibinfo {author} {\bibfnamefont {M.}~\bibnamefont {Barbatti}}, \
  and\ \bibinfo {author} {\bibfnamefont {W.}~\bibnamefont {Thiel}},\ }\href
  {\doibase 10.1021/acs.jpclett.8b02469} {\bibfield  {journal} {\bibinfo
  {journal} {J. Phys. Chem. Lett.}\ }\textbf {\bibinfo {volume} {9}},\ \bibinfo
  {pages} {5660} (\bibinfo {year} {2018})}\BibitemShut {NoStop}%
\bibitem [{\citenamefont {Westermayr}\ and\ \citenamefont
  {Marquetand}(2020)}]{west-marq20mlst}%
  \BibitemOpen
  \bibfield  {author} {\bibinfo {author} {\bibfnamefont {J.}~\bibnamefont
  {Westermayr}}\ and\ \bibinfo {author} {\bibfnamefont {P.}~\bibnamefont
  {Marquetand}},\ }\href {\doibase 10.1088/2632-2153/ab9c3e} {\bibfield
  {journal} {\bibinfo  {journal} {Mach. Learn.: Sci. Technol.}\ }\textbf
  {\bibinfo {volume} {1}},\ \bibinfo {pages} {043001} (\bibinfo {year}
  {2020})}\BibitemShut {NoStop}%
\bibitem [{\citenamefont {Hjorth~Larsen}\ \emph {et~al.}(2017)\citenamefont
  {Hjorth~Larsen}, \citenamefont {J{\o}rgen~Mortensen}, \citenamefont
  {Blomqvist}, \citenamefont {Castelli}, \citenamefont {Christensen},
  \citenamefont {Du{\l}ak}, \citenamefont {Friis}, \citenamefont {Groves},
  \citenamefont {Hammer}, \citenamefont {Hargus}, \citenamefont {Hermes},
  \citenamefont {Jennings}, \citenamefont {Bjerre~Jensen}, \citenamefont
  {Kermode}, \citenamefont {Kitchin}, \citenamefont {Leonhard~Kolsbjerg},
  \citenamefont {Kubal}, \citenamefont {Kaasbjerg}, \citenamefont {Lysgaard},
  \citenamefont {Bergmann~Maronsson}, \citenamefont {Maxson}, \citenamefont
  {Olsen}, \citenamefont {Pastewka}, \citenamefont {Peterson}, \citenamefont
  {Rostgaard}, \citenamefont {Schi{\o}tz}, \citenamefont {Sch{\"u}tt},
  \citenamefont {Strange}, \citenamefont {Thygesen}, \citenamefont {Vegge},
  \citenamefont {Vilhelmsen}, \citenamefont {Walter}, \citenamefont {Zeng},\
  and\ \citenamefont {Jacobsen}}]{hjor+17jpcm}%
  \BibitemOpen
  \bibfield  {author} {\bibinfo {author} {\bibfnamefont {A.}~\bibnamefont
  {Hjorth~Larsen}}, \bibinfo {author} {\bibfnamefont {J.}~\bibnamefont
  {J{\o}rgen~Mortensen}}, \bibinfo {author} {\bibfnamefont {J.}~\bibnamefont
  {Blomqvist}}, \bibinfo {author} {\bibfnamefont {I.~E.}\ \bibnamefont
  {Castelli}}, \bibinfo {author} {\bibfnamefont {R.}~\bibnamefont
  {Christensen}}, \bibinfo {author} {\bibfnamefont {M.}~\bibnamefont
  {Du{\l}ak}}, \bibinfo {author} {\bibfnamefont {J.}~\bibnamefont {Friis}},
  \bibinfo {author} {\bibfnamefont {M.~N.}\ \bibnamefont {Groves}}, \bibinfo
  {author} {\bibfnamefont {B.}~\bibnamefont {Hammer}}, \bibinfo {author}
  {\bibfnamefont {C.}~\bibnamefont {Hargus}}, \bibinfo {author} {\bibfnamefont
  {E.~D.}\ \bibnamefont {Hermes}}, \bibinfo {author} {\bibfnamefont {P.~C.}\
  \bibnamefont {Jennings}}, \bibinfo {author} {\bibfnamefont {P.}~\bibnamefont
  {Bjerre~Jensen}}, \bibinfo {author} {\bibfnamefont {J.}~\bibnamefont
  {Kermode}}, \bibinfo {author} {\bibfnamefont {J.~R.}\ \bibnamefont
  {Kitchin}}, \bibinfo {author} {\bibfnamefont {E.}~\bibnamefont
  {Leonhard~Kolsbjerg}}, \bibinfo {author} {\bibfnamefont {J.}~\bibnamefont
  {Kubal}}, \bibinfo {author} {\bibfnamefont {K.}~\bibnamefont {Kaasbjerg}},
  \bibinfo {author} {\bibfnamefont {S.}~\bibnamefont {Lysgaard}}, \bibinfo
  {author} {\bibfnamefont {J.}~\bibnamefont {Bergmann~Maronsson}}, \bibinfo
  {author} {\bibfnamefont {T.}~\bibnamefont {Maxson}}, \bibinfo {author}
  {\bibfnamefont {T.}~\bibnamefont {Olsen}}, \bibinfo {author} {\bibfnamefont
  {L.}~\bibnamefont {Pastewka}}, \bibinfo {author} {\bibfnamefont
  {A.}~\bibnamefont {Peterson}}, \bibinfo {author} {\bibfnamefont
  {C.}~\bibnamefont {Rostgaard}}, \bibinfo {author} {\bibfnamefont
  {J.}~\bibnamefont {Schi{\o}tz}}, \bibinfo {author} {\bibfnamefont
  {O.}~\bibnamefont {Sch{\"u}tt}}, \bibinfo {author} {\bibfnamefont
  {M.}~\bibnamefont {Strange}}, \bibinfo {author} {\bibfnamefont {K.~S.}\
  \bibnamefont {Thygesen}}, \bibinfo {author} {\bibfnamefont {T.}~\bibnamefont
  {Vegge}}, \bibinfo {author} {\bibfnamefont {L.}~\bibnamefont {Vilhelmsen}},
  \bibinfo {author} {\bibfnamefont {M.}~\bibnamefont {Walter}}, \bibinfo
  {author} {\bibfnamefont {Z.}~\bibnamefont {Zeng}}, \ and\ \bibinfo {author}
  {\bibfnamefont {K.~W.}\ \bibnamefont {Jacobsen}},\ }\href {\doibase
  10.1088/1361-648X/aa680e} {\bibfield  {journal} {\bibinfo  {journal} {J.
  Phys.: Condens. Matter}\ }\textbf {\bibinfo {volume} {29}},\ \bibinfo {pages}
  {273002} (\bibinfo {year} {2017})}\BibitemShut {NoStop}%
\bibitem [{\citenamefont {Kapil}\ \emph {et~al.}(2019)\citenamefont {Kapil},
  \citenamefont {Rossi}, \citenamefont {Marsalek}, \citenamefont {Petraglia},
  \citenamefont {Litman}, \citenamefont {Spura}, \citenamefont {Cheng},
  \citenamefont {Cuzzocrea}, \citenamefont {Mei{\ss}ner}, \citenamefont
  {Wilkins}, \citenamefont {Helfrecht}, \citenamefont {Juda}, \citenamefont
  {Bienvenue}, \citenamefont {Fang}, \citenamefont {Kessler}, \citenamefont
  {Poltavsky}, \citenamefont {Vandenbrande}, \citenamefont {Wieme},
  \citenamefont {Corminboeuf}, \citenamefont {K{\"u}hne}, \citenamefont
  {Manolopoulos}, \citenamefont {Markland}, \citenamefont {Richardson},
  \citenamefont {Tkatchenko}, \citenamefont {Tribello}, \citenamefont
  {Van~Speybroeck},\ and\ \citenamefont {Ceriotti}}]{kapi+19cpc}%
  \BibitemOpen
  \bibfield  {author} {\bibinfo {author} {\bibfnamefont {V.}~\bibnamefont
  {Kapil}}, \bibinfo {author} {\bibfnamefont {M.}~\bibnamefont {Rossi}},
  \bibinfo {author} {\bibfnamefont {O.}~\bibnamefont {Marsalek}}, \bibinfo
  {author} {\bibfnamefont {R.}~\bibnamefont {Petraglia}}, \bibinfo {author}
  {\bibfnamefont {Y.}~\bibnamefont {Litman}}, \bibinfo {author} {\bibfnamefont
  {T.}~\bibnamefont {Spura}}, \bibinfo {author} {\bibfnamefont
  {B.}~\bibnamefont {Cheng}}, \bibinfo {author} {\bibfnamefont
  {A.}~\bibnamefont {Cuzzocrea}}, \bibinfo {author} {\bibfnamefont {R.~H.}\
  \bibnamefont {Mei{\ss}ner}}, \bibinfo {author} {\bibfnamefont {D.~M.}\
  \bibnamefont {Wilkins}}, \bibinfo {author} {\bibfnamefont {B.~A.}\
  \bibnamefont {Helfrecht}}, \bibinfo {author} {\bibfnamefont {P.}~\bibnamefont
  {Juda}}, \bibinfo {author} {\bibfnamefont {S.~P.}\ \bibnamefont {Bienvenue}},
  \bibinfo {author} {\bibfnamefont {W.}~\bibnamefont {Fang}}, \bibinfo {author}
  {\bibfnamefont {J.}~\bibnamefont {Kessler}}, \bibinfo {author} {\bibfnamefont
  {I.}~\bibnamefont {Poltavsky}}, \bibinfo {author} {\bibfnamefont
  {S.}~\bibnamefont {Vandenbrande}}, \bibinfo {author} {\bibfnamefont
  {J.}~\bibnamefont {Wieme}}, \bibinfo {author} {\bibfnamefont
  {C.}~\bibnamefont {Corminboeuf}}, \bibinfo {author} {\bibfnamefont {T.~D.}\
  \bibnamefont {K{\"u}hne}}, \bibinfo {author} {\bibfnamefont {D.~E.}\
  \bibnamefont {Manolopoulos}}, \bibinfo {author} {\bibfnamefont {T.~E.}\
  \bibnamefont {Markland}}, \bibinfo {author} {\bibfnamefont {J.~O.}\
  \bibnamefont {Richardson}}, \bibinfo {author} {\bibfnamefont
  {A.}~\bibnamefont {Tkatchenko}}, \bibinfo {author} {\bibfnamefont {G.~A.}\
  \bibnamefont {Tribello}}, \bibinfo {author} {\bibfnamefont {V.}~\bibnamefont
  {Van~Speybroeck}}, \ and\ \bibinfo {author} {\bibfnamefont {M.}~\bibnamefont
  {Ceriotti}},\ }\href {\doibase 10.1016/j.cpc.2018.09.020} {\bibfield
  {journal} {\bibinfo  {journal} {Comput. Phys. Commun.}\ }\textbf {\bibinfo
  {volume} {236}},\ \bibinfo {pages} {214} (\bibinfo {year}
  {2019})}\BibitemShut {NoStop}%
\bibitem [{\citenamefont {Tribello}\ \emph {et~al.}(2014)\citenamefont
  {Tribello}, \citenamefont {Bonomi}, \citenamefont {Branduardi}, \citenamefont
  {Camilloni},\ and\ \citenamefont {Bussi}}]{trib+14cpc}%
  \BibitemOpen
  \bibfield  {author} {\bibinfo {author} {\bibfnamefont {G.~A.}\ \bibnamefont
  {Tribello}}, \bibinfo {author} {\bibfnamefont {M.}~\bibnamefont {Bonomi}},
  \bibinfo {author} {\bibfnamefont {D.}~\bibnamefont {Branduardi}}, \bibinfo
  {author} {\bibfnamefont {C.}~\bibnamefont {Camilloni}}, \ and\ \bibinfo
  {author} {\bibfnamefont {G.}~\bibnamefont {Bussi}},\ }\href {\doibase
  10.1016/j.cpc.2013.09.018} {\bibfield  {journal} {\bibinfo  {journal}
  {Comput. Phys. Commun.}\ }\textbf {\bibinfo {volume} {185}},\ \bibinfo
  {pages} {604} (\bibinfo {year} {2014})}\BibitemShut {NoStop}%
\bibitem [{\citenamefont {Sidky}\ \emph {et~al.}(2018)\citenamefont {Sidky},
  \citenamefont {Col{\'o}n}, \citenamefont {Helfferich}, \citenamefont
  {Sikora}, \citenamefont {Bezik}, \citenamefont {Chu}, \citenamefont
  {Giberti}, \citenamefont {Guo}, \citenamefont {Jiang}, \citenamefont
  {Lequieu}, \citenamefont {Li}, \citenamefont {Moller}, \citenamefont
  {Quevillon}, \citenamefont {Rahimi}, \citenamefont {{Ramezani-Dakhel}},
  \citenamefont {Rathee}, \citenamefont {Reid}, \citenamefont {Sevgen},
  \citenamefont {Thapar}, \citenamefont {Webb}, \citenamefont {Whitmer},\ and\
  \citenamefont {{\noopsort{pablo}}{de Pablo}}}]{sidk+18jcp}%
  \BibitemOpen
  \bibfield  {author} {\bibinfo {author} {\bibfnamefont {H.}~\bibnamefont
  {Sidky}}, \bibinfo {author} {\bibfnamefont {Y.~J.}\ \bibnamefont
  {Col{\'o}n}}, \bibinfo {author} {\bibfnamefont {J.}~\bibnamefont
  {Helfferich}}, \bibinfo {author} {\bibfnamefont {B.~J.}\ \bibnamefont
  {Sikora}}, \bibinfo {author} {\bibfnamefont {C.}~\bibnamefont {Bezik}},
  \bibinfo {author} {\bibfnamefont {W.}~\bibnamefont {Chu}}, \bibinfo {author}
  {\bibfnamefont {F.}~\bibnamefont {Giberti}}, \bibinfo {author} {\bibfnamefont
  {A.~Z.}\ \bibnamefont {Guo}}, \bibinfo {author} {\bibfnamefont
  {X.}~\bibnamefont {Jiang}}, \bibinfo {author} {\bibfnamefont
  {J.}~\bibnamefont {Lequieu}}, \bibinfo {author} {\bibfnamefont
  {J.}~\bibnamefont {Li}}, \bibinfo {author} {\bibfnamefont {J.}~\bibnamefont
  {Moller}}, \bibinfo {author} {\bibfnamefont {M.~J.}\ \bibnamefont
  {Quevillon}}, \bibinfo {author} {\bibfnamefont {M.}~\bibnamefont {Rahimi}},
  \bibinfo {author} {\bibfnamefont {H.}~\bibnamefont {{Ramezani-Dakhel}}},
  \bibinfo {author} {\bibfnamefont {V.~S.}\ \bibnamefont {Rathee}}, \bibinfo
  {author} {\bibfnamefont {D.~R.}\ \bibnamefont {Reid}}, \bibinfo {author}
  {\bibfnamefont {E.}~\bibnamefont {Sevgen}}, \bibinfo {author} {\bibfnamefont
  {V.}~\bibnamefont {Thapar}}, \bibinfo {author} {\bibfnamefont {M.~A.}\
  \bibnamefont {Webb}}, \bibinfo {author} {\bibfnamefont {J.~K.}\ \bibnamefont
  {Whitmer}}, \ and\ \bibinfo {author} {\bibfnamefont {J.~J.}\ \bibnamefont
  {{\noopsort{pablo}}{de Pablo}}},\ }\href {\doibase 10.1063/1.5008853}
  {\bibfield  {journal} {\bibinfo  {journal} {The Journal of Chemical Physics}\
  }\textbf {\bibinfo {volume} {148}},\ \bibinfo {pages} {044104} (\bibinfo
  {year} {2018})}\BibitemShut {NoStop}%
\bibitem [{\citenamefont {Zhang}\ \emph {et~al.}(2021)\citenamefont {Zhang},
  \citenamefont {Xia},\ and\ \citenamefont {Jiang}}]{zhan+21prl}%
  \BibitemOpen
  \bibfield  {author} {\bibinfo {author} {\bibfnamefont {Y.}~\bibnamefont
  {Zhang}}, \bibinfo {author} {\bibfnamefont {J.}~\bibnamefont {Xia}}, \ and\
  \bibinfo {author} {\bibfnamefont {B.}~\bibnamefont {Jiang}},\ }\href
  {\doibase 10.1103/PhysRevLett.127.156002} {\bibfield  {journal} {\bibinfo
  {journal} {Phys. Rev. Lett.}\ }\textbf {\bibinfo {volume} {127}},\ \bibinfo
  {pages} {156002} (\bibinfo {year} {2021})}\BibitemShut {NoStop}%
\bibitem [{\citenamefont {Artrith}\ \emph {et~al.}(2011)\citenamefont
  {Artrith}, \citenamefont {Morawietz},\ and\ \citenamefont
  {Behler}}]{artr+11prb}%
  \BibitemOpen
  \bibfield  {author} {\bibinfo {author} {\bibfnamefont {N.}~\bibnamefont
  {Artrith}}, \bibinfo {author} {\bibfnamefont {T.}~\bibnamefont {Morawietz}},
  \ and\ \bibinfo {author} {\bibfnamefont {J.}~\bibnamefont {Behler}},\ }\href
  {\doibase 10.1103/PhysRevB.83.153101} {\bibfield  {journal} {\bibinfo
  {journal} {Phys. Rev. B}\ }\textbf {\bibinfo {volume} {83}},\ \bibinfo
  {pages} {153101} (\bibinfo {year} {2011})}\BibitemShut {NoStop}%
\bibitem [{\citenamefont {Bereau}\ \emph {et~al.}(2018)\citenamefont {Bereau},
  \citenamefont {DiStasio}, \citenamefont {Tkatchenko},\ and\ \citenamefont
  {{\noopsort{lilienfeld}}{von Lilienfeld}}}]{bere+18jcp}%
  \BibitemOpen
  \bibfield  {author} {\bibinfo {author} {\bibfnamefont {T.}~\bibnamefont
  {Bereau}}, \bibinfo {author} {\bibfnamefont {R.~A.}\ \bibnamefont
  {DiStasio}}, \bibinfo {author} {\bibfnamefont {A.}~\bibnamefont
  {Tkatchenko}}, \ and\ \bibinfo {author} {\bibfnamefont {O.~A.}\ \bibnamefont
  {{\noopsort{lilienfeld}}{von Lilienfeld}}},\ }\href {\doibase
  10.1063/1.5009502} {\bibfield  {journal} {\bibinfo  {journal} {The Journal of
  Chemical Physics}\ }\textbf {\bibinfo {volume} {148}},\ \bibinfo {pages}
  {241706} (\bibinfo {year} {2018})}\BibitemShut {NoStop}%
\bibitem [{\citenamefont {Ghasemi}\ \emph {et~al.}(2015)\citenamefont
  {Ghasemi}, \citenamefont {Hofstetter}, \citenamefont {Saha},\ and\
  \citenamefont {Goedecker}}]{ghas+15prb}%
  \BibitemOpen
  \bibfield  {author} {\bibinfo {author} {\bibfnamefont {S.~A.}\ \bibnamefont
  {Ghasemi}}, \bibinfo {author} {\bibfnamefont {A.}~\bibnamefont {Hofstetter}},
  \bibinfo {author} {\bibfnamefont {S.}~\bibnamefont {Saha}}, \ and\ \bibinfo
  {author} {\bibfnamefont {S.}~\bibnamefont {Goedecker}},\ }\href {\doibase
  10.1103/PhysRevB.92.045131} {\bibfield  {journal} {\bibinfo  {journal} {Phys.
  Rev. B}\ }\textbf {\bibinfo {volume} {92}},\ \bibinfo {pages} {045131}
  (\bibinfo {year} {2015})}\BibitemShut {NoStop}%
\bibitem [{\citenamefont {Niblett}\ \emph {et~al.}(2021)\citenamefont
  {Niblett}, \citenamefont {Galib},\ and\ \citenamefont {Limmer}}]{nibl+21jcp}%
  \BibitemOpen
  \bibfield  {author} {\bibinfo {author} {\bibfnamefont {S.~P.}\ \bibnamefont
  {Niblett}}, \bibinfo {author} {\bibfnamefont {M.}~\bibnamefont {Galib}}, \
  and\ \bibinfo {author} {\bibfnamefont {D.~T.}\ \bibnamefont {Limmer}},\
  }\href {\doibase 10.1063/5.0067565} {\bibfield  {journal} {\bibinfo
  {journal} {J. Chem. Phys.}\ }\textbf {\bibinfo {volume} {155}},\ \bibinfo
  {pages} {164101} (\bibinfo {year} {2021})}\BibitemShut {NoStop}%
\bibitem [{\citenamefont {Gao}\ and\ \citenamefont
  {Remsing}(2022)}]{gao-rems22nc}%
  \BibitemOpen
  \bibfield  {author} {\bibinfo {author} {\bibfnamefont {A.}~\bibnamefont
  {Gao}}\ and\ \bibinfo {author} {\bibfnamefont {R.~C.}\ \bibnamefont
  {Remsing}},\ }\href {\doibase 10.1038/s41467-022-29243-2} {\bibfield
  {journal} {\bibinfo  {journal} {Nat Commun}\ }\textbf {\bibinfo {volume}
  {13}},\ \bibinfo {pages} {1572} (\bibinfo {year} {2022})}\BibitemShut
  {NoStop}%
\bibitem [{\citenamefont {Grisafi}\ and\ \citenamefont
  {Ceriotti}(2019)}]{gris-ceri19jcp}%
  \BibitemOpen
  \bibfield  {author} {\bibinfo {author} {\bibfnamefont {A.}~\bibnamefont
  {Grisafi}}\ and\ \bibinfo {author} {\bibfnamefont {M.}~\bibnamefont
  {Ceriotti}},\ }\href {\doibase 10.1063/1.5128375} {\bibfield  {journal}
  {\bibinfo  {journal} {J. Chem. Phys.}\ }\textbf {\bibinfo {volume} {151}},\
  \bibinfo {pages} {204105} (\bibinfo {year} {2019})}\BibitemShut {NoStop}%
\bibitem [{\citenamefont {Grisafi}\ \emph {et~al.}(2021)\citenamefont
  {Grisafi}, \citenamefont {Nigam},\ and\ \citenamefont
  {Ceriotti}}]{gris+21cs}%
  \BibitemOpen
  \bibfield  {author} {\bibinfo {author} {\bibfnamefont {A.}~\bibnamefont
  {Grisafi}}, \bibinfo {author} {\bibfnamefont {J.}~\bibnamefont {Nigam}}, \
  and\ \bibinfo {author} {\bibfnamefont {M.}~\bibnamefont {Ceriotti}},\ }\href
  {\doibase 10.1039/D0SC04934D} {\bibfield  {journal} {\bibinfo  {journal}
  {Chem. Sci.}\ }\textbf {\bibinfo {volume} {12}},\ \bibinfo {pages} {2078}
  (\bibinfo {year} {2021})}\BibitemShut {NoStop}%
\bibitem [{\citenamefont {Qiao}\ \emph {et~al.}(2020)\citenamefont {Qiao},
  \citenamefont {Welborn}, \citenamefont {Anandkumar}, \citenamefont {Manby},\
  and\ \citenamefont {Miller}}]{qiao+20jcp}%
  \BibitemOpen
  \bibfield  {author} {\bibinfo {author} {\bibfnamefont {Z.}~\bibnamefont
  {Qiao}}, \bibinfo {author} {\bibfnamefont {M.}~\bibnamefont {Welborn}},
  \bibinfo {author} {\bibfnamefont {A.}~\bibnamefont {Anandkumar}}, \bibinfo
  {author} {\bibfnamefont {F.~R.}\ \bibnamefont {Manby}}, \ and\ \bibinfo
  {author} {\bibfnamefont {T.~F.}\ \bibnamefont {Miller}},\ }\href {\doibase
  10.1063/5.0021955} {\bibfield  {journal} {\bibinfo  {journal} {J. Chem.
  Phys.}\ }\textbf {\bibinfo {volume} {153}},\ \bibinfo {pages} {124111}
  (\bibinfo {year} {2020})}\BibitemShut {NoStop}%
\bibitem [{\citenamefont {Fabrizio}\ \emph {et~al.}(2022)\citenamefont
  {Fabrizio}, \citenamefont {Briling},\ and\ \citenamefont
  {Corminboeuf}}]{fabr+22dd}%
  \BibitemOpen
  \bibfield  {author} {\bibinfo {author} {\bibfnamefont {A.}~\bibnamefont
  {Fabrizio}}, \bibinfo {author} {\bibfnamefont {K.~R.}\ \bibnamefont
  {Briling}}, \ and\ \bibinfo {author} {\bibfnamefont {C.}~\bibnamefont
  {Corminboeuf}},\ }\href {\doibase 10.1039/D1DD00050K} {\bibfield  {journal}
  {\bibinfo  {journal} {Digital Discovery}\ ,\ \bibinfo {pages}
  {10.1039.D1DD00050K}} (\bibinfo {year} {2022})}\BibitemShut {NoStop}%
\bibitem [{\citenamefont {Brockherde}\ \emph {et~al.}(2017)\citenamefont
  {Brockherde}, \citenamefont {Vogt}, \citenamefont {Li}, \citenamefont
  {Tuckerman}, \citenamefont {Burke},\ and\ \citenamefont
  {M{\"u}ller}}]{broc+17nc}%
  \BibitemOpen
  \bibfield  {author} {\bibinfo {author} {\bibfnamefont {F.}~\bibnamefont
  {Brockherde}}, \bibinfo {author} {\bibfnamefont {L.}~\bibnamefont {Vogt}},
  \bibinfo {author} {\bibfnamefont {L.}~\bibnamefont {Li}}, \bibinfo {author}
  {\bibfnamefont {M.~E.}\ \bibnamefont {Tuckerman}}, \bibinfo {author}
  {\bibfnamefont {K.}~\bibnamefont {Burke}}, \ and\ \bibinfo {author}
  {\bibfnamefont {K.~R.}\ \bibnamefont {M{\"u}ller}},\ }\href {\doibase
  10.1038/s41467-017-00839-3} {\bibfield  {journal} {\bibinfo  {journal} {Nat.
  Commun.}\ }\textbf {\bibinfo {volume} {8}},\ \bibinfo {pages} {872} (\bibinfo
  {year} {2017})}\BibitemShut {NoStop}%
\bibitem [{\citenamefont {Kalita}\ \emph {et~al.}(2021)\citenamefont {Kalita},
  \citenamefont {Li}, \citenamefont {McCarty},\ and\ \citenamefont
  {Burke}}]{kali+21acr}%
  \BibitemOpen
  \bibfield  {author} {\bibinfo {author} {\bibfnamefont {B.}~\bibnamefont
  {Kalita}}, \bibinfo {author} {\bibfnamefont {L.}~\bibnamefont {Li}}, \bibinfo
  {author} {\bibfnamefont {R.~J.}\ \bibnamefont {McCarty}}, \ and\ \bibinfo
  {author} {\bibfnamefont {K.}~\bibnamefont {Burke}},\ }\href {\doibase
  10.1021/acs.accounts.0c00742} {\bibfield  {journal} {\bibinfo  {journal}
  {Acc. Chem. Res.}\ ,\ \bibinfo {pages} {acs.accounts.0c00742}} (\bibinfo
  {year} {2021})}\BibitemShut {NoStop}%
\bibitem [{\citenamefont {Kirkpatrick}\ \emph {et~al.}(2021)\citenamefont
  {Kirkpatrick}, \citenamefont {McMorrow}, \citenamefont {Turban},
  \citenamefont {Gaunt}, \citenamefont {Spencer}, \citenamefont {Matthews},
  \citenamefont {Obika}, \citenamefont {Thiry}, \citenamefont {Fortunato},
  \citenamefont {Pfau}, \citenamefont {Castellanos}, \citenamefont {Petersen},
  \citenamefont {Nelson}, \citenamefont {Kohli}, \citenamefont
  {{Mori-S{\'a}nchez}}, \citenamefont {Hassabis},\ and\ \citenamefont
  {Cohen}}]{kirk+21science}%
  \BibitemOpen
  \bibfield  {author} {\bibinfo {author} {\bibfnamefont {J.}~\bibnamefont
  {Kirkpatrick}}, \bibinfo {author} {\bibfnamefont {B.}~\bibnamefont
  {McMorrow}}, \bibinfo {author} {\bibfnamefont {D.~H.~P.}\ \bibnamefont
  {Turban}}, \bibinfo {author} {\bibfnamefont {A.~L.}\ \bibnamefont {Gaunt}},
  \bibinfo {author} {\bibfnamefont {J.~S.}\ \bibnamefont {Spencer}}, \bibinfo
  {author} {\bibfnamefont {A.~G. D.~G.}\ \bibnamefont {Matthews}}, \bibinfo
  {author} {\bibfnamefont {A.}~\bibnamefont {Obika}}, \bibinfo {author}
  {\bibfnamefont {L.}~\bibnamefont {Thiry}}, \bibinfo {author} {\bibfnamefont
  {M.}~\bibnamefont {Fortunato}}, \bibinfo {author} {\bibfnamefont
  {D.}~\bibnamefont {Pfau}}, \bibinfo {author} {\bibfnamefont {L.~R.}\
  \bibnamefont {Castellanos}}, \bibinfo {author} {\bibfnamefont
  {S.}~\bibnamefont {Petersen}}, \bibinfo {author} {\bibfnamefont {A.~W.~R.}\
  \bibnamefont {Nelson}}, \bibinfo {author} {\bibfnamefont {P.}~\bibnamefont
  {Kohli}}, \bibinfo {author} {\bibfnamefont {P.}~\bibnamefont
  {{Mori-S{\'a}nchez}}}, \bibinfo {author} {\bibfnamefont {D.}~\bibnamefont
  {Hassabis}}, \ and\ \bibinfo {author} {\bibfnamefont {A.~J.}\ \bibnamefont
  {Cohen}},\ }\href {\doibase 10.1126/science.abj6511} {\bibfield  {journal}
  {\bibinfo  {journal} {Science}\ }\textbf {\bibinfo {volume} {374}},\ \bibinfo
  {pages} {1385} (\bibinfo {year} {2021})}\BibitemShut {NoStop}%
\end{thebibliography}
\end{document}